\documentclass[%
 reprint,
 amsmath,amssymb,
 aps,
floatfix,
]{revtex4-2}

\usepackage[utf8]{inputenc}
\usepackage{graphicx}
\usepackage{dcolumn}
\usepackage{bm}
\usepackage{bbold}
\usepackage{physics}
\usepackage{subcaption}
\usepackage{caption}

\begin{document}

\preprint{APS/123-QED}
\title{Non-monotonic flow variations in a TASEP-based traffic model featuring cars searching for parking}

\author{Valentin ANFRAY}
\email{valentin.anfray@gmail.com}
\author{Alexandre NICOLAS}%
 \email{alexandre.nicolas@cnrs.fr}
\affiliation{%
Institut Lumi\`ere Mati\`ere, CNRS, Univ. Lyon 1, 6 rue Ada Byron, F-69622 Villeurbanne, France.
}%

\date{\today}

\begin{abstract}
The Totally Asymmetric Simple Exclusion Process (TASEP) is a paradigm of out-of-equilibrium Statistical Physics that serves as a simplistic model for one-way vehicular traffic. Since traffic is perturbed by cars cruising for parking in many metropolises, we introduce a variant of TASEP, dubbed SFP, in which particles are initially cruising at a slower speed and aiming to park on one of the sites adjacent to the main road, described by a unidimensional lattice. After parking, they pull out at a finite rate and move at a normal speed. 

We show that this model, which breaks many of the conservation rules applicable in other TASEP variants, exhibits singular features, in particular non-monotonic variations of the steady-state current with the injection rate and re-entrant transitions in the phase diagram, for some range of parameters. These features are robust to variations in the update rule and the boundary conditions.

Neither the slow speed of cruising cars nor the perturbation of the flow due to pull-out maneuvers, taken in isolation, can rationalize these observations. Instead, they originate in a  cramming (or `paper jam') effect which results from the coupling of these mechanisms: 
injecting too many cars into the system saturates the first sites of the road, which prevents parked cars from pulling out, thus forcing cruising cars to travel farther along the road.

These strong discrepancies with even the qualitative trends of the baseline TASEP model highlight the importance of considering the effect of perturbations on traffic.
\end{abstract}

\maketitle


\section{Introduction}

An inseparable, but often cumbersome companion of urban life, vehicular traffic appeals to the physicist because of the original combination of 
regular, predictable features and haphazard perturbations that govern the motion of its microscopic constituents (i.e., cars) and the collective effects that emerge out of this combination, such as congestion, traffic jams, or stop-and-go waves \cite{kerner1999physics,schadschneider_stochastic_2010}. 
It is an inherently out-of-equilibrium system made of interacting constituents \cite{schutz_exactly_2000, schadschneider_stochastic_2010} and, as such, it has been studied using
a wide variety of statistical physical approaches \cite{wolff_collective_1989, chowdhury_statistical_2000, schadschneider_traffic_2002, seo_traffic_2017}, either operating on macroscopic quantities as hydrodynamics models \cite{lighthill_kinematic_1955, kerner_cluster_1993, helbing_improved_1995} or resorting to a microscopic description of cars, for instance, optimal velocity models \cite{bando_dynamical_1995} or particle-hopping models \cite{nagel_cellular_1992, schreckenberg_discrete_1995, nagel_particle_1996, rickert_two_1996} among many others. 

\begin{figure}
    \centering
    \includegraphics[width=\columnwidth]{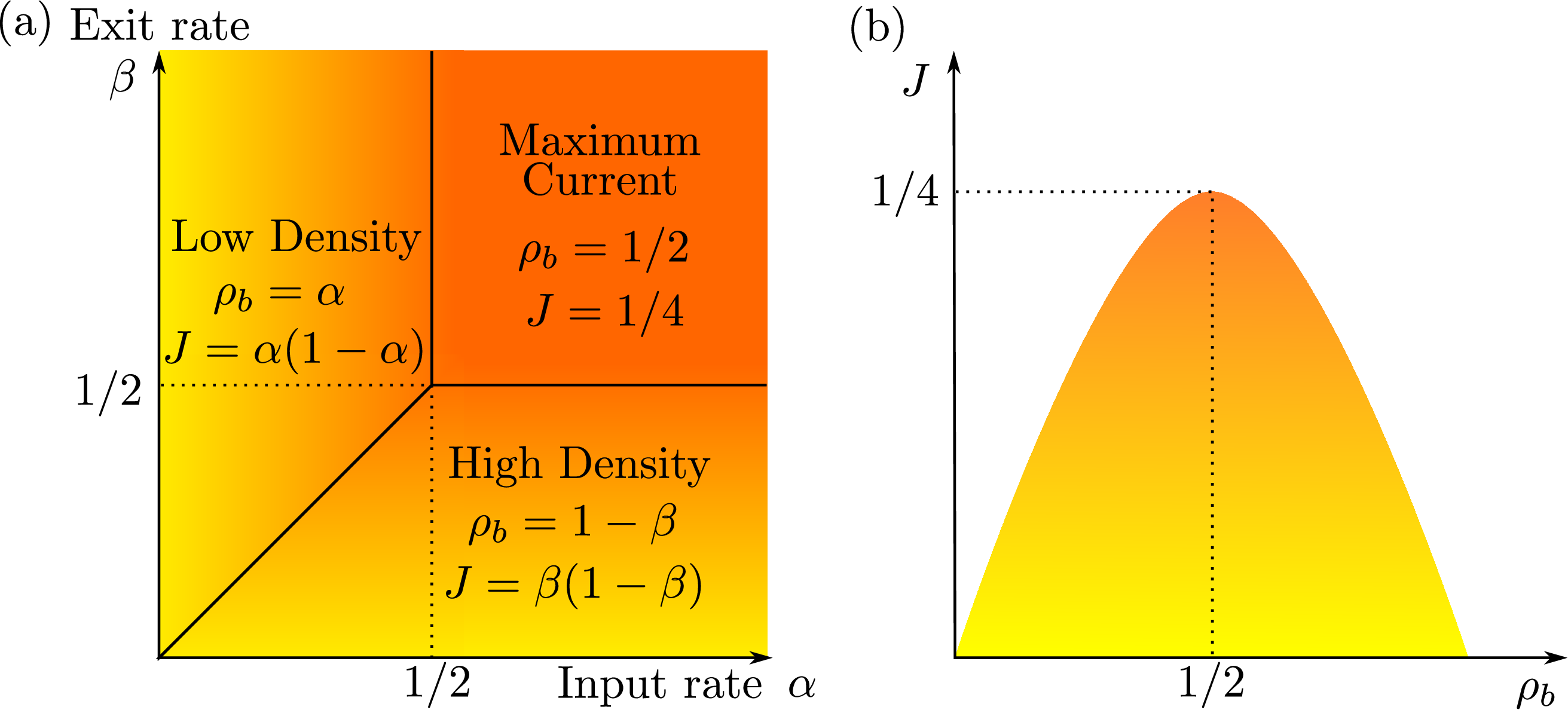}
    \caption{(a) Phase diagram of the classical TASEP model; adapted from \cite{derrida_exact_1993}. The three phases are low density (LD), high density (HD), and maximum current (MC), with different relations between the current $J$ and the density $\rho_b$ in the bulk. The color overlay represents the current, as in the right panel. (b) Fundamental diagram $J=J(\rho_b)$ of TASEP.}
    \label{fig:DiagPhaseTASEP}
\end{figure}

Among the microscopic models, the Totally Asymmetric Exclusion Process (TASEP) can be regarded as the analogue  of the `ideal gas' for traffic.
It is a minimal stochastic model wherein particles move in a given direction on a one-dimensional lattice and can only hop to the neighbouring site if it is empty. Nonetheless, it qualitatively captures some salient collective features of car traffic, such as the existence of backward-propagating waves and the general form of the fundamental diagram (Fig.~\ref{fig:DiagPhaseTASEP}b) relating the flow rate $J$ to the density $\rho$, which rises more or less linearly at low density and has a peak at a critical density, past which it decays because of congestion. Thanks to its minimalism, it is amenable to analytical resolution through various methods \cite{derrida_exact_1992, derrida_exact_1993, schutz_phase_1993, rajewsky_asymmetric_1997, de_gier_bethe_2005}. Unlike an extended system at equilibrium, its properties in the bulk (far from the edges) sensitively depend on the boundary conditions, which are given the entry rate $\alpha$ of injected particles at one end and exit rate $\beta$ of removed particles at the other end, for open boundary conditions. In particular, its `phase' diagram (Fig.~\ref{fig:DiagPhaseTASEP}a) depends solely on $\alpha$ and $\beta$; it consists of a
low density (LD) phase separated from a high density (HD) phase by a first-order transition and from a maximum current (MC) phase by a second-order transition.

To bring this `ideal-world' description closer to real scenarios, a large number of 
 variations of TASEP have been put forward and their impact on the flow properties have been inspected in detail. These variations include multi-lane systems \cite{brankov_totally_2004, pronina_two-channel_2004, mitsudo_synchronization_2005, pronina_spontaneous_2007, tsekouras_inhomogeneous_2008}, systems with a shortcut \cite{yuan_totally_2007, bunzarova_asymmetric_2014, xiao_shortcut_2017}, with particles creation and annihilation \cite{evans_shock_2003, jiang_two-lane_2007, wang_effects_2007, gupta_asymmetric_2014, botto_dynamical_2019}, with many species \cite{evans_asymmetric_1995, schutz_critical_2003, muhuri_collective_2008, ayyer_classes_2010, crampe_matrix_2016, bottero_analysis_2017, bonnin_two-species_2022} or with inhomogeneities \cite{kolomeisky_asymmetric_1998, ha_queuing_2003, chou_clustered_2004, pierobon_bottleneck-induced_2006, dong_towards_2007, greulich_phase_2008, schmidt_defect-induced_2015}. 

 Among the perturbations that affect urban traffic, the cruising traffic, made of cars in search of a parking space, 
is widely believed to play a substantial role in many large cities: studies in various locations over the years found proportions of cruising cars in the total through-traffic amounting to anywhere between 8 and 74 percent \cite{shoup_cruising_2006}, 
contributing to congestion and pollution and increasing travel times (e.g., by some ten percent in  \cite{dowling2019modeling}). 
While parking garages have already been considered in variants of TASEP \cite{ha_macroscopic_2002, adams_far--equilibrium_2008, cook_feedback_2009, cook_competition_2009, greulich_mixed_2012}, often playing the role of a reservoir of cars, in this paper the focus is put on curbside parking. Our aim is to determine whether emulating the process of searching for parking on the curb can \emph{qualitatively} alter predictions based on TASEP, or whether this source of perturbations only affects the \emph{quantitative} output.

To this end, we introduce and study a new TASEP variant, which features cars searching for parking (SFP) on the curb of the main street. As a matter of fact, parking search is actually a complex process involving a number of
determinants related to the parking supply in the city as well as to the driver's search behavior \cite{levy_exploring_2013, xiao_how_2018, fulman_modeling_2020, lubrich_analysis_2023, dutta_parking_2023}; the many investigations incorporating this phenomenon, notably numerically, have fallen short of unveiling its basic effect on the traffic dynamics. Here, in the spirit of a gradual departure from an `ideal' (non-perturbed) traffic theory, SFP is reduced to its core: slow (S) cars searching for parking aim to park (\emph{pull in}) along a linear road, and parked (P) cars may leave their space (\emph{pull out}), then turning into fast (F) cars that will move forward but not park anymore. It is noteworthy that these extensions already undermine common theoretical methods based on the strictly linear geometry (e.g., in multi-lane TASEP \cite{brankov_totally_2004, pronina_two-channel_2004, mitsudo_synchronization_2005, pronina_spontaneous_2007, tsekouras_inhomogeneous_2008}) or the conservation of the number of each particle species (e.g. in TASEP with two  species \cite{evans_asymmetric_1995, schutz_critical_2003, muhuri_collective_2008, ayyer_classes_2010,crampe_matrix_2016, bonnin_two-species_2022, bottero_analysis_2017}).
In particular, compared to interesting recent developments in TASEP variants endowing each site with a finite `side-pocket' to store particles, either motivated by biological applications \cite{humenyuk_separation_2020} or mimicking parking spots or alleys \cite{bhatia_far_2022}, a key feature here is the distinct behavior of particles before they park (S) and after they leave their parking spot (F).
Within SFP, we find an intriguing non-monotonic variation of the flow rate $J$ with the injection rate $\alpha$, contrasting the always monotonic trends found in TASEP:  in a range of parameters, the higher the rate of possible injections of cars, the lower the flow. This leads to a re-entrant transition into the low-density phase of the phase diagram. We ascribe these numerical observations to a `cramming effect' which prevents parked cars from getting back on the road. 
(This mechanism strongly differs from those at play in the few previous reports of re-entrant transitions,  or  back-and-forth effects, in TASEP variants, which involve a shock phase \cite{brackley_multiple_2012, verma_stochastic_2019, banerjee_smooth_2020, s_role_2022, s_multiple_2023}.)


 The rest of the paper is organized as follows. After exposing the rules of the SFP model in Sec.~\ref{section:model} and flashing the singularities of the behavior of the steady state of this model compared to TASEP in Sec.~\ref{section:novel_behavior}, we detail the methods that were used in Sec.~\ref{section:methodology}, namely, kinetic Monte-Carlo simulation, exact diagonalization of the stochastic matrix and solving the set of differential equations within a mean-field approximation, as well as variants of implementation of the model. Finally, in Sec.~\ref{section:explanation}, we present theoretical arguments and numerical explanations that help understand the singular behavior that is observed and, more broadly, delineate the steady-state properties of the SFP model in general. 

\section{Model}
\label{section:model}
\begin{figure}
    \centering
    \includegraphics[width=0.95\columnwidth]{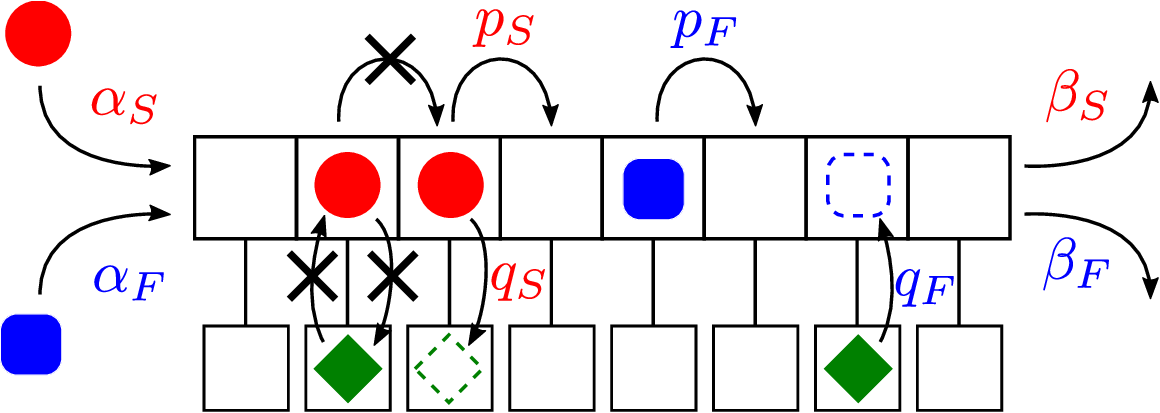}
    \caption{Schematic diagram of the dynamics of the SFP model. Red disks represent particles of species $S$, blue squares are particles of species $F$ and green diamonds are particles of species $P$. Dashed edges denote configurations after a possible transition. }
    \label{fig:Schema_parking_ABC}
\end{figure}

The SFP model is based on a one-dimensional lattice of $L \gg 1$ sites, which represents a road (e.g., secondary road, or local road, or sub-local road), populated by two distinct species denoted as $S$ (cars searching for parking) and $F$ (faster cars driving towards the exit). Adjacent to each lattice site is an additional site corresponding to a parking space, as depicted in Fig.~\ref{fig:Schema_parking_ABC}. The particles occupying these parking spaces are referred to as species $P$.  Each road site is either vacant or occupied by an $S$ or an $F$  particle. The parking sites are either vacant or occupied by a $P$ (parked) particle. The $S$ and $F$ species differ by their hopping rates on the road ($p_S, p_F$), the `injection' rates ($\alpha_S, \alpha_F$) at which they are injected into the first site if vacant and the `exit' rates ($\beta_S, \beta_F$) at which they are removed from the last road site. As we always assume $p_S \leq p_F$, particles $S$ are referred as slow particles and particles $F$ as fast ones. Moreover, $S$ particles can jump with a rate $q_S$ on the neighbouring parking place if vacant. The particle then turns into $P$. $P$  particles can jump with rate $q_F$ on the adjacent road site if vacant (hence, a mean parking duration $q_F^{-1}$), in which case they turn into $F$. We have used the random update rule i.e. at each time step a particle is selected at random and may move with some probability, if possible. 

The time unit is set such that $p_F=1$.  We also set $\beta_S=\beta_F=\beta$ (equal exit rates) and $\alpha_F = 0$ (all injected cars belong to species $S$) since our focus is on the novel steady-state dynamics related to cars cruising for parking. For large system sizes $L$, only $F$ particles will leave the system, making $\beta_S$ irrelevant; the exit rate
is thus simply referred to as $\beta$. All in all, the proposed model involves five independent variables: $p_S,\, q_S,\, q_F, \,\alpha_S$, and $\beta$.

\section{In a nutshell: Main results about the singular behavior of the SFP model}
\label{section:novel_behavior}

This concise section flashes the new behaviors that are observed within SFP and that set it apart from both the traditional TASEP and its existing variants. For clarity, the broader exploration of the model and the theoretical rationalizations of our observations are deferred to Sec.~\ref{section:explanation}.

\subsection{Non-monotonic current as a function of the input rate}

\begin{figure}
    \centering
    \includegraphics[width =0.8\columnwidth]{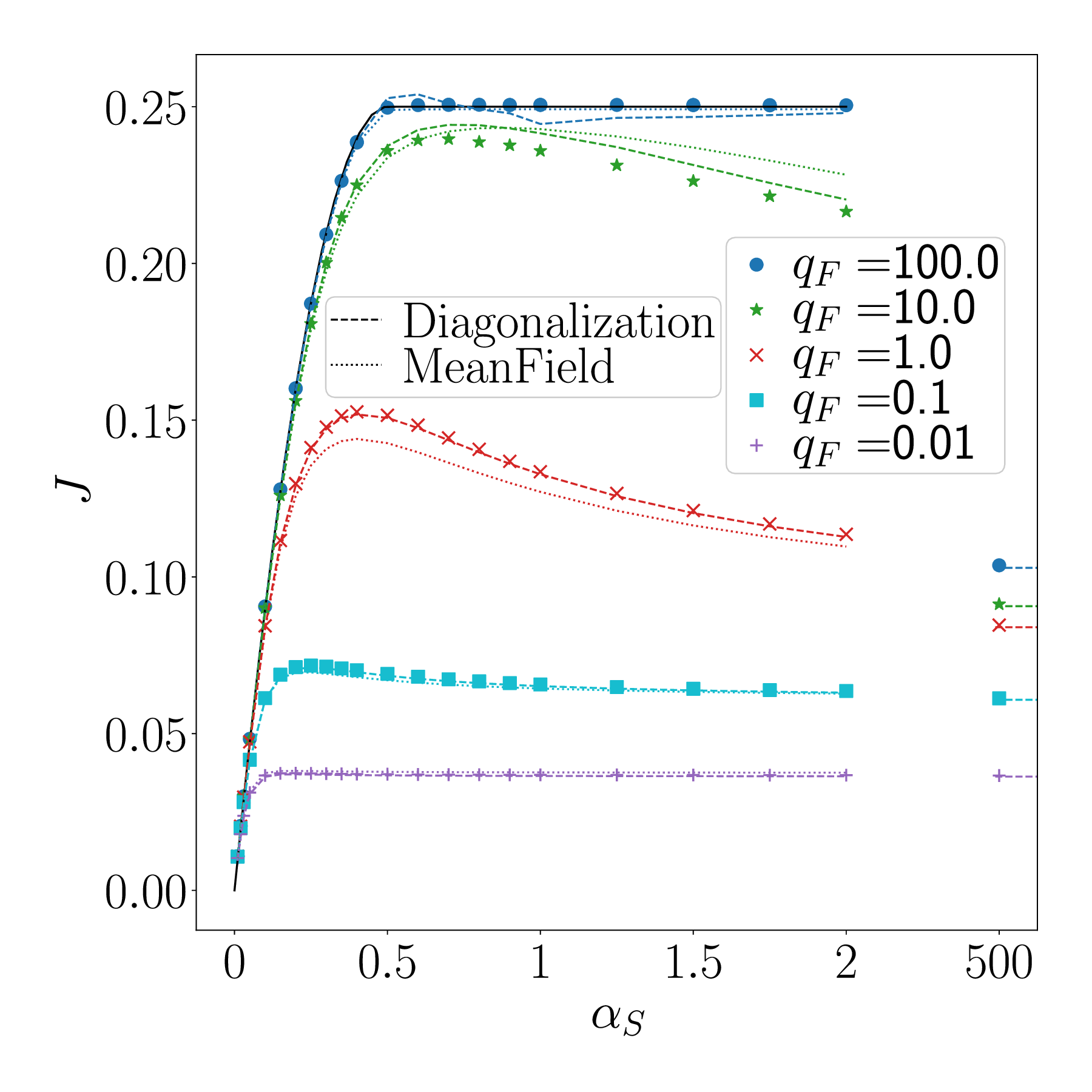}
    \caption{Variations of the current $J$ with the input rate $\alpha_S$ for different leaving rates $q_F$ (different colors). Exact diagonalization results (dashed lines) with $L=6$ fully agree with the kinetic Monte-Carlo simulations (symbols) with $L=1000$, except close to the Maximum Current phase. Mean-field results (dotted lines) for $L=500$ display stronger deviations. The classical TASEP current is shown in black, as a benchmark. Fixed parameters :  $p_S = 0.1$, $q_S = \infty, \beta = 0.6$.}
    \label{fig:Courant_fonctionAlphaA_MC_MF_ED_ParkingABC_pa_0-10_qaINF_betaB_0-60}
\end{figure}

One of the most striking properties of the SFP model is the non-monotonic behavior of the steady-state current $J$ as a function of the input rate $\alpha_S$, as shown in Fig.~\ref{fig:Courant_fonctionAlphaA_MC_MF_ED_ParkingABC_pa_0-10_qaINF_betaB_0-60}: at some point, increasing the input rate paradoxically reduces the current, although the TASEP dynamics on the road would suggest monotonic variations with the input rate.

The observed non-monotonicity has several implications. First, it implies that the current is maximized at at finite input rate $\alpha^*$.
Secondly, it leads to re-entrant transitions in the phase diagram. Indeed, in the thermodynamic limit $L \to \infty$, all particles $S$ will eventually turn into $F$ (if the rates are non-zero and system size independent). Then, sufficiently far from the entrance, the system corresponds to TASEP with an effective input rate $\alpha_{eff}$ of particles $F$. As the current $J$ is conserved, when the system is in the low density phase, this effective input rate yields a current

 \begin{equation}
     J = \alpha_{eff}(1-\alpha_{eff})\text{, hence } \alpha_{eff} = \frac{1-\sqrt{1-4J}}{2}.
 \end{equation}

\begin{figure}
\centering
\begin{subfigure}{.48\textwidth}
  \centering
    \includegraphics[width=0.8\textwidth]{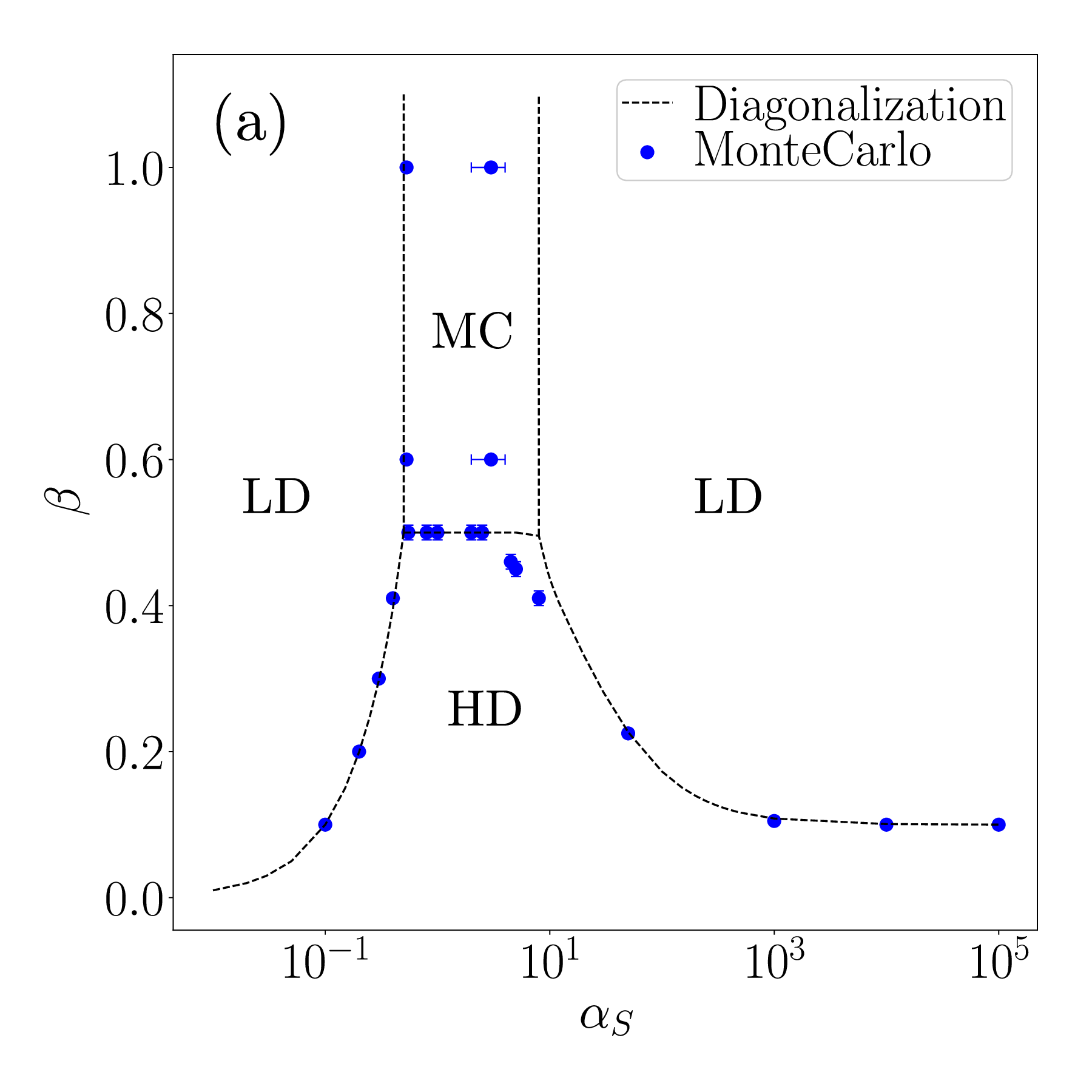}
    \phantomsubcaption
    \label{fig:DiagPhase_MC_ED_ParkingABC_qaINF_pa_0-10_qb_100-00_a}
\end{subfigure}
\begin{subfigure}{.48\textwidth}
  \centering
  \includegraphics[width=0.8\textwidth]{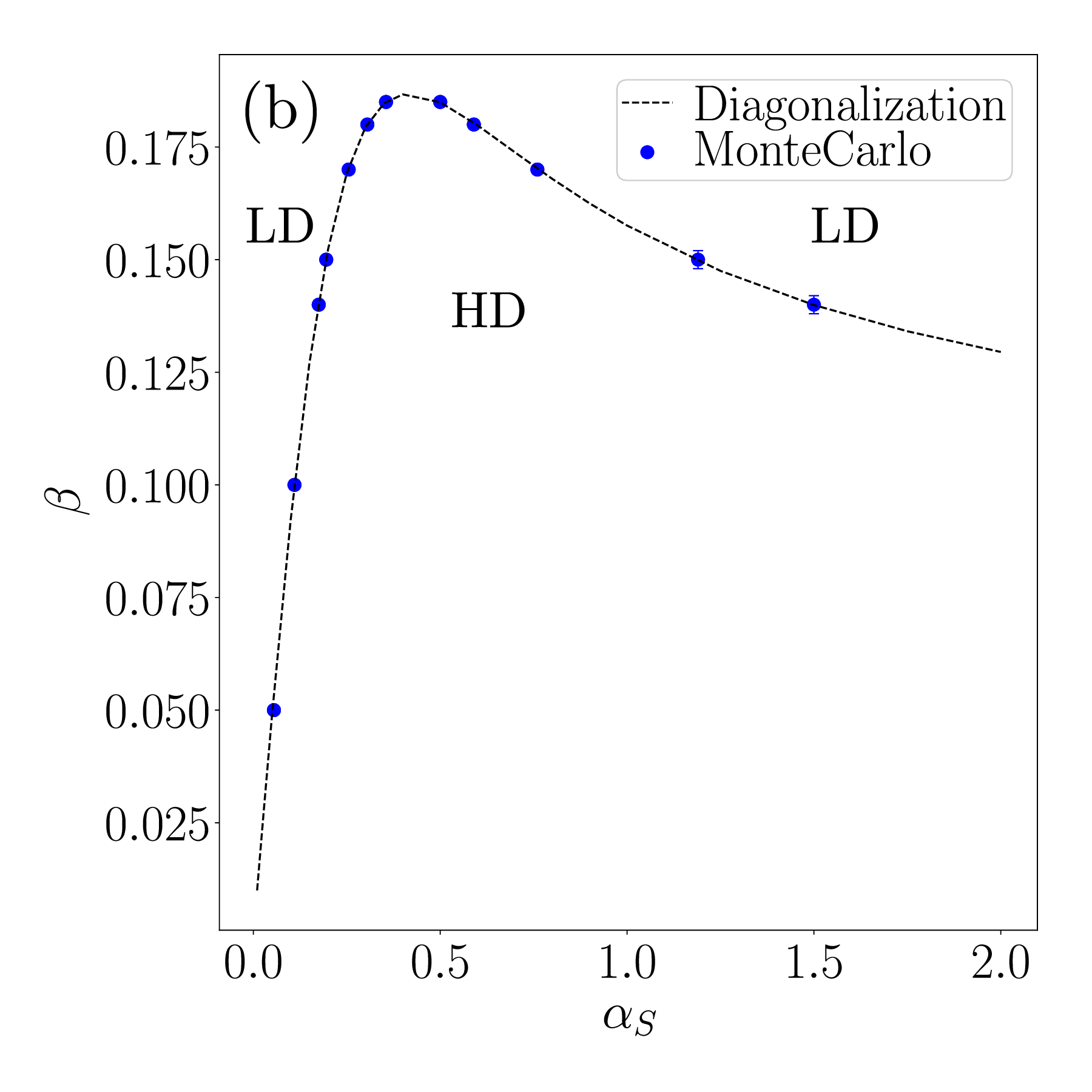}
  \phantomsubcaption
  \label{fig:DiagPhase_MC_ED_ParkingABC_qaINF_pa_0-10_qb_100-00_b}
\end{subfigure}
\hfill
\caption{SFP phase diagrams with a re-entrant transition, obtained for $q_S \rightarrow \infty$, $p_S = 0.1$ and (a) $q_F = 100$ (b) $q_F = 1$. Monte-Carlo simulation (blue dots) were performed with $L=1000$ and exact diagonalization (black dashed lines) with $L=6$.
}
    \label{fig:DiagPhase_MC_ED_ParkingABC_qaINF_pa_0-10_qb_100-00}
\end{figure}

A non-monotonic current $J(\alpha_S)$ gives rise to a non-monotonic $\alpha_{eff}(\alpha_S)$, hence the possibility of a strong departure from the TASEP phase diagram (Fig.~\ref{fig:DiagPhaseTASEP}a) consisting of a low-density (LD) phase, a high-density (HD) phase, and a maximum current (MC) phase. Starting from LD and increasing $\alpha_S$, in a first stage  $\alpha_{eff}$ may increase so that one crosses the LD-HD boundary, but then $\alpha_{eff}$ may decrease so that one crosses back from HD to LD. This is the hallmark of a LD-HD-LD re-entrant transition, illustrated in Fig.~\ref{fig:DiagPhase_MC_ED_ParkingABC_qaINF_pa_0-10_qb_100-00_b}. For some parameters, an additional re-entrant transition, from LD to MC and back to LD, may be found if one first encounters the LD-MC boundary as $\alpha_S$ increases (Fig. \ref{fig:DiagPhase_MC_ED_ParkingABC_qaINF_pa_0-10_qb_100-00_a}). 
The nature and origin of these re-entrant transitions markedly differ from
previous reports of such transitions, or  `back-and-forth' effects, in TASEP variants, either with limited resources or with two-lanes Langmuir-kinetics under specific settings \cite{brackley_multiple_2012, verma_stochastic_2019, banerjee_smooth_2020, s_role_2022, s_multiple_2023}. The latter involve a shock phase, whereeas re-entrant transitions in SFP are obtained between LD and MC or HD. 

\begin{figure*}
    \centering
    \includegraphics[width=0.7\textwidth]{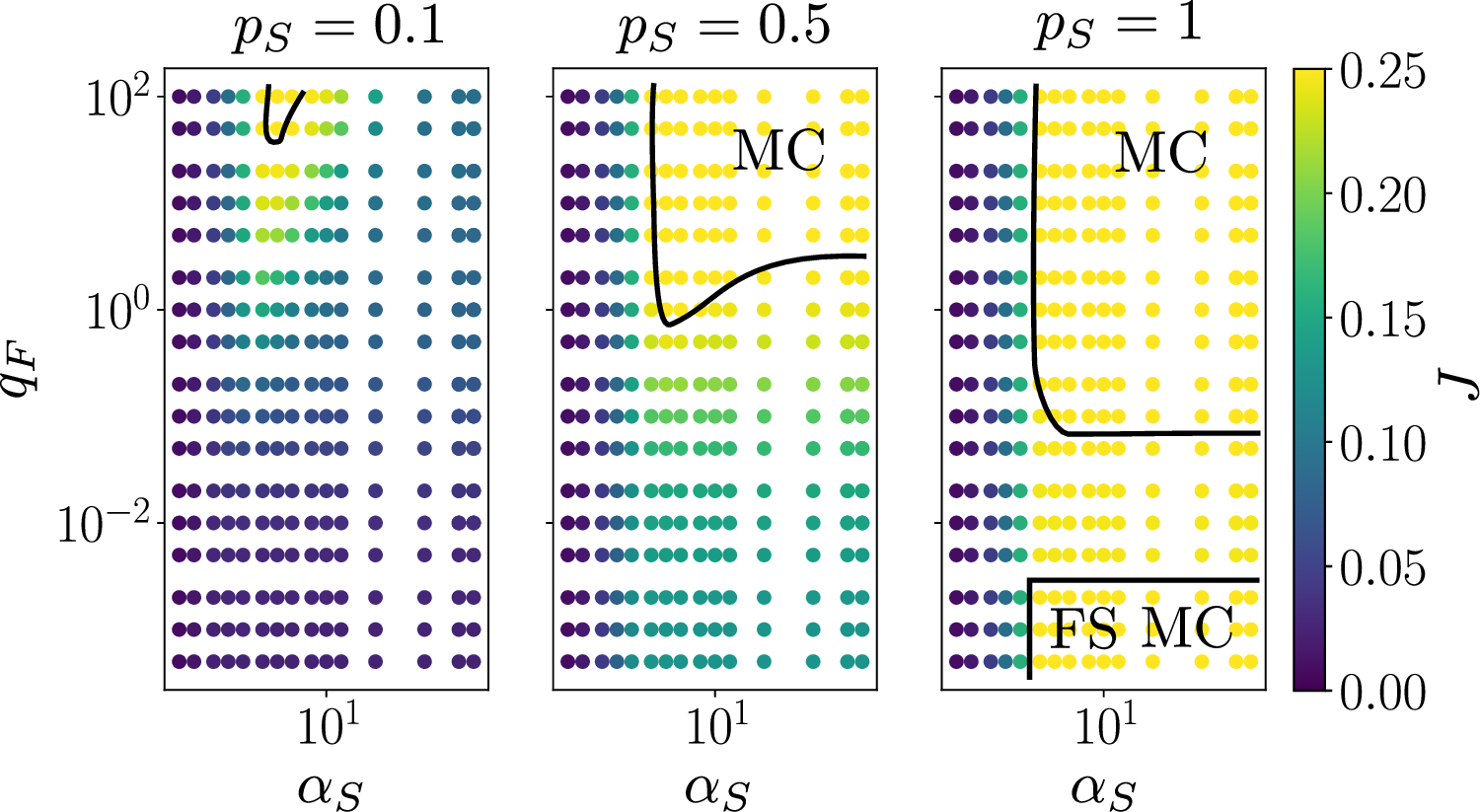}
    \caption{Heat map of the total current $J$ depending on $\alpha_S$ and $q_F$ for different hopping rate $p_S$, computed using Monte-Carlo simulations with $L = 200, q_S = +\infty$ and $\beta = 0.6$. When $p_S < 0.5$, the MC phase disappears for large $\alpha_S$ leading to re-entrant phase transition LD-MC-LD. For $p_S = 1$ and $\alpha_S > 0.5$, MC vanishes as $q_F$ gets lower and is only restored because of the finite size of the system (FS-MC). Boundaries between the LD and the MC phases (black line) were plotted by roughly tracking the iso-current line $J=0.25$. }
    \label{fig:ColorPlot_Courant_fct_alphaA_qb_plusieurs_pa_qaINF_betaB_0-60_L_200}
\end{figure*}

Interestingly, the non-monotonic behavior of the current is not limited to specific values of the hopping rate $p_S$. For instance, in the case of $p_S < 0.5$, while there exists no MC phase as $\alpha_S \rightarrow \infty$, such a phase is restored for intermediate $\alpha_S$ values, provided that the parking and leaving rates $q_S$ and $q_F$ are large, but finite (an equivalence with TASEP can then be found). This gives rise to an island of MC phase in the $\alpha_S - q_F$ plane for $\beta > 0.5$ and $q_S \gg 1$ as shown in Fig. \ref{fig:ColorPlot_Courant_fct_alphaA_qb_plusieurs_pa_qaINF_betaB_0-60_L_200} for $p_S < 0.5$. More generally, re-entrant phase transitions with respect to the input rate $\alpha_S$ are observed for a wide range of (finite) values of $q_S$ and $q_F$. Note, however, that this range depends on the hopping rate $p_S$: as $p_S$ gets close to $p_F=1$ (slow particles drive almost as fast as fast ones), this range shrinks.

\subsection{Non-monotonic current as a function of the parking and leaving rates}

Fig.~\ref{fig:ColorPlot_Courant_fct_alphaA_qb_plusieurs_pa_qaINF_betaB_0-60_L_200} shows that transitions can also be induced by adjusting the inverse parking duration $q_F$ while keeping the input rate constant, even for $p_S = 1$. In this scenario, for $\alpha_S \geq 0.5$, MC vanishes at low $q_F$ and pops back up when $q_F = 0$. For finite-size systems, MC is reinstated when $q_F$ is roughly less than $\frac{1}{2L}$. We term this behavior as the "Finite Size Maximum Current Phase" (FS-MC), for reasons that we will explain in Sec.~\ref{section:explanation}.

\begin{figure}
\centering
\begin{subfigure}{.48\textwidth}
  \centering
    \includegraphics[width=0.8\textwidth]{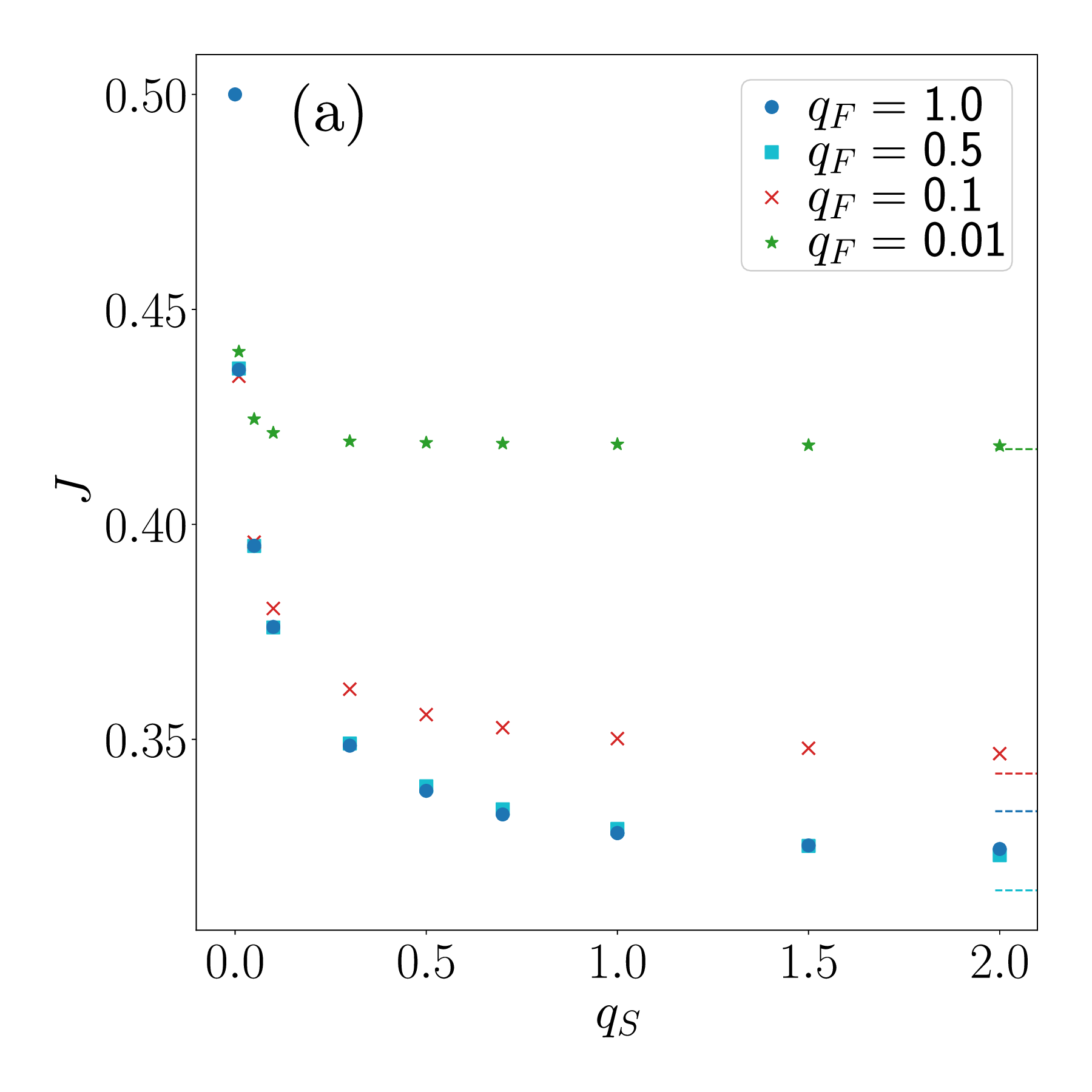}
    \phantomsubcaption
    \label{fig:Current_parallel_fct_qA_different_qb_MC_ParkingABC_pa_1-00_alphaA_1-00_beta_1-00_a}
\end{subfigure}
\begin{subfigure}{.48\textwidth}
  \centering
  \includegraphics[width=0.8\textwidth]{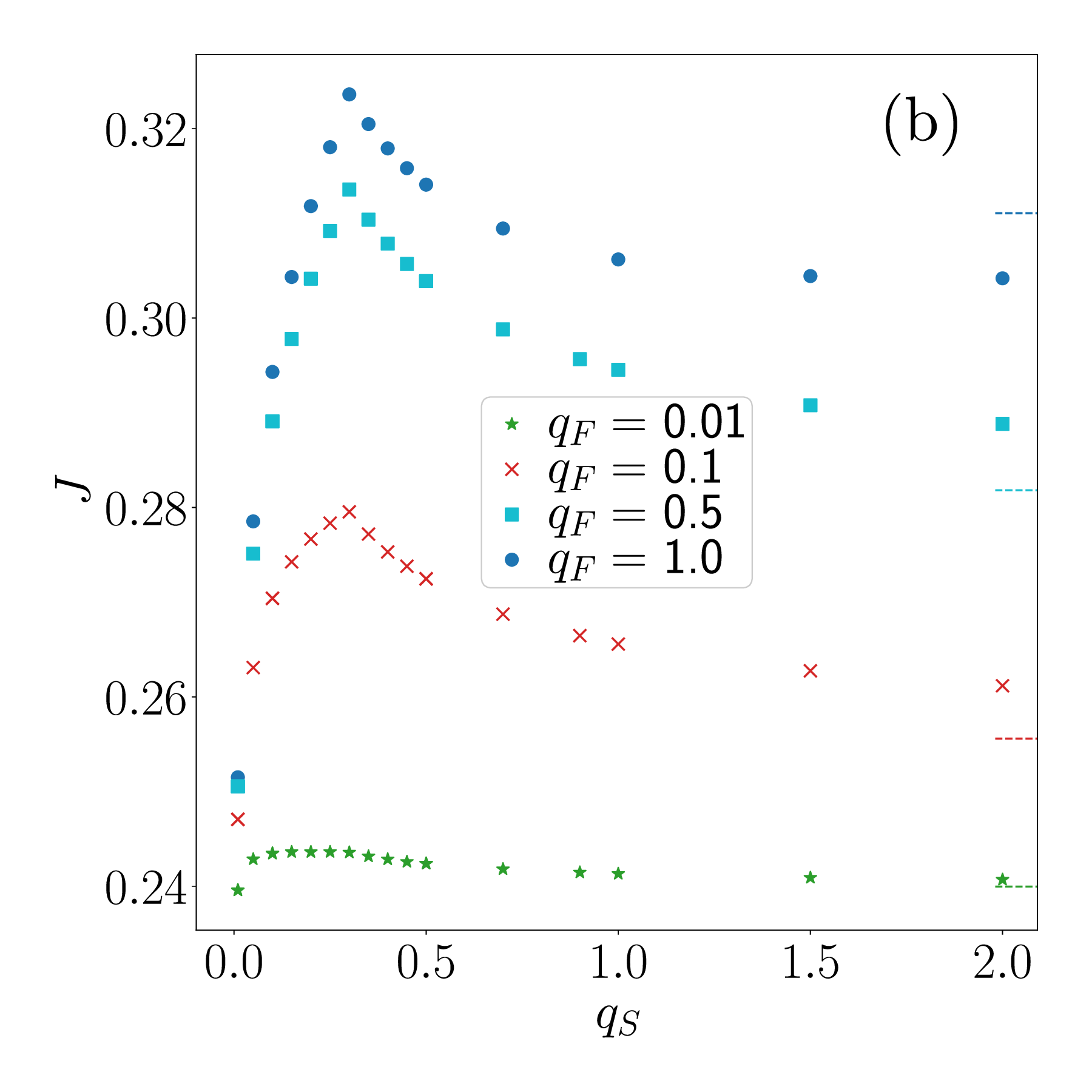}
  \phantomsubcaption
  \label{fig:Current_parallel_fct_qA_different_qb_MC_ParkingABC_pa_1-00_alphaA_1-00_beta_1-00_b}
\end{subfigure}
\hfill
\caption{Variations of the current $J$ with the parking propensity $q_S$, with the parallel update rule, for (a) $p_S = 1$ (note that non-monotonicity can then be analytically predicted for $q_F = 1$) and (b) $p_S = 0.7$, where a maximum current exists at an intermediate value $q_S$ for all inverse parking durations $q_F$. All data were obtained from Monte-Carlo simulations with $L=1000$, $\alpha_S = 1, \beta = 1$. }
    \label{fig:Current_parallel_fct_qA_different_qb_MC_ParkingABC_pa_1-00_alphaA_1-00_beta_1-00}
\end{figure}

At fixed input rate, non-monotonic variations of the current are also obtained with respect to the parking propensity $q_S$ for different inverse parking durations $q_F$ and hopping rate $p_S$, as shown in Fig. \ref{fig:Current_parallel_fct_qA_different_qb_MC_ParkingABC_pa_1-00_alphaA_1-00_beta_1-00}, using a different update rule (the parallel update exposed in Sec.~\ref{sub:parallel_update}). Switching to parallel updates here serves two purposes. Firstly, it demonstrates that the non-monotonic current behavior is not limited to a specific update rule. Secondly, for specific parameter values, the dynamics become entirely deterministic, enabling the analytical prediction of the non-monotonic behavior, as we will discuss in Sec.~\ref{section:parallelUpdate}.

\subsection{Bump in the density profile}

\begin{figure}
\centering
\begin{subfigure}{.48\textwidth}
  \centering
    \includegraphics[width=0.8\textwidth]{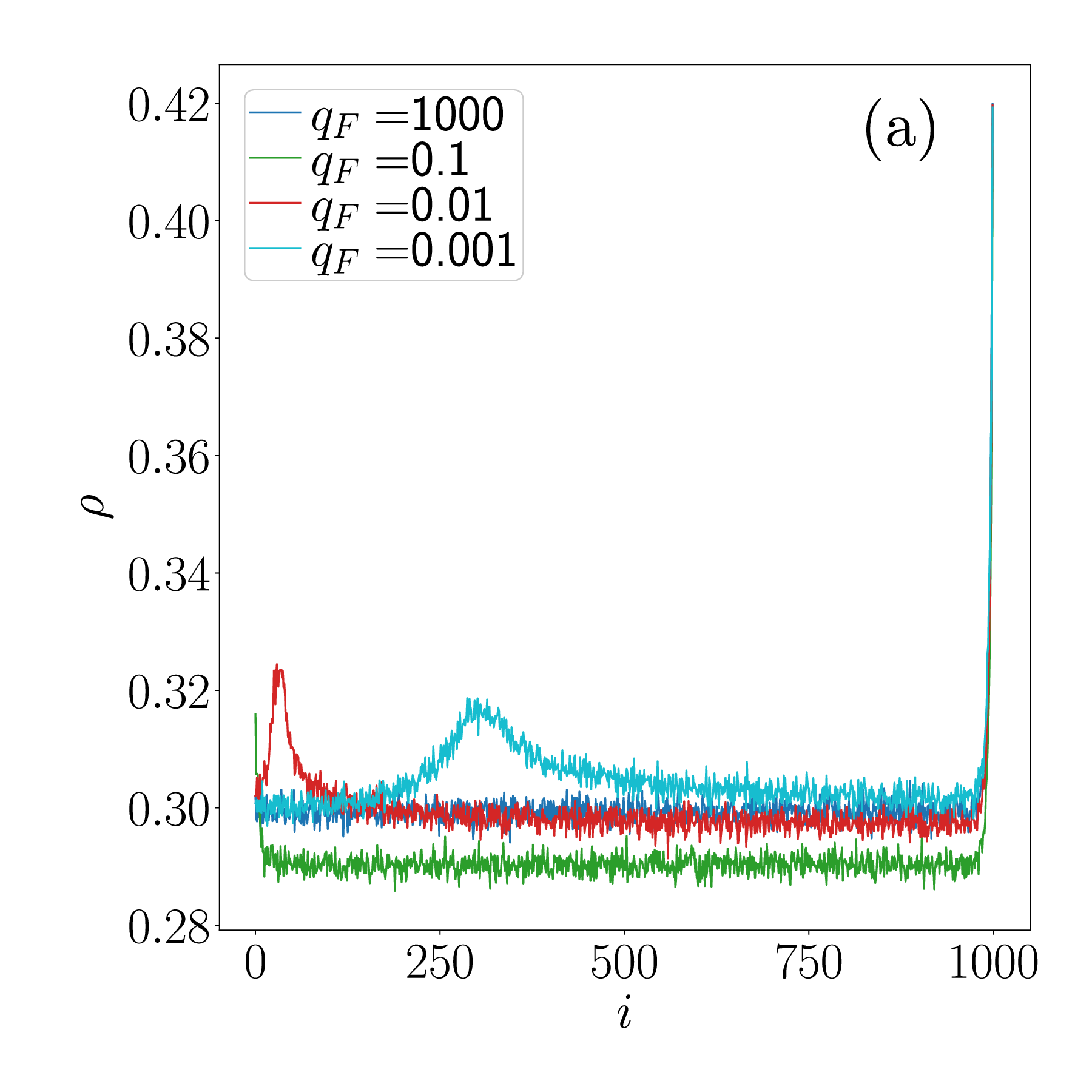}
    \phantomsubcaption
    \label{fig:Densite_fct_qb_qaINF_pa_1_LD_a}
\end{subfigure}
\begin{subfigure}{.48\textwidth}
  \centering
  \includegraphics[width=0.8\textwidth]{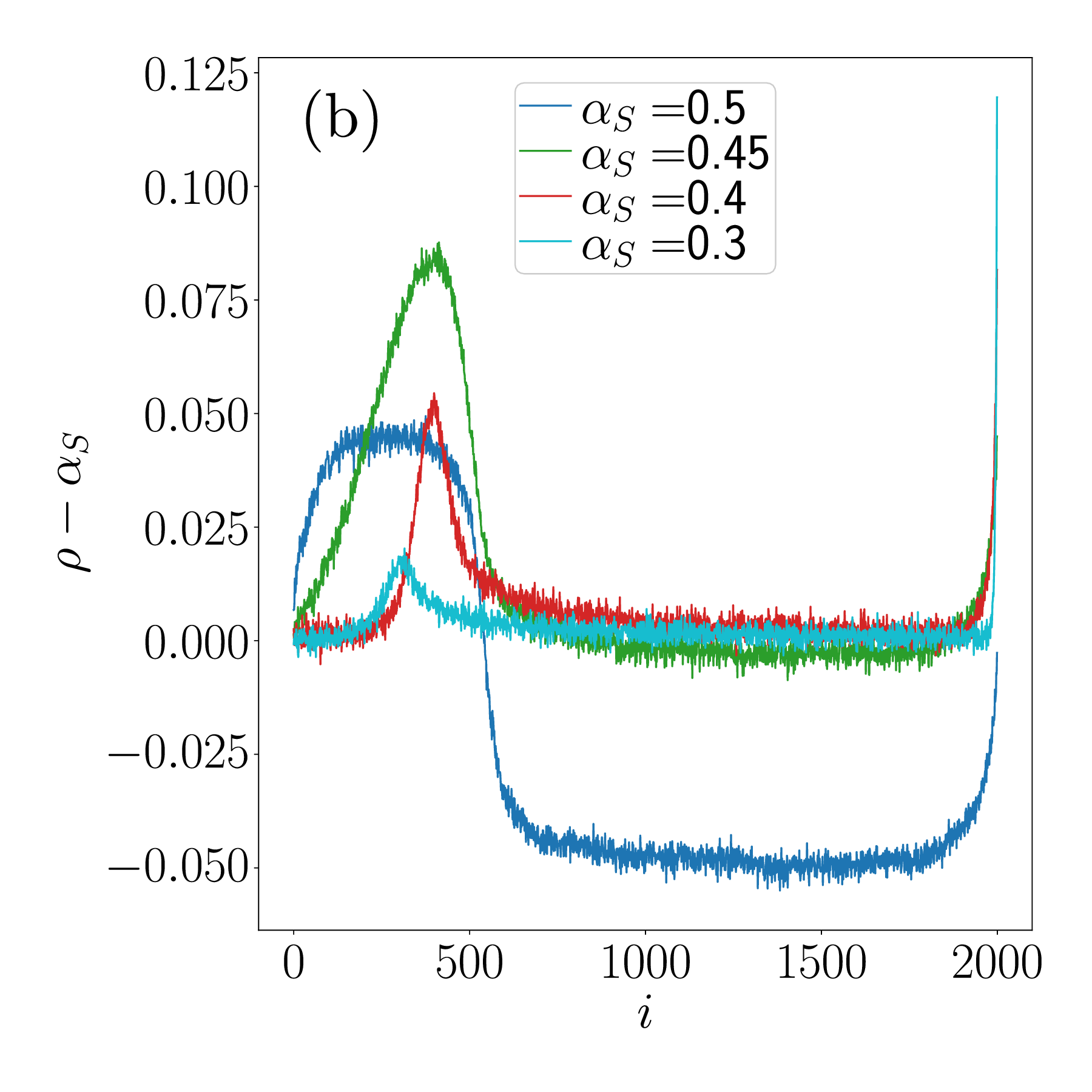}
  \phantomsubcaption
  \label{fig:Densite_fct_qb_qaINF_pa_1_LD_b}
\end{subfigure}
\hfill
\caption{SFP density profiles at $p_S = 1, q_S \rightarrow \infty$ and $\beta = 0.5$, showing an unusual bump. (a) In the LD phase with $\alpha_S = 0.3$; the bump shifts to the right when $q_F$ decreases; as it gets far away from the boundary, the expected current $J=\alpha_S(1-\alpha_S)$ is restored. (b) At fixed $q_F = 0.001$; the height and width of the bump in the rescaled density $\rho-\alpha_S$ depend on $\alpha_S$. }
    \label{fig:Densite_fct_qb_qaINF_pa_1_LD}
\end{figure}

Switching from the perspective of the current $J$ to the density, 
we note on Fig.~\ref{fig:Densite_fct_qb_qaINF_pa_1_LD_a} that SFP also displays non-trivial features in the density profile ($\rho^S + \rho^F$), such as the presence of a bump influenced by the inverse parking duration $q_F$. The bump gradually shifts to the right and diffuses as $q_F$ increases. When the bump is located far enough from the left boundary, the current becomes insensitive to variations in $q_F$.
Together with the disappearance of the MC phase with $q_F$ in Fig.~\ref{fig:ColorPlot_Courant_fct_alphaA_qb_plusieurs_pa_qaINF_betaB_0-60_L_200}, these phenomena can be ascribed to the influence of parked particles returning to the main road. As they pull out, the main flow is perturbed and the induced perturbations can propagate backward (see Sec.~\ref{par:propagation_perturbation}). Should they reach the left boundary, they will then block the injection process before they dissipate, consequently leading to a reduction in the current.

\subsection{Influence of the topology}

Admittedly, the rate values considered thus far to disclose singular behaviors might seem unrealistic in the context of vehicular traffic, where the rates $q_S$, $q_F$, and $\alpha_S$  are generally comparable or higher than the hopping rate. 
As an example, suppose that sites are 5 meters long and that $F$ cars drive at $30\mathrm{km/h}$. From Fig.~\ref{fig:Courant_fonctionAlphaA_MC_MF_ED_ParkingABC_pa_0-10_qaINF_betaB_0-60}, one observes a clear-cut non-monotonic behavior only when $p_F/q_F \sim 1$ (even if the $S$ cruising speed $p_S$ is as low as one tenth of $F$), which would correspond to an unrealistically low average parking duration of $q_F^{-1}\simeq 0.6$ seconds. It would even be too short to model drop-off parking. 

\begin{figure}
\centering
\begin{subfigure}{.48\textwidth}
  \centering
    \includegraphics[width=0.8\textwidth]{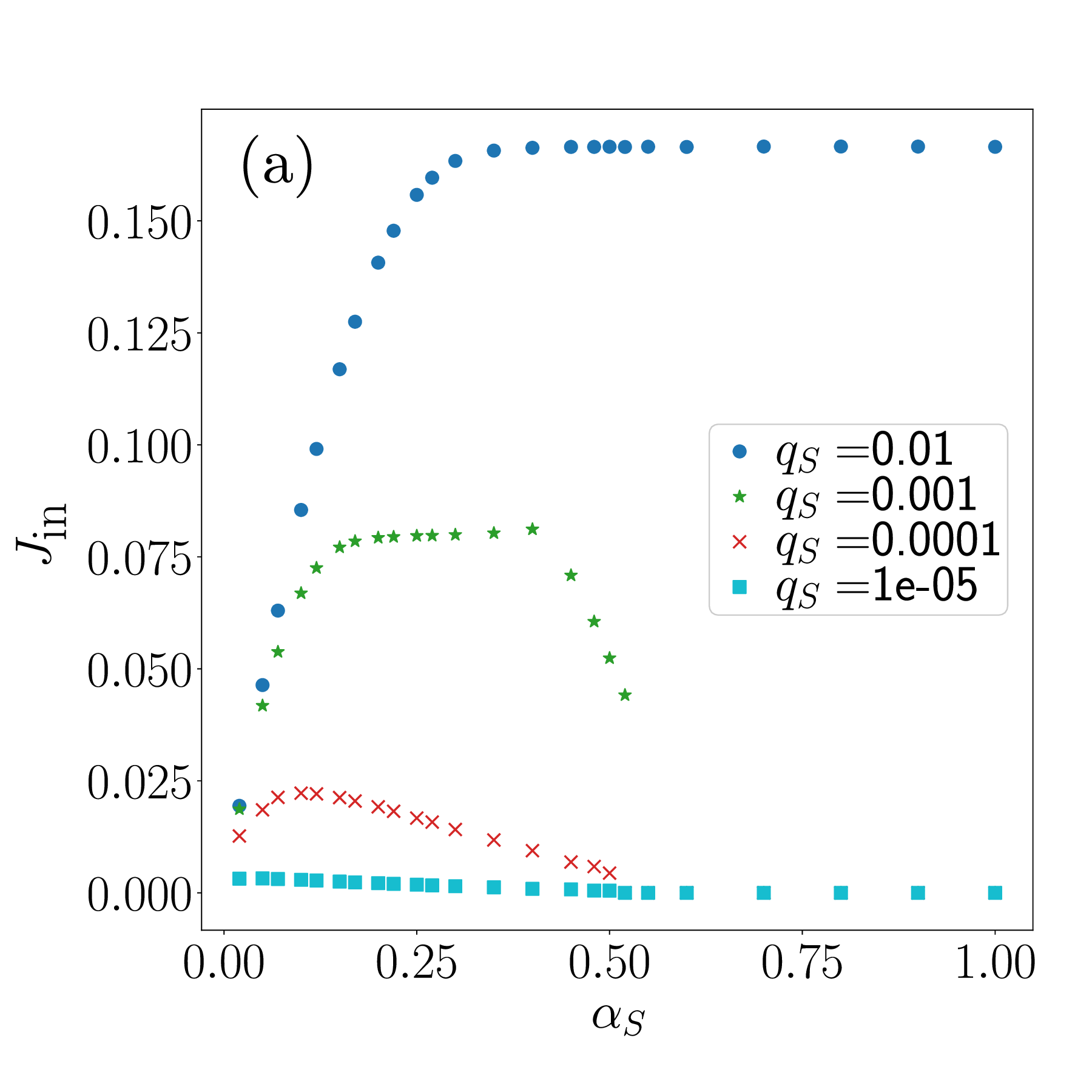}
    \phantomsubcaption
    \label{fig:Courant_fonctionAlphaA_MC_plusieurs_qa_qb_0-001_pa_0-70_betaB_0-60_RingGeometry_L_200_a}
\end{subfigure}
\begin{subfigure}{.48\textwidth}
  \centering
  \includegraphics[width=0.8\textwidth]{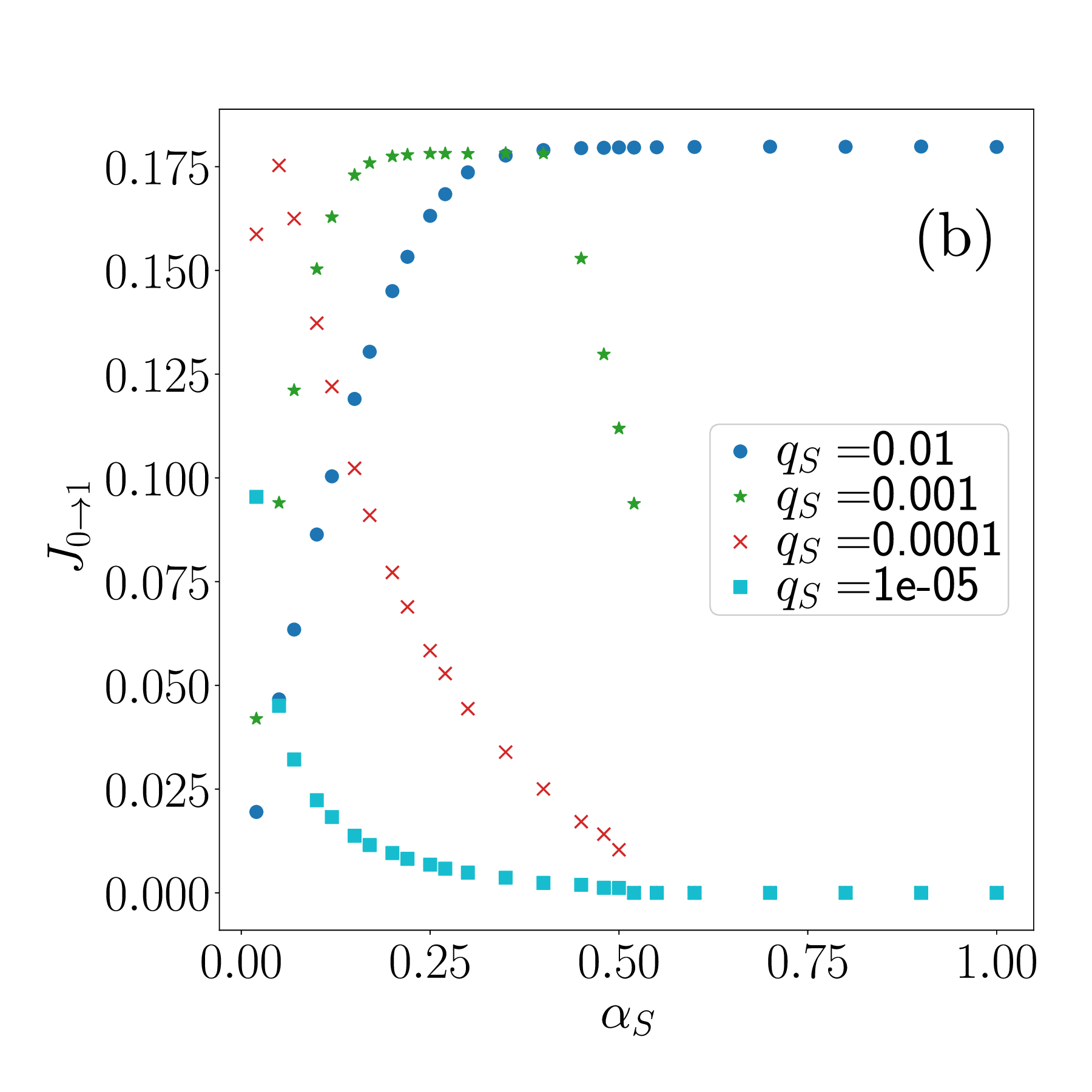}
  \phantomsubcaption
  \label{fig:Courant_fonctionAlphaA_MC_plusieurs_qa_qb_0-001_pa_0-70_betaB_0-60_RingGeometry_L_200_b}
\end{subfigure}
\hfill
\caption{ Variations of the SFP current, for `periodic' boundary conditions, i.e. the ring geometry shown in Fig.~ \ref{fig:Schema_parking_ABC_ring_topology}. (a) Input current $J_{in}$ and (b) current through the first site $J_{0\to 1}$. In this topology, non-monotonicity is observed even for both low parking propensity $q_S$ and long parking durations $q_F^{-1}$. Differences between the two figures highlight  that cruising particles $S$ circle around, waiting to find a parking spot. Around $\alpha_S \sim 0.5$, a blocking state can be obtained for $q_S \leq 0.001$.
Parameters : $p_S=0.7, \beta=0.6, q_F=0.001, L=200$.}
    \label{fig:Courant_fonctionAlphaA_MC_plusieurs_qa_qb_0-001_pa_0-70_betaB_0-60_RingGeometry_L_200}
\end{figure}

However, the topology of the system alters these results in a quantitative way. As one switches from a linear topology to a ring geometry from which only $F$ particles can exit (see Sec.~\ref{sub:topology}), tentatively describing  the circling behavior of drivers in search of parking \cite{shoup_cruising_2006},
much longer parking durations $q_F^{-1}$ (and lower parking propensities $q_S$) become compatible with non-monotonicity. In the example shown in  Fig.~\ref{fig:Courant_fonctionAlphaA_MC_plusieurs_qa_qb_0-001_pa_0-70_betaB_0-60_RingGeometry_L_200_a}, for $S$ cars driving 70\% as fast as $F$, the ratio $p_F/q_F \sim 10 ^{-3}$ gives rise to non-monotonic variations, corresponding to a parking duration around 10 minutes. This is substantially closer to a realistic estimate.
\section{Methods}
\label{section:methodology}

This section exposes the numerical and analytical framework used to study SFP. It introduces three distinct methodologies for its analysis.
Direct simulations of the model were performed using a kinetic Monte-Carlo  (kMC) algorithm, for efficiency.
Independently, in an analytical endeavor, we derive mean-field equations to capture the system's dynamics. Alternatively, we performed an exact diagonalization of the stochastic matrix of the model to compute the total current; this method was notably applied to the first sites of the lattice, using a splitting technique to decompose the lattice into two blocks.

\subsection{Kinetic Monte-Carlo (kMC) algorithm}

The most direct numerical simulations of our model are performed with a kinetic Monte-Carlo algorithm \cite{bortz_new_1975, sokal_monte_1997}. Given that the transitions are controlled by Poisson processes with known rates, this enables us to  speed up the calculation, by allowing the possibility to reach a new configuration at every simulation step. Furthermore, rates are naturally encoded such that they can take values in $[0, +\infty]$. For instance if $\alpha_S = +\infty$ and all the others rates are finite, then, as soon as a particle leaves the first site, a new particles is immediately introduced such that the first site is always occupied by a $S$ particle. Different system size $L$ have been used, from $L=200$ to $L=10000$. $10^6$ Monte-Carlo steps have been performed independently of the size knowing that one Monte-Carlo step corresponds to $2L$ time steps. The first $5\%$ have been ignored to remove the transient and focus on the steady-state.

\subsection{Master equation and mean-field approximation}

\subsubsection{Master equations}

In an endeavor to solve the problem in a more analytical fashion, we derive a master equation for the evolution of the average site occupation (denoted as $\langle ... \rangle$) for the three species, more precisely, for species $S$ and $F$ on lattice sites indexed by $i = 2, ..., L-1$, and species $P$ 
on sites $i=1,...,L$, viz.,
\begin{align} 
    \frac{d \langle n_i^S \rangle}{dt} =& p_S \langle n_{i-1}^S(1-n_i^S-n_i^F) \rangle - p_S \langle n_i^S(1-n_{i+1}^S-n_{i+1}^F) \rangle \nonumber\\ &- q_S \langle n_i^S(1-n_i^P) \rangle \nonumber \\
    \frac{d\langle n_i^F \rangle}{dt} =& p_F \langle n_{i-1}^F(1-n_i^S-n_i^F) \rangle - p_F \langle n_i^F(1-n_{i+1}^S-n_{i+1}^F) \rangle \nonumber\\ &+ q_F \langle n_i^P(1-n_i^S-n_i^F) \rangle \nonumber \\
    \frac{d\langle n_i^P \rangle}{dt} =& q_S \langle n_i^S(1-n_i^P) \rangle - q_F \langle n_i^P(1-n_i^S-n_i^F) \rangle 
    \label{eq:Master_eq_1}
\end{align}
The boundary conditions are given by:
\begin{align*}
    \frac{d\langle n_1^S \rangle}{dt} =& \alpha_S(1-\langle n_1^S \rangle-\langle n_1^F \rangle) - p_S\langle n_1^S(1-n_2^S-n_2^F) \rangle \\ &- q_S \langle n_1^S(1-n_1^P) \rangle \\
    \frac{d\langle n_1^F \rangle}{dt} =& \alpha_F(1-\langle n_1^S \rangle-\langle n_1^F \rangle) - p_F\langle n_1^F(1-n_2^S-n_2^F) \rangle \\ &+ q_F \langle n_1^P(1-n_1^S-n_1^F) \rangle\\
    \frac{d\langle n_L^S \rangle}{dt} =& -\beta_S \langle n_L^S \rangle + p_S\langle n_{L-1}^S(1-n_L^S-n_L^F) \rangle \\ &- q_S \langle n_L^S(1-n_L^P) \rangle \\
    \frac{d\langle n_L^F \rangle}{dt} =& -\beta_F \langle n_L^F \rangle + p_F\langle n_{L-1}^F(1-n_L^S-n_L^F) \rangle \\ &+ q_F \langle n_L^P(1-n_L^S-n_L^F) \rangle
\end{align*}

Both limits $q_S, q_F \rightarrow \infty$ have been studied. In these cases, the master equations need to be redefined. This is done in Appendix \ref{annexe:master_equation_qF} and \ref{annexe:master_equation_qS}.

\subsubsection{Mean-field solution}

Because Eq.~\ref{eq:Master_eq_1} involves correlations between the occupations on adjacent sites, the system of equations cannot be solved directly. The mean-field approximation consists in factoring out all possible correlations, which are thus dismissed, viz., $\langle n_i^X n_j^Y \rangle = \langle n_i^X\rangle \langle n_i^Y\rangle$ with $X,Y = S,F$ or $P$. Then, we coarse grain the lattice with the lattice constant $\epsilon = 1/L$. If $L$ is large enough, the rescaled position $x= i/L$ and time $t'=t/L$ are quasi-continuous. Introducing the average lattice density of the specie $Y=S,F,P$, $\rho^Y(x,t') = \langle n_i^Y(t) \rangle$, we arrive at
\begin{align*}
    \frac{d\rho^S}{dt'} =& - p_S\left[1-2\rho^S-\rho^F \right]\partial_x \rho^S + p_S \rho^S \partial_x \rho^F \\ &+ \frac{\epsilon}{2}p_S(1-\rho^F)\partial_x^2\rho^S + \frac{\epsilon}{2}p_S\rho^S\partial_x^2\rho^F - L q_S \rho^S(1-\rho^P)\\
    \frac{d\rho^F}{dt'} =& - p_F\left[1-2\rho^F-\rho^S \right]\partial_x \rho^F +  p_F \rho^F \partial_x \rho^S \\ &+ \frac{\epsilon}{2}p_F(1-\rho^S)\partial_x^2\rho^F + \frac{\epsilon}{2}p_F\rho^F\partial_x^2\rho^S + L q_F \rho^P(1-\rho^S-\rho^F) \\
    \frac{d\rho^P}{dt'} =&  L q_S \rho^S(1-\rho^P)- L q_F \rho^P(1-\rho^S-\rho^F).
\end{align*}

Note that, if particles were allowed to visit parking spaces multiple times,  as in \cite{bhatia_far_2022}, then the dynamics would be fully controlled by the exchange with the parking spot because of the associated $L$ factor in the equation. This would
also hold  for the Langmuir kinetics, where particles can always be removed or added. On the contrary, here, cars can only pull
in once in their entire journey from the entrance to the exit; the gradual attrition of cruising cars makes the first two equations important.

\subsection{Exact diagonalization}
\label{section:methodology_exactDiag}

To parry the limitations of the mean-field approximation when correlations are present, we introduce a method based on the exact diagonalization of the stochastic matrix of the problem, which is tightly connected with the finite-segment mean field theory \cite{chou_clustered_2004}.
The computational cost of the diagonalization prohibits
this method for large lattices ($L \gg 1$), but it is well suited for the low-density phase, where only the first few parking sites are ever occupied, and more generally it can be coupled to the splitting process detailed below (Sec.~\ref{sub:splitting}), where
only the first segment of the lattice is exactly diagonalized, while the rest is effectively described by a classical TASEP model.

 Technically speaking, the stochastic matrix of the problem is constructed by drawing on an analogy with the Quantum formalism, that is to say, by writing an equivalent Quantum Hamiltonian \cite{schutz2001exactly}. The state of a site $i$ on the road  is given by $\ket{k}_r^i$ with $k=0$ if the site is empty and $k=1\,(2)$ if it is occupied by a $S$ ($F$) particle. The raising (lowering) operator acting on road site $i$ is denoted $\sigma_{i,r}^{+(-)}$ and it has the following properties 
$$ \sigma_{i,r}^{+}\ket{k}_r^i = \left\{
    \begin{array}{ll}
        \ket{k+1}_r^i & \mbox{if } k=0,1 \\
        0 & \mbox{if } k=2
    \end{array}
\right., $$ 
$$ \sigma_{i,r}^{-}\ket{k}_r^i = \left\{
    \begin{array}{ll}
        \ket{k-1}_r^i & \mbox{if } k=1,2 \\
        0 & \mbox{if } k=0
    \end{array}
\right.,  .$$

Similarly, the state of parking spot $i$ is given by $\ket{k}_p^i$ with $k=0$ if empty and $k=1$ if occupied by a $P$ particle and the properties of the associated raising (lowering) operator $\sigma_{i,p}^{+(-)}$ are easily found.

All the desired operators can be obtained as combinations of the raising and lowering operators. For instance, the operator that creates a $S$ particle on the site $i$ if it is empty is given by 
$$ \sigma^{i,r}_{0\rightarrow S} = \sigma_{i,r}^{-}\sigma_{i,r}^{+}\sigma_{i,r}^{+}$$
which has the desired properties 
$$ \sigma^{i,r}_{0\rightarrow S}\ket{k}_r^i = \left\{
    \begin{array}{ll}
        \ket{1}_r^i & \mbox{if } k=0 \\
        0 & \mbox{if } k=1,2
    \end{array}
\right. , $$
$$ \sigma^{i,r}_{0\rightarrow S}\ket{k}_u^j = \ket{k}_u^j  \mbox{ if } j\neq i \mbox{ or } i=j, u=p$$

All global operators assume the form 
$$\mathcal{O} = X_{1,r} \otimes X_{1,p} \otimes ... \otimes X_{L,r} \otimes X_{L,p} $$
where the dimension of the $X_{i,r}$ operators is $3\times3$ and $X_{i,p}$ is $2\times 2$. The $\sigma^{i,r}_{0\rightarrow S}$ operator read 
$$\sigma^i_{0\rightarrow S} = \mathbb{1}_2^{i-1}\otimes \mathbb{1}_3^{i-1}\otimes \sigma_{0\rightarrow S}\otimes \mathbb{1}_2^{L-i+1}\otimes \mathbb{1}_3^{L-i}$$
with $\mathbb{1}_d$ the identity matrix of size $d\times d$.

Finally, the stochastic matrix of our model is given by 
\begin{equation}
    M = h_{1,r} + h_{1,L} + \sum_{i=1}^{L-1} h_{i,i+1,r} + \sum_{i=1}^{L}h_{i, p}
\end{equation}
with 
\begin{align*}
    h_{i,i+1,r} =& p_S \left(\sigma^{i,r}_{S\rightarrow 0}\sigma^{i+1,r}_{0\rightarrow S} -  \sigma^{i,r}_{S\rightarrow S}\sigma^{i+1,r}_{0\rightarrow 0}\right) \\ &+ p_F \left(\sigma^{i,r}_{F\rightarrow 0}\sigma^{i+1,r}_{0\rightarrow F} - \sigma^{i,r}_{F\rightarrow F}\sigma^{i+1,r}_{0\rightarrow 0}\right)  \\
    h_{i,p} =& q_S \left(\sigma^{i,r}_{S\rightarrow 0}\sigma^{i,p}_{0\rightarrow P} -  \sigma^{i,r}_{S\rightarrow S}\sigma^{i,p}_{0\rightarrow 0}\right)  \\ &+ q_F \left(\sigma^{i,p}_{P\rightarrow 0}\sigma^{i,r}_{0\rightarrow F} -  \sigma^{i,p}_{P\rightarrow P}\sigma^{i,r}_{0\rightarrow 0}\right)  \\
    h_{1, r} =& \alpha_S \left(\sigma^{1,r}_{0\rightarrow S}-\sigma^{1,r}_{0\rightarrow 0}\right)+\alpha_F \left(\sigma^{1,r}_{0\rightarrow F}-\sigma^{1,r}_{0\rightarrow 0}\right) \\
    h_{L,r} =& \beta_S \left(\sigma^{L,s}_{S\rightarrow 0}-\sigma^{L,s}_{S\rightarrow S}\right)+\beta_F \left(\sigma^{L,s}_{F\rightarrow 0}-\sigma^{L,s}_{F\rightarrow F}\right).
\end{align*}
Each minus sign guarantees that the matrix is stochastic, i.e., that each column sums to zero.

The dimension of this matrix is $6^{L_1} \times 6^{L_1}$. It soon becomes intractable as $L$ increases. 
Even if we are only interested in the eigenstate with the eigenvalue $0$, which corresponds to the steady state, the matrix cannot be diagonalized for system sizes $L$ significantly larger than a few sites. However, the contrivance introduced just below, which splits the system into two segments, only the first of which requires exact diagonalization, extends the possible range of
application of this method to much larger system sizes.

\subsection{Decomposition of the system into two connected blocks}
\label{sub:splitting}

\begin{figure}
    \centering
    \includegraphics[width=0.9\columnwidth]{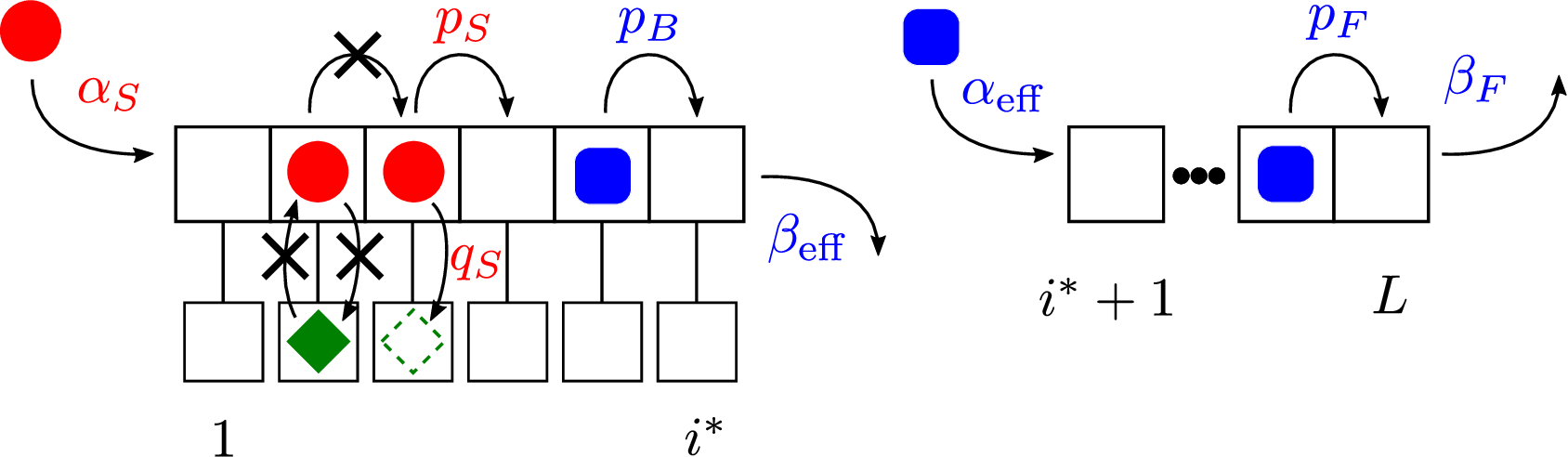}
    \caption{Schematic diagram of the splitting technique, decomposing the SFP lattice into a first segment with an effective exit rate $\beta_{\mathrm{eff}}$ and SFP dynamics, and a second segment, with an effective input rate $\alpha_{\mathrm{eff}}$ and standard single-species TASEP dynamics.}
    \label{fig:Schema_parking_ABC_simplified}
\end{figure}

Since the parking propensity  $q_S$ and the inverse parking duration $q_F$ do not scale with system size $L$ in SFP, contrary to other TASEP variants, whenever $q_S$ is finite and the system is large,  there exists a site $i^*$ past which virtually only particles $F$ are present. This observation
inspired the idea of splitting the system into two blocks, as illustrated in Fig.~\ref{fig:Schema_parking_ABC_simplified}. 
The first segment, of fixed length $L_1$ (which ideally is set larger than $i^{\star}$), contains the full SFP phenomenology and will be diagonalized exactly, using the above method, whereas the second segment ($i>L_1$) reduces to a classical TASEP with a single species, $F$, and an effective entry rate $\alpha_{eff}$. The current as well as the dynamics are primarily controlled by $\alpha_{eff}$, which controls the coupling between the segments and whose dependency on $\alpha_S$ thus deserves closer inspection.

To determine $\alpha_{eff}$, we diagonalize the first segment, as exposed above, and deduce the total current $J$. In doing so, attention must be 
paid to the exit rate $\beta$ out of this first segment. Indeed, finite-size effects
are enhanced if too na\"ive an estimate is used for $\beta$. Instead, 
we resort to an iterative diagonalization process: After each iteration the exit rate is updated to :
\begin{equation}
    \beta = 1 - \rho_{L_1}^S - \rho_{L_1}^F,
\end{equation}
until the difference between two successive $\beta$ is less than 0.01. 
This choice is motivated by the exit is equivalent to an additional site occupied with probability $\rho_{L_1+1}$ such that $\beta=p_F (1-\rho_{L_1+1})$, if most particles found near $L_1$ belong to the $F$ species; since the density is flat in the steady-state LD phase, we assume that $\rho_{L_1+1} \simeq \rho_{L_1}^S + \rho_{L_1}^F$. The fact that the `true' $\beta$ does not enter these expressions is bolstered by
the insensitivity of the TASEP current $J$ to $\beta$ in LD.
This assumption (and, hence, the method) will fail in HD, but also in MC (where the density decreases as a power law near the entrance \cite{derrida_exact_1993}).
Finally, having found $J$ and making use of the equality \cite{derrida_exact_1992, derrida_exact_1993, schutz_phase_1993}
$$ J = \alpha_{eff}(1-\alpha_{eff})$$
(valid in the LD phase of TASEP i.e. when $\beta_F > \alpha_{eff}$ and $\alpha_{eff} < 1/2$), we obtain the value of $\alpha_{eff}$ that should be used as input into the second block. 

\subsection{Parallel update}
\label{sub:parallel_update}

So far, we have considered a random sequential update scheme, where only one particle is updated per infinitesimal time step, $dt \rightarrow 0$. 

For vehicular traffic flow, the alternative option of parallel update is very popular \cite{nagel_cellular_1992, biham_self-organization_1992, schreckenberg_discrete_1995}.
In this scheme, all particles are updated simultaneously at each time step, corresponding to the case $dt = 1$; the rate parameters are then actually probabilities for each possible move. Changing the update rule entails major differences in the output of TASEP: for parallel update, MC is reached only for $\alpha = \beta = 1$ and the current is then twice larger than with random sequential update, i.e., $J=1/2$; otherwise, LD and HD phases are observed on the two sides of the first bisector  $\alpha = \beta $, with respective currents $J= \frac{\alpha}{1+\alpha}$ and $J=\frac{\beta}{1+\beta}$ \cite{de_gier_exact_1999}. Therefore, it will be worth testing the impact of parallel update on our main results.

To do so, we ought to explain how conflicting moves arising in the parallel update process are solved. Firstly, the sum of rates governing a particle's possible moves (e.g., $p_S+q_S$ for cruising cars) may be larger than one; in this case, probabilities of motion are obtained by renormalizing the rates with this sum, viz., substituting $p_S$ with  $\frac{p_S}{p_S + q_S}$.
Secondly, there may be conflicts for the occupation of a given site on the road. Should such a conflict occur (between a car moving forward and one pulling out), following traffic laws, priority will be given to the main road; we thus update particles $S$ and $F$ first and then update particles $P$. It follows that cars cannot pull in and out in the same time step.

\section{Investigation and rationalization of the model properties}
\label{section:explanation}

This Section aims to account for the singular properties of SFP, compared to classical TASEP, succinctly outlined in Sec.~\ref{section:novel_behavior} and explore the model in a broader sense. Since the number of parameters featured in the model and their complex interplay make an exhaustive exploration of parameter space elusive, this will not be achieved by an independent exploration of the effect of each parameter in each case.
Instead, this Section will shed light on the main mechanisms affecting the SFP steady-state current $J$, as a function of the model parameters. To this end, we will largely draw parallels with other variants of TASEP and study limit cases of SFP, which can be rationalized in detail.

An SFP particle differs from its TASEP counterpart at two separate stages: (i) before it parks, while it moves slowly as an $S$ particle ($p_S < p_F$), (ii) when it pulls out of its parking space, which may disturb the cars behind on the road. Thus, we identify two distinct flow-altering mechanisms specific to  SFP , namely (i) the survival of slow $S$ vehicles near the entrance, (ii) the pulling out of parked cars into the road. Let us study the cases when neither of these mechanisms is at play, when only one of them is effective, and ultimately the effect of the coupling between these mechanisms.

\subsection{Case Zero: no slowing down near the injection point and no disruptive pull-outs}
\label{section:Connections_models}

In the limits $q_S=0$ (cars never park) and $q_F=0$ (cars stay parked forever), the SFP model boils down to a situation in which parking  plays no role on the flow (after a transient, for the case $q_F=0$).

Nevertheless, even in these limits and even if $p_S=p_F$, there is more than meets the eye: the model would exhibit interesting properties, which are not captured by naive mean-field arguments, if both $S$ and $F$ particles were injected into the system (which is not the case in SFP). Indeed, one species may for instance have a much lower exit rate, leading to temporary clogs near the exit, which would depart from the phenomenology of the single-species TASEP model, but not exhibit the singular behaviour of SFP, as shown by \cite{bonnin_two-species_2022}. 
Similarly when $p_S \neq p_F$ with equal exit rates $\beta$, simple mean-field approximation fails but the model can be described as an effective single specie TASEP with different particle size \cite{bottero_analysis_2017}. Nonetheless, it does not feature properties of SFP. 

This reference situation, in which neither of the flow-altering mechanisms is at play, also holds if cars \emph{almost} never park. In particular, if $q_S \sim 1/L$ and $q_F \sim 1/L$ but cars are allowed to park many times (i.e., turn back to $S$ when it pulls out), then one recovers  the model with pockets (here, of size one) studied by \cite{bhatia_far_2022}, which displays the very same phenomenology as TASEP.

\subsection{Survival of slow cars near the injection point ($q_F \to \infty$)}

Let us now explore the first flow-altering mechanism, whereby the injected $S$ particles perturb the entrance flow because they are slower than their $F$ counterparts, $p_S < p_F=1$ (see Appendix~\ref{app:Identical_transmutation} for the case $p_S=p_F$). To expose this mechanism in its purest form, we consider a limit-case in which the second mechanism is not at play, by setting $q_F \to \infty$ (hence, $q_F \gg \alpha$ and $q_F \gg p_S$) so that cars are immediately pull out after pulling in, as though the $S \to F$ transmutation took place on the road, at a rate $q_S$.

\begin{figure}[h]
    \centering
    \includegraphics[width=0.8\columnwidth]{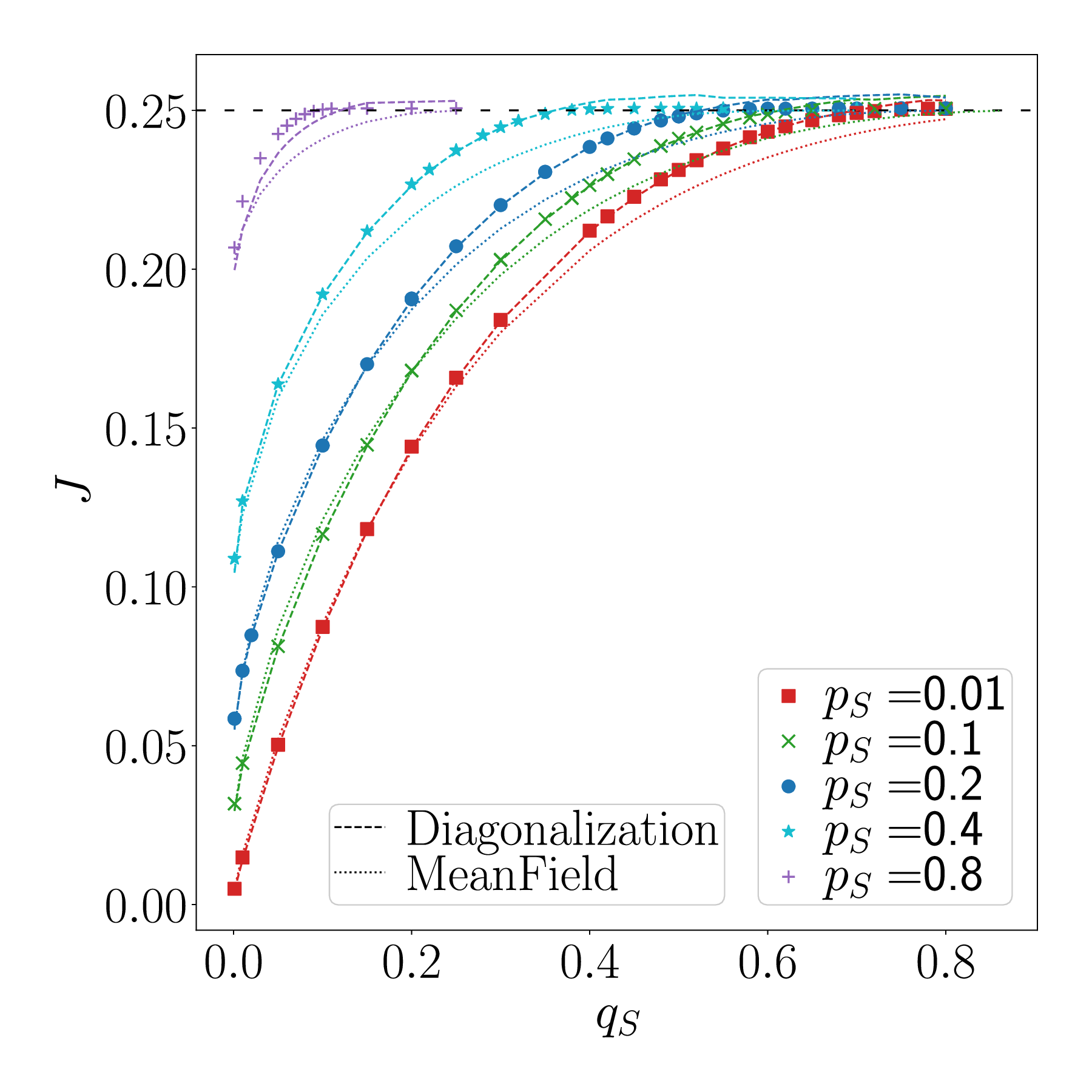}
    \caption{Current as a function of the parking propensity $q_S$ for different
    hopping rates $p_S$. The MC current $J=1/4$ (horizontal long dashed line) can be reached even when slow $S$ particles $p_S < 1$, provided that $q_S$ is large enough. Symbols: Monte-Carlo simulations for $L=1000$; dash lines: exact diagonalization with $L=10$; dotted lines: mean-field solution for $L=500$. Fixed parameters: $\alpha_S = 1, \beta = 0.6, q_F \to \infty$.
    }
    \label{fig:EvolutionJ_qa_differents_pa_MC_MF_ED_ParkingABC_qbINF_alphaA_1-00_betaB_0-60}
\end{figure}

\begin{figure}[h]
\centering
\begin{subfigure}{.45\textwidth}
  \centering
  \includegraphics[width=0.8\textwidth]{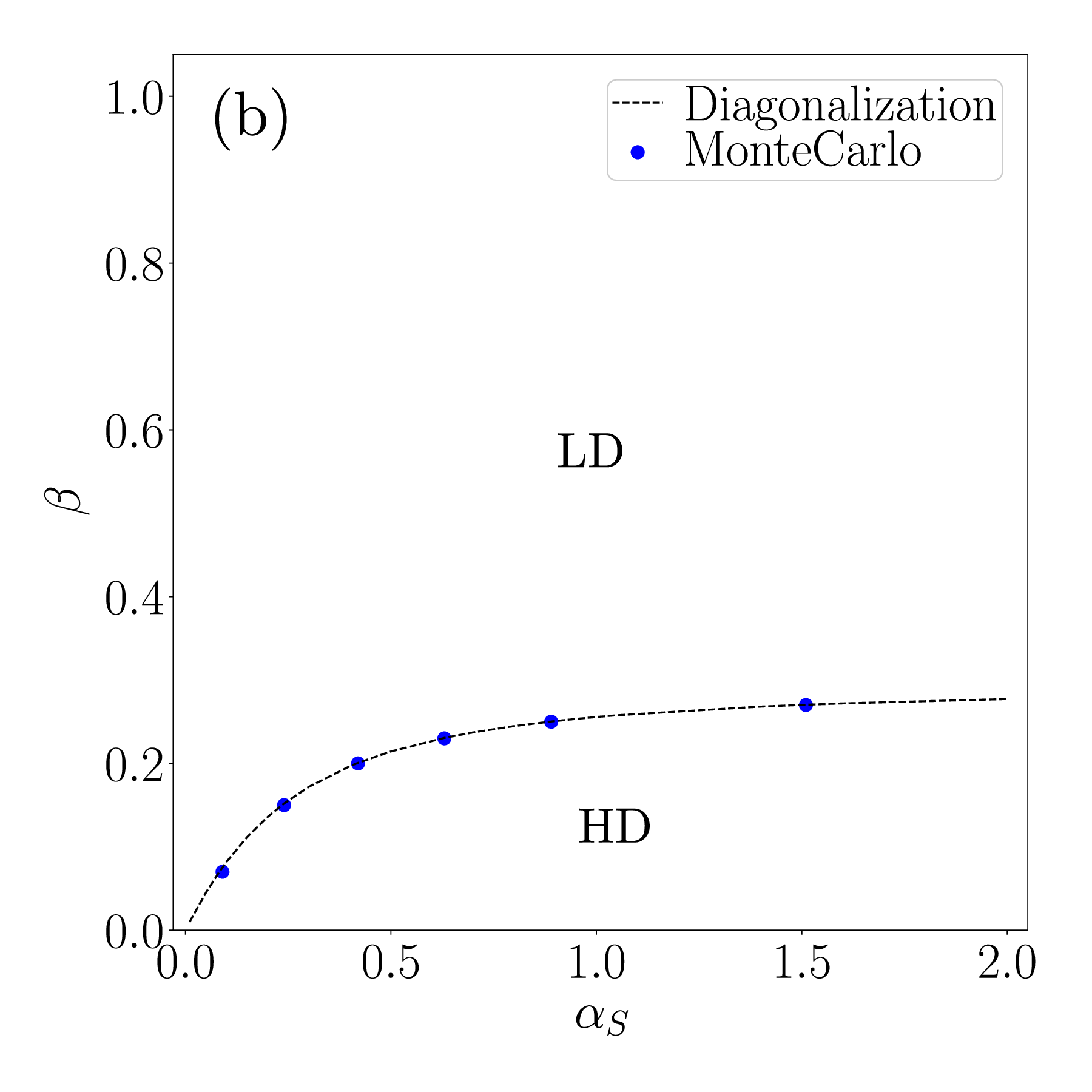}
  \phantomsubcaption
  \label{fig:DiagPhase_MC_ED_ParkingABC_qbINF_pa_0-20_qa_0-20_qa_1-00_a}
\end{subfigure}
\begin{subfigure}{.45\textwidth}
  \centering
  \includegraphics[width=0.8\textwidth]{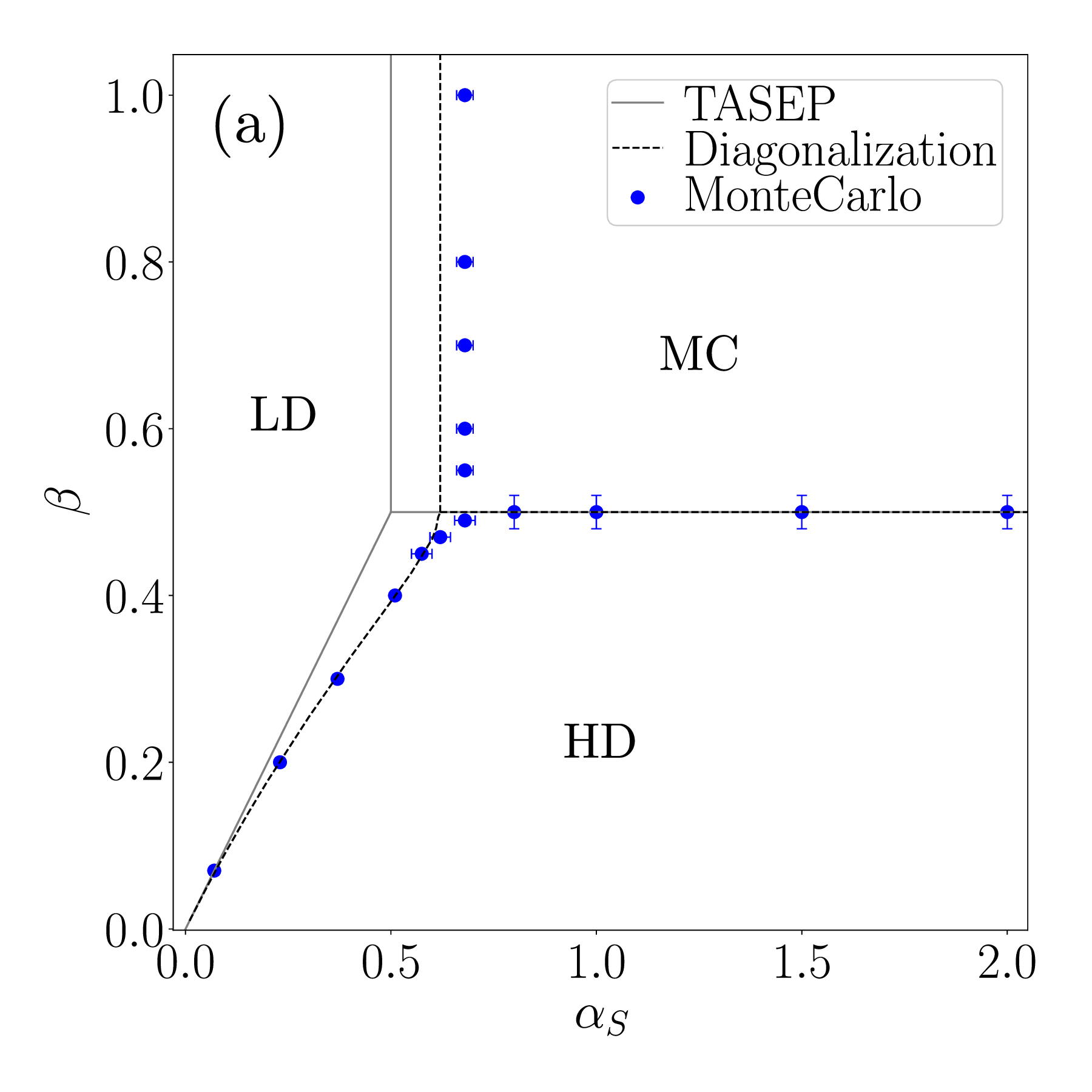}
  \phantomsubcaption
  \label{fig:DiagPhase_MC_ED_ParkingABC_qbINF_pa_0-20_qa_0-20_qa_1-00_b}
\end{subfigure}
\hfill
\caption{Phase diagrams of SFP in the asymptotic limit $q_F \to \infty$ (vanishing parking duration), for $p_S = 0.2$, displaying no re-entrant transition. (a) At $q_S=0.2$; for low $q_S$ values, no MC phase is observed. (b) At $q_S=1$; for larger $q_S$, an MC phase exits but appears for higher input rate $\alpha_S$ than TASEP (grey lines). Monte-Carlo simulations (blue dots) are generally in line with the predictions based on an exact diagonalization (dashed lines). }
\label{fig:DiagPhase_MC_ED_ParkingABC_qbINF_pa_0-20_qa_0-20_qa_1-00}
\end{figure}

Two distinct limits ($q_S \ll 1/L$ or $q_S \to \infty$)  reduce to single-species ($S$ or $F$, respectively) TASEP models, with their own maximum current phase (at $J=p_S/4$ or $J=1/4$, respectively). Otherwise, slow particles generate a bottleneck near the entrance, where the density is thus higher than in the bulk. It naturally follows that, for $p_S<1$, a maximum current phase $J=1/4$ may exist in SFP, contrary to TASEP with slow bond \cite{greulich_phase_2008}, but only exist for large enough $q_S$ (if cars do not dither any moment before parking), with a threshold $q_S^{\star}$ that will decrease as $p_S$ gets closer to 1. Fig.~\ref{fig:EvolutionJ_qa_differents_pa_MC_MF_ED_ParkingABC_qbINF_alphaA_1-00_betaB_0-60} confirms these expectations with Monte-Carlo simulations. They have been performed for $\alpha_S = 1$ and $\beta = 0.6$. Larger injection rate $\alpha_S$ (keeping $q_F \gg \alpha$) decreases the threshold $q_S^{\star}$, but is otherwise inconsequential. Therefore, this first mechanism may reduce the flow by creating a bottleneck at the entry that shift and possibly suppress the MC phase (see phase diagram in Fig.~\ref{fig:DiagPhase_MC_ED_ParkingABC_qbINF_pa_0-20_qa_0-20_qa_1-00}), 
but it hardly introduces any singularly new phenomenology into play compared to TASEP. In particular, the current increases monotonically with $\alpha_S$.

We defer the comparison between Monte-Carlo results, Exact Diagonalization and mean-field calculation to Annexe \ref{annexe:comparaison_MF_ED}. In a nutshell, discrepancies using the Exact Diagonalization occur when the total density is still not flat after $\sim 10$ sites as in the MC phase. Discrepancies using the mean-field approximation indicate correlations that have to be taken into account. Note that the effective MF theory of \cite{bottero_analysis_2017} is also inapplicable, because the number of  $S$ particles is not conserved.

\subsection{Disruptive effect of pull-outs}
We turn to the second flow-altering mechanism in the SFP model, namely, pull-outs (recall that in SFP pulling in does not cause any additional delay). To lay bare this, and only this, effect, we consider the case $p_S = p_F$ where cruising cars drive as fast as the other cars.

When a car pulls out, it fills a gap on the road. This may prevent a particle behind from moving forward, generating some congestion locally that notably triggers a backward-propagating congestion wave. But the current will only be affected by this perturbation if the waves reaches the entrance and hinders the injection of a vehicle; otherwise, it may be damped along the way, with no effect on the steady-state flow.
From the literature, it is known that if a slowdown is persistent and occurs always at the same location, then this `slow bond' will annihilate the MC phase even if the speed reduction is infinitesimal \cite{basu_last_2016}. In the same vein, and perhaps even more closely related, injection of particles in the bulk of a TASEP model via an `on-ramp' destroys the MC phase upstream of the ramp, even if the injection rate is very small \cite{xiao_investigation_2013}. One could therefore expect the MC phase to disappear from SFP as soon as $q_F < \infty$.

\begin{figure}[h]
\centering
\begin{subfigure}{.48\textwidth}
  \centering
    \includegraphics[width=0.8\textwidth]{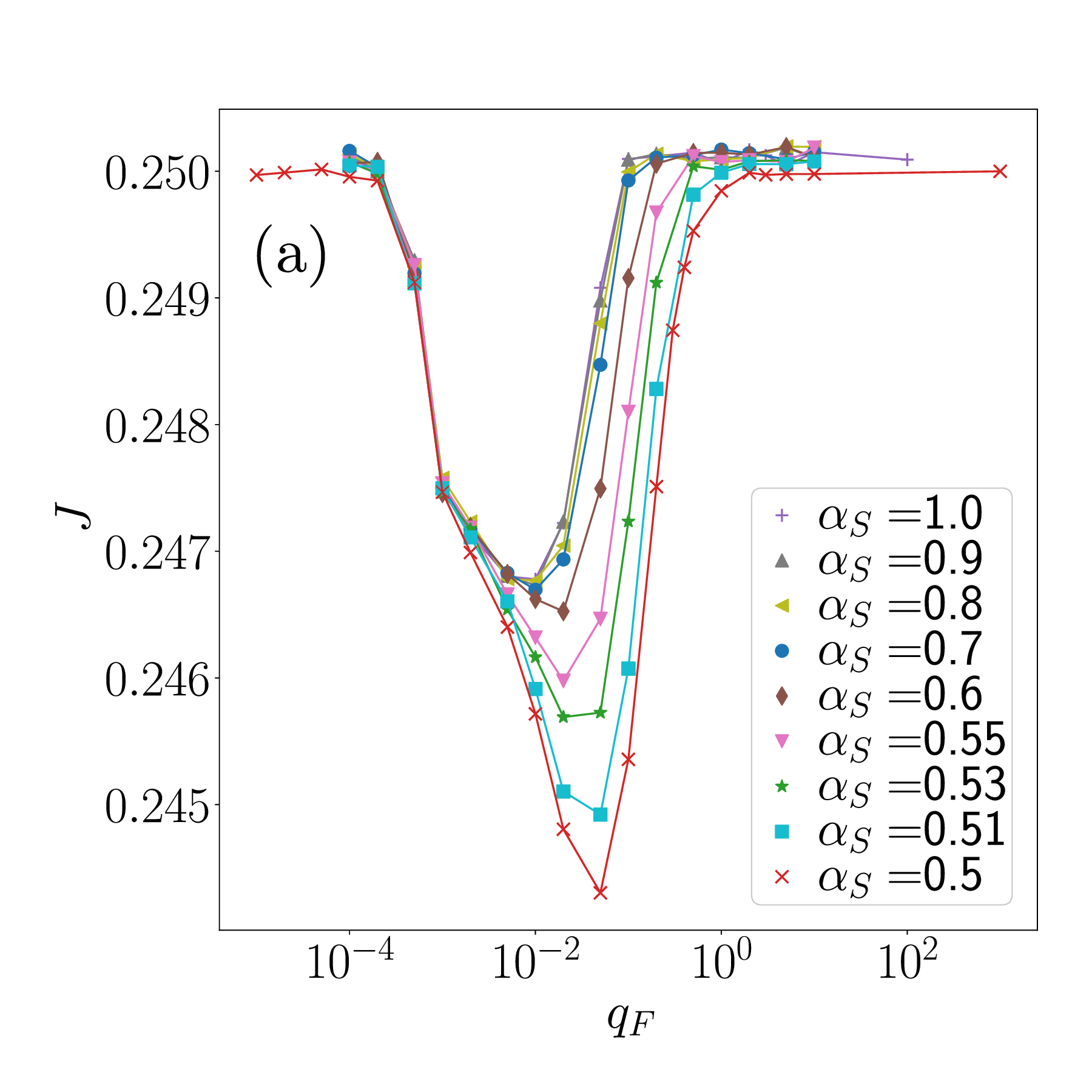}
    \phantomsubcaption
    \label{fig:Courant_fct_qb_qaINF_pa_1_a}
\end{subfigure}
\begin{subfigure}{.48\textwidth}
  \centering
  \includegraphics[width=0.8\textwidth]{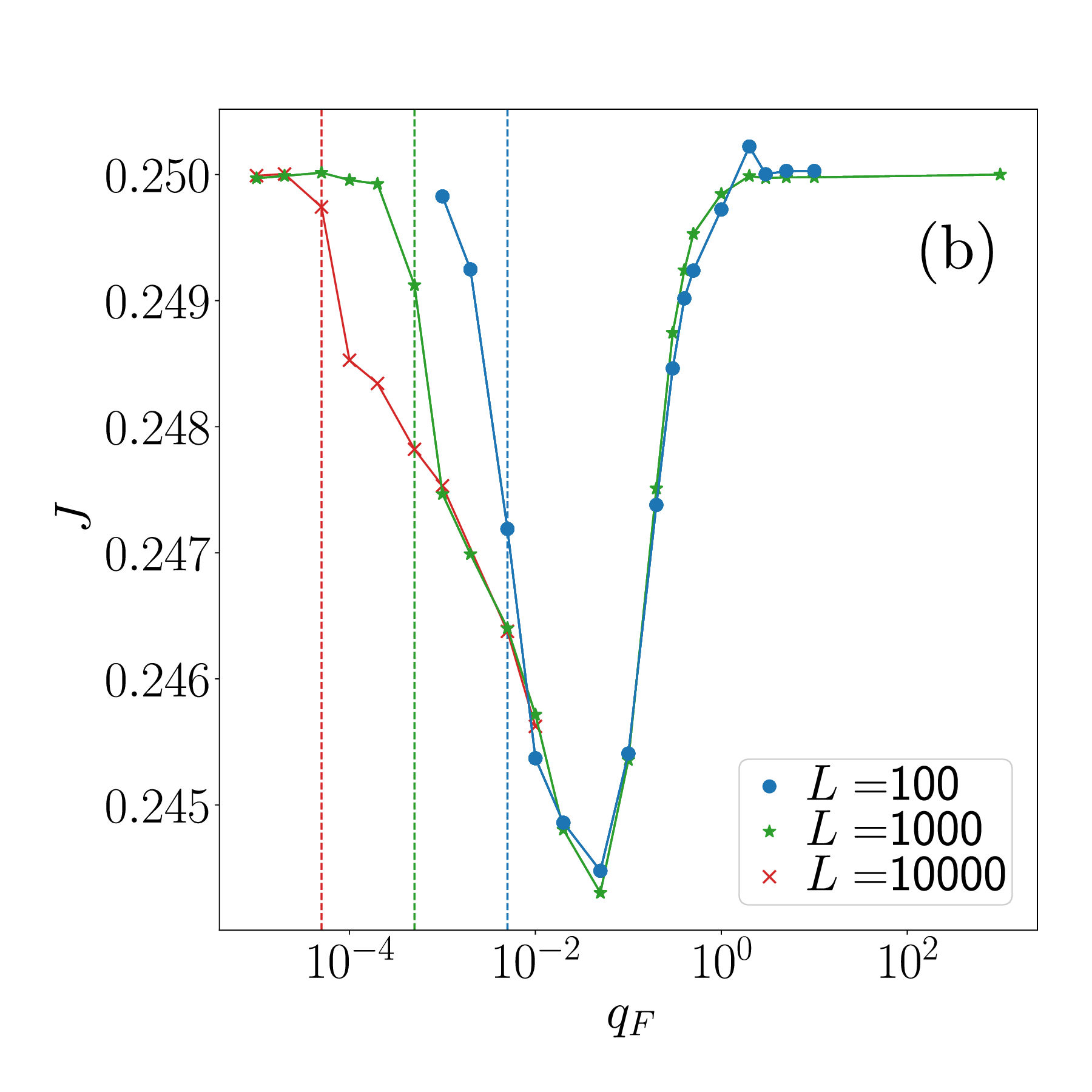}
    \phantomsubcaption
    \label{fig:Courant_fct_qb_qaINF_pa_1_b}
\end{subfigure}
\hfill
\caption{Impact of the inverse parking duration $q_F$ on the current $J$, computed with kMC, for $p_S=1$ (slow cars are as fast as fast ones), for (a) different input rates $\alpha_S$ at $L=1000$, (b) different system sizes $L$ at $\alpha_S = 0.5$. Pull-outs perturb the flow and suppress the MC phase $J=1/4$ for low enough $q_F$, except at
extremely low $q_F$, where MC is recovered due to finite-size effects.
The vertical dashed lines in panel (b), based on Eq.~\ref{eq:critical_qb_qaINF_finite_size}, represent the theoretical expectations of where such finite-size effects should play a role.
 Fixed parameters : $\beta = 0.6, q_S \to \infty$. }
    \label{fig:Courant_fct_qb_qaINF_pa_1}
\end{figure}

It turns out that this expectation is not fully justified at all $q_F$, because the random occurrence of pull-outs and the conservation of the number of particles undermine the foregoing analogies. Indeed, Fig.~\ref{fig:Courant_fct_qb_qaINF_pa_1} shows that  at $q_S \rightarrow \infty$ (i.e., cars park in the first vacant space) an MC phase  (and the full TASEP phenomenology) is
obtained if the pull-out rate is much greater than the hopping rate, viz. $q_F \gg p_S$, or vanishingly small, $q_F \to 0$. This last limit corresponds to the situation in which cars stay parked so long that
driving cars may travel through the whole system without encountering any pull-out. As $q_S\to \infty$, we can approximate the density of parked particles to be close to $1$ before a distance noted $L^*$ and $0$ after. Knowing that after $L^*$ there is no $S$ particles and equating the sum of the local current $q_F n_i^P (1-n_i^S-n_i^F)$ up to $i=L^*$ to the total current leads to 
$$ J = q_F L^*(1-\rho)$$
under the assumption that the total density in the road is roughly constant, i.e. $\rho = n_i^S+n_i^F$. Assuming an MC phase exists, then $J = 1/4$ and $\rho = 1/2$. This provides a relation ship between the leaving rate $q_F$ and the distance $L^*$:
\begin{equation}
    q_F^* = \frac{1}{2L^*}.
    \label{eq:critical_qb_qaINF_finite_size}
\end{equation}
Then, if the system size $L$ is lower than $L^*$, some $S$ particles can leave the system. For a given system size $L$, $q_F^*(L)$ provides an estimate below which new behaviors can occur due to the finite system size.
$q_F^*(L)$ corresponds to the dotted vertical line in Fig. \ref{fig:Courant_fct_qb_qaINF_pa_1_b}. This estimate correlates well with the value of $q_F$ at which the total current significantly deviates from that of larger system sizes.

Leaving these asymptotic limits of $q_F$, we notice that the density may surge locally following a pull-out maneuver. However, while pull-outs occur anywhere there are $P$ particles, these surges are observed mostly at a specific intermediate position (Fig.~\ref{fig:Densite_fct_qb_qaINF_pa_1_LD}), especially in the limit $q_S \to \infty$. This position corresponds to the point $i^{\star}$ where the road traffic gets depleted in $S$ cars, i.e., the end of the parking zone. Indeed, while parked cars at $i \ll i^{\star}$ also pull out, the excess density thus created is soon annihilated by the pulling in of an $S$ particle upstream, which creates a void. Only at $i \approx i^{\star}$ are there frequent pull-outs that can not be immediately screened as most of the particles behind are $F$ particles.
Having established the existence and location of density surges, it now behooves us to ascertain if the effect of this perturbation will propagate all the way up the road to the entrance in order to determine if this perturbation induces a difference with respect to the TASEP current.

\begin{figure}[h]
\centering
\begin{subfigure}{.48\textwidth}
  \centering
    \includegraphics[width=0.8\textwidth]{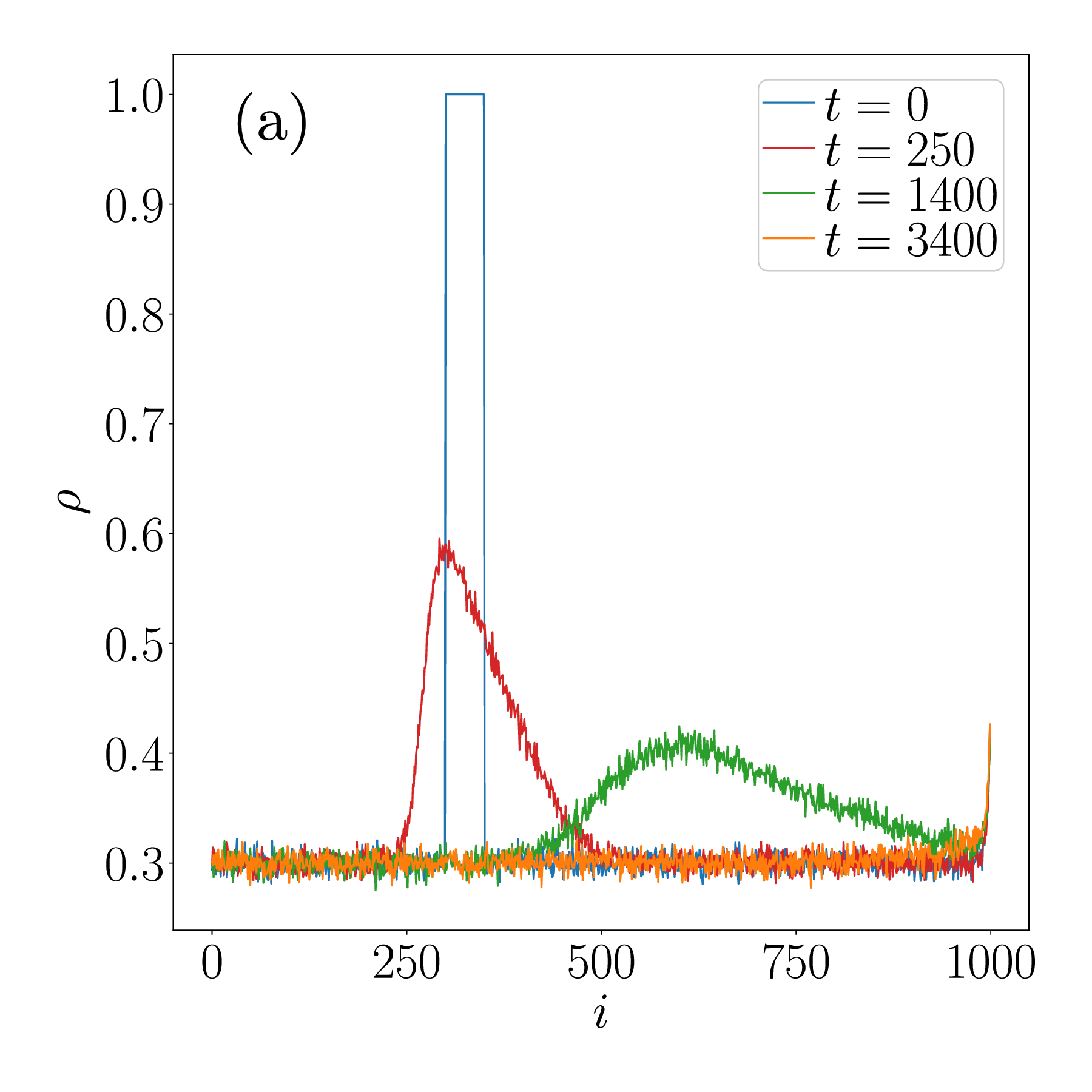}
    \phantomsubcaption
    \label{fig:EvolutionTemporelleDensite_pa_1_alphaA_0-5_beta_0-50_perturbation_500_L_1000_posPertub_300_timeStep_s_10-00000_blocSize_50_a}
\end{subfigure}
\begin{subfigure}{.48\textwidth}
  \centering
  \includegraphics[width=0.8\textwidth]{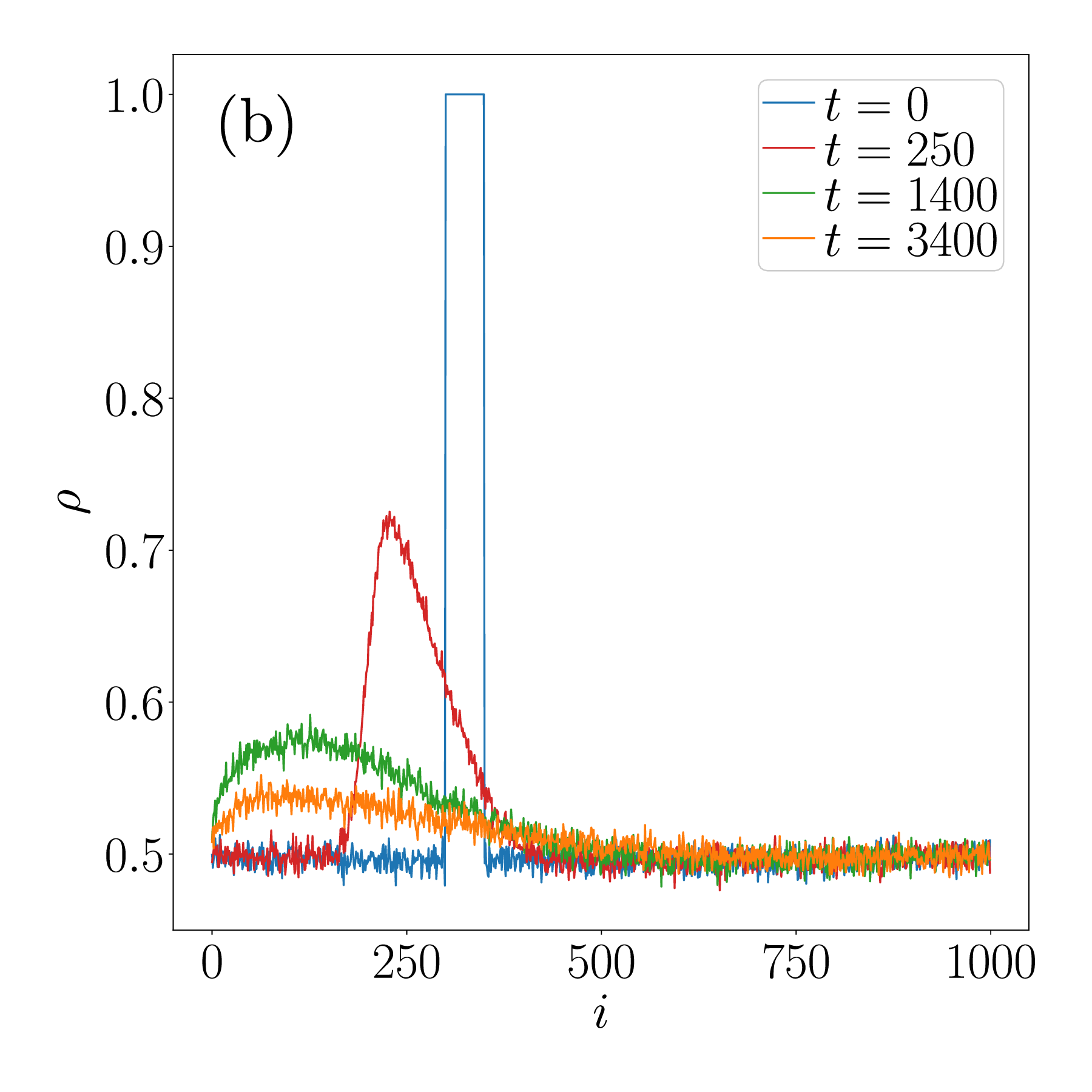}
  \phantomsubcaption
  \label{fig:EvolutionTemporelleDensite_pa_1_alphaA_0-5_beta_0-50_perturbation_500_L_1000_posPertub_300_timeStep_s_10-00000_blocSize_50_b}
\end{subfigure}
\hfill
\caption{Effect of an artificially inserted jam on the density profile of a standard TASEP flow ($\beta=0.5$) initially at steady state, at different times after the jam was created. The jam is created by suddenly filling all sites between $i=300$ and $i=349$). (a) In the LD phase, at $\alpha = 0.3$. The density bump first propagates backward (red), then frontward (green) and finally vanishes (orange).  (b) In the MC phase, at $\alpha=0.5$.  The density bump propagates only backward and has a long life time close to the left boundary. The results that are shown were averaged over 4000 independent realizations. }
    \label{fig:EvolutionTemporelleDensite_pa_1_alphaA_0-5_beta_0-50_perturbation_500_L_1000_posPertub_300_timeStep_s_10-00000_blocSize_50}
\end{figure}

\paragraph*{Propagation of a perturbation.}
\label{par:propagation_perturbation}
To settle this question, we perform `fictitious' auxiliary simulations in which a density excess is artificially introduced in the bulk of a steady-state standard TASEP flow. More precisely, at some $t_0$ in the steady flow, we fill all sites from $x_0$ to $x_1$, creating a rectangular bump in the density profile, and study the time evolution of this perturbation.
The results are averaged over 4000 independent realizations.  Fig.~\ref{fig:EvolutionTemporelleDensite_pa_1_alphaA_0-5_beta_0-50_perturbation_500_L_1000_posPertub_300_timeStep_s_10-00000_blocSize_50} shows one density wave traveling backward and sometimes one wave traveling forward.
The extent of the backward propagation of the perturbation strongly depends on the phase of the system. In the MC phase (Fig.~\ref{fig:EvolutionTemporelleDensite_pa_1_alphaA_0-5_beta_0-50_perturbation_500_L_1000_posPertub_300_timeStep_s_10-00000_blocSize_50_b}), 
only the backward wave is observed but it propagates over an extensive distance and reaches the leftmost boundary, thus interfering with injection.
This is in accordance with the aforementioned observation that the MC phase can be suppressed by perturbations due to pull-outs. By contrast, in the LD phase (Fig.~\ref{fig:EvolutionTemporelleDensite_pa_1_alphaA_0-5_beta_0-50_perturbation_500_L_1000_posPertub_300_timeStep_s_10-00000_blocSize_50_a}), the wave travels may temporarily move backward but it will eventually forward and the excess density is absorbed upstream by existing gaps or at the exit. 

The foregoing results can be quantitatively rationalized using the theory of domain walls (more details in Appendix~~\ref{annexe:rectangular_bump}). The artificial density bump creates two domain walls. The left domain wall separates a region of low density ($\rho_1$) to the left from a high-density one ($\rho_2>\rho_1$) to the right; this wall moves at a speed
$$ v_{\mathrm{DW}} = \frac{J(\rho_2) - J (\rho_1)}{\rho_2 - \rho_1},$$
where $J(\rho) = \rho(1-\rho)$. This formula is known for TASEP \cite{kolomeisky_asymmetric_1998}, but more generally it corresponds to the backward propagation speed of jams in first-order traffic models \cite{daganzo_cell_1994}. 
While the right domain wall separates a region of high density to the left from a low density one to the right. With time, it stretches linearly \cite{corwin_limit_2010}, causing a gradual decay of the density bump. 
Clearly, in MC phase, since $J(\rho_1)$ is maximal, the left wall necessarily moves to the left independently of $\rho_2$ except when $\rho_2 = \rho_1$.
By contrast, if the system is in LD, as the head of the jam sprawls and the density $\rho_2$ in the bump decays, one reaches a stage when $J(\rho_2)>J (\rho_1)$ and the left wall reverses its course from backward to forward. Thus, in the MC phase and in the LD phase, only if the left wall reaches the leftmost boundary, will the perturbation caused by pull-outs reach the entrance and alter the current. Furthermore, even under these conditions, the impact of these (instantaneous) maneuvers on the current tends to remain modest (compare the deviation of the LD bulk density observed  in Fig.~\ref{fig:Densite_fct_qb_qaINF_pa_1_LD} to $\alpha_S$ and see the modest current reduction in Fig.~\ref{fig:Courant_fct_qb_qaINF_pa_1_a}).

It is worth noting that, once the density adopts a given profile, its further evolution in TASEP is independent of the way in which it was created; thus, the foregoing analysis of backward (and forward) traveling waves has bearing on the impact of perturbations on the TASEP flow in general. 

\subsection{Effect of the coupling between slowing down at the entrance and re-insertion on the road}

\label{section:qsINF}

So far, neither of the two flow-altering mechanisms inspected in isolation above explains our singular observations of Sec.~\ref{section:novel_behavior} in their full extent. Therefore, we must explore the coupling of these mechanisms.
For simplicity, we investigate this coupling in the limit $q_S \to \infty$, i.e., when cars park as soon as they can.

Restoring the two mechanisms has the interesting consequence that the possibilities for cars to pull in or out can now impact the upstream traffic by enabling or not the following cars to park and thus affecting their driving speed. In other words, the longer survival of cruising cars if parked cars are prevented from pulling out because of the cruising traffic must be taken into account.

At a very qualitative level, one readily understands that this coupling effect can give rise to the singular non-monotonic behaviours evidenced above, in particular the non-monotonic variation of the current $J$ with the injection rate $\alpha_S$. Indeed, at high enough global densities on the road, if the injection rate is further increased, the density of cars on the first site(s) will get even higher, thus preventing parked cars from pulling out, especially if $\alpha_S \gg q_F$. As a consequence, the parking space availability will be reduced, forcing cruising cars to travel farther along the road before they park and turn into faster $F$ cars.

The existence of a non-monotonic behavior can also be established by a more quantitative study of the asymptotic regimes for the injection rate, when $p_S \ll q_F$. At low injection rates $\alpha \ll q_F$, an injected car will almost always pull in on the first site (recall that $q_S \to \infty$ here), stay parked for about $q_F^{-1}$, pull out and turn into $F$ (thus freeing the first parking spot), before the next particle is injected ($\alpha \ll q_F$). Thus, crossing the first site will typically take a time $T_1(\alpha \to 0)=q_F^{-1} + p_F^{-1}$, irrespective of $p_S$.
At the other extreme, for very high injection rates $\alpha \gg q_F$, the first parking spot will almost always be occupied, so that almost all injected cars will remain in the $S$ state at least on the first site. Crossing the first site thus typically takes a time $T_1(\alpha \to \infty)=p_S^{-1}$, regardless of $p_F$. 

\begin{figure}[!htbp]
\centering
\begin{subfigure}{.48\textwidth}
  \centering
    \includegraphics[width=0.8\textwidth]{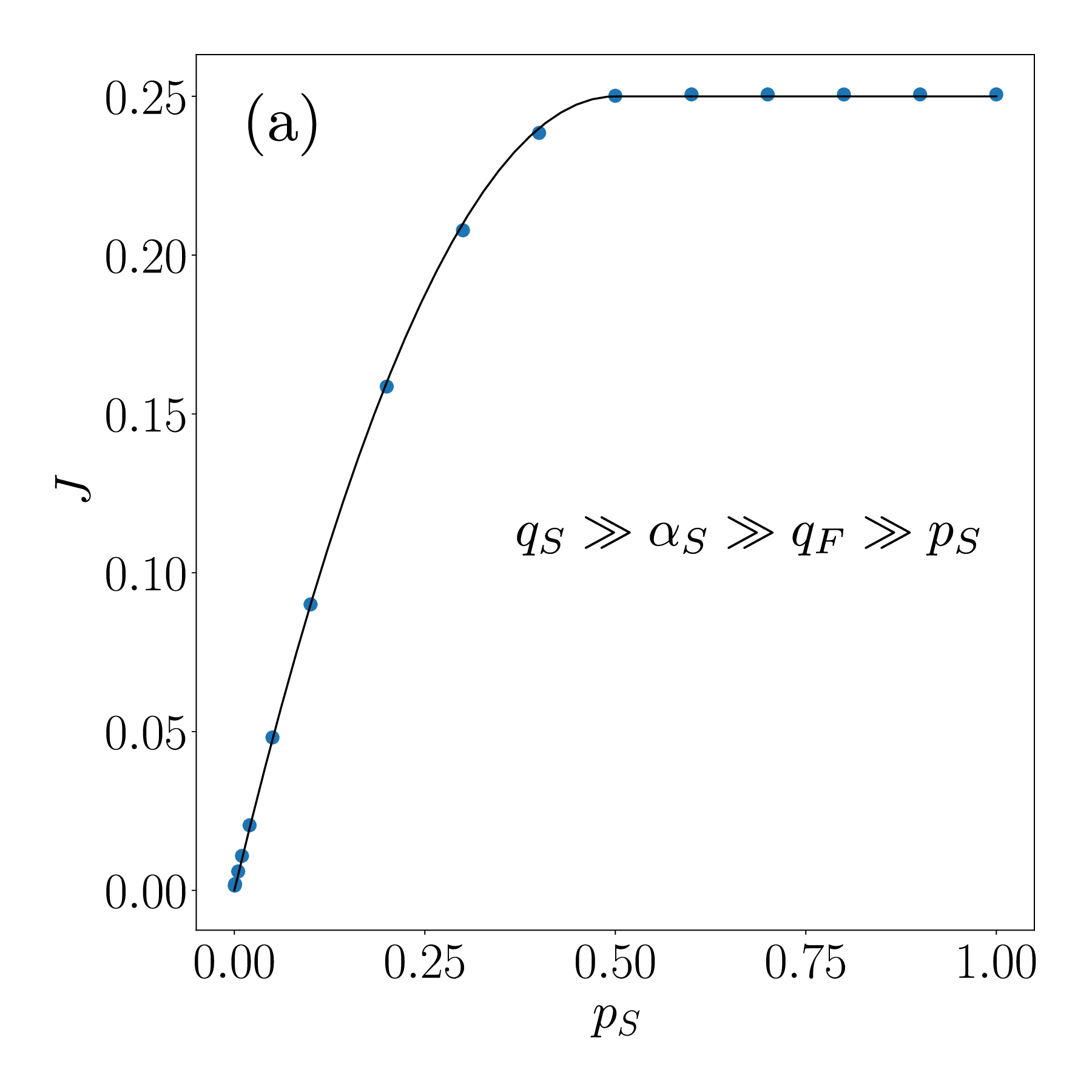}
    \phantomsubcaption
    \label{fig:Courant_fct_pa_ParkingABC_qb_10-00_qaINF_betaB_0-60_alpha100000_L_1000_a}
\end{subfigure}
\begin{subfigure}{.48\textwidth}
  \centering
  \includegraphics[width=0.8\textwidth]{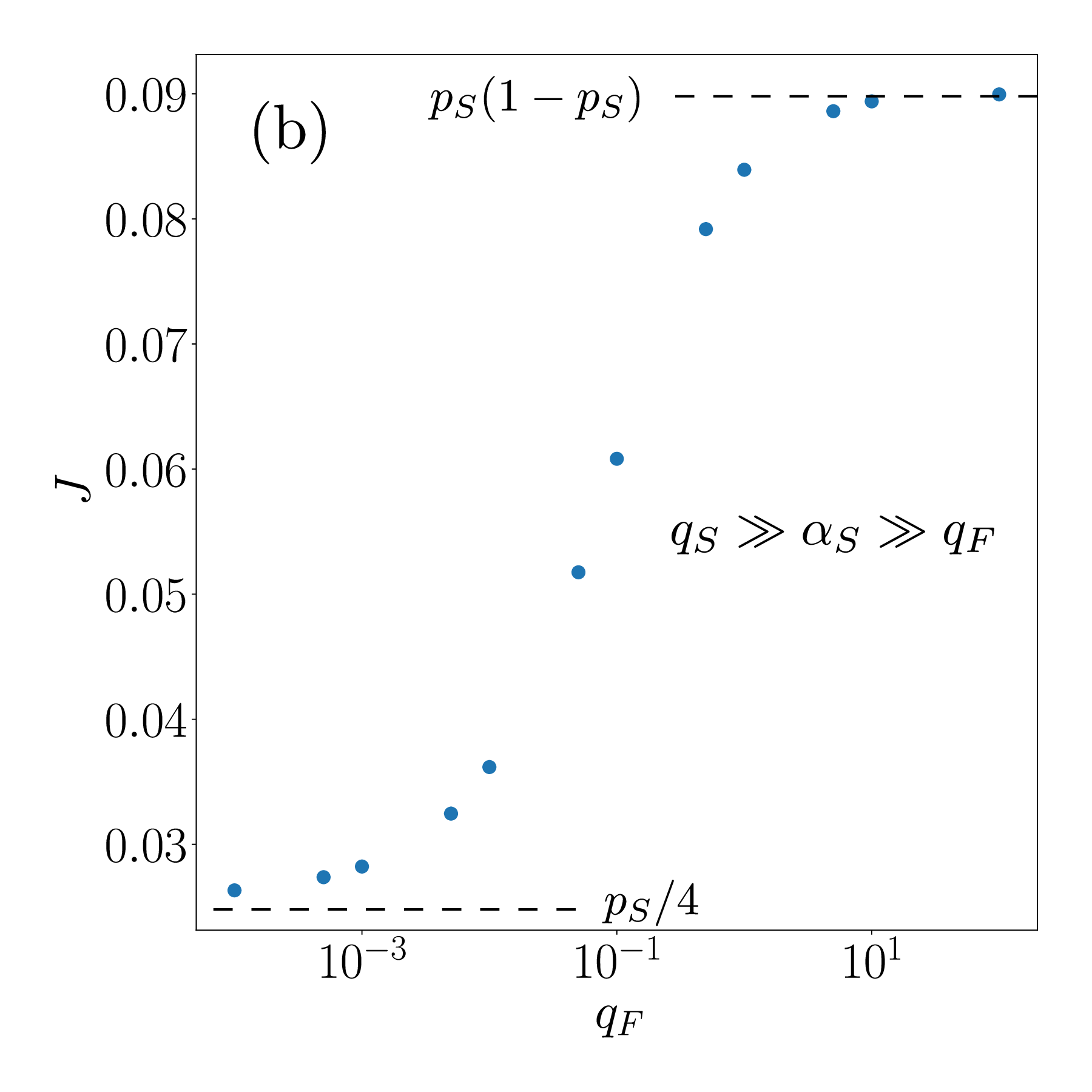}
  \phantomsubcaption
  \label{fig:Courant_fct_pa_ParkingABC_qb_10-00_qaINF_betaB_0-60_alpha100000_L_1000_b}
\end{subfigure}
\hfil
\caption{Comparison of the SFP current $J$ to theoretical predictions based on TASEP with a slow bond, for drivers extremely eager to park ($q_S= \infty$) and high but finite input rates relative to the parking duration, i.e., $\alpha_S \gg q_F$. (a) Current as a function of the (small) hopping rate $p_S$, for $p_S \ll q_F = 10$. The slow-bond prediction, $J=p_S(1-p_S)$, shown as a solid line, perfectly replicates the kMC results (dots).
(b) Current as a function of the inverse parking duration $q_F$, for $p_S=0.1$. The current interpolates between the theoretical predictions (dashed lines)
for a single-$S$-species TASEP at vanishing $q_F$ (cars stay parked forever) and for a slow-bond TASEP at large $q_F$ (cars pull out immediately).
 Fixed parameters : $\beta = 0.6, L=1000$. }
    \label{fig:Courant_fct_pa_ParkingABC_qb_10-00_qaINF_betaB_0-60_alpha100000_L_1000}
\end{figure}

It follows that, if cruising cars are slow ($p_S < p_F$), then particles remain longer in the first site for $\alpha \to \infty$ than for $\alpha \to 0$  (i.e., $T_1(\alpha \to 0)<T_1(\alpha \to \infty)=p_S^{-1}$), so fewer cars can \emph{effectively} be injected for $\alpha \to \infty$. More rigorously, the current for $\alpha \to 0$ is precisely that of a TASEP model featuring a slow first bond with a hopping rate $p_{\mathrm{slow}}=T_1(\alpha \to 0)^{-1}$, whereas the current for $\alpha \to \infty$ is bounded above by that of a TASEP model featuring a slow first bond with a hopping rate $p_{\mathrm{slow}}=p_S$ (this upper bound is reached if all $S$ particles have already turned into $F$ on the second site).  In the limit $\alpha \to \infty$, for a slow \emph{first} bond, the maximum current can be shown to be $p_{\mathrm{slow}}(1-p_{\mathrm{slow}})$, consistently with the asymptotic value obtained when $q_F \gg p_S$ in  Fig~\ref{fig:Courant_fct_pa_ParkingABC_qb_10-00_qaINF_betaB_0-60_alpha100000_L_1000_b}.
(More generally, using mean-field arguments, the maximum current in slow-bond TASEP is $p_{\mathrm{slow}}/(p_{\mathrm{slow}}+1)^2$ if the slow bond is in the bulk \cite{kolomeisky_asymmetric_1998}, but (positive or negative) edge effects are found when the slow bond gets close to the boundaries \cite{greulich_phase_2008}.)
Then, the SFP current can be smaller for $\alpha \gg q_F$ than for $p_S < \alpha \ll q_F$, which establishes the possibility of non-monotonicity. Figure~\ref{fig:Courant_fonctionAlphaA_MC_MF_ED_ParkingABC_pa_0-10_qaINF_betaB_0-60} illustrates this result for $p_S = 0.1,  \beta = 0.6$. In particular, for $q_F = 100$ (dark blue dots), when $\alpha_S < 2 \ll q_F$,  the current is identical to TASEP (black curve). Whereas when $\alpha_S = 500 > q_F$, the current is lower than $1/4$ and is ultimately bounded by $p_S(1-p_S)$ for $\alpha_S \gg q_F$. Note that this bound is not reached in Fig.~\ref{fig:Courant_fonctionAlphaA_MC_MF_ED_ParkingABC_pa_0-10_qaINF_betaB_0-60} as $\alpha_S=500$ is not large enough compared to $q_F = 100$. 

These results are valid when $\beta$ is large enough so that the exit does not act as a significant bottleneck. Given that the exit bottleneck limits the current to $\beta(1-\beta)$ when $\beta < 1/2$, transitions are found when $J(\alpha) = \beta(1-\beta)$. 
This results in interesting re-entrant transitions, LD-MC-LD as well as HD-MC-HD, when  $J(\alpha_S)$ itself depends non-monotonically on $\alpha_S$,
as reported in Fig.~\ref{fig:DiagPhase_MC_ED_ParkingABC_qaINF_pa_0-10_qb_100-00_a}
 for $q_F = 100$ and $p_S = 0.1$. For longer parking durations (smaller $q_F$), the MC phase no longer exists and only the LD-HD-LD re-entrant transition is observed (Fig.~\ref{fig:DiagPhase_MC_ED_ParkingABC_qaINF_pa_0-10_qb_100-00_b}).
 What further underlines the relation between the non-monotonic flow variations and these re-entrant transitions is that, when the cruising speed $p_S$ is increased above $1/2$, MC phase can be restored in the limit $\alpha \to \infty$ as the bound $p_S(1-p_S)>1/4$ no longer hold. Nonetheless, it happens only for large enough $q_F$ and for intermediate values, re-entrant transitions are maintained as depicted in Fig.~\ref{fig:ColorPlot_Courant_fct_alphaA_qb_plusieurs_pa_qaINF_betaB_0-60_L_200} but shrinks as $p_S\to 1$.


\paragraph*{Parallel update algorithm.}
\label{section:parallelUpdate}
Complementary insight into the origin of the observed non-monotonicity can be gained by changing the random sequential update rule of the algorithm for a parallel update scheme. For specific values of the parameter, the parallel update algorithm becomes deterministic, yielding a periodic evolution of the system. Besides, while the change of algorithm to quantitatively impact the lines of transition, we robustly observed non-monotonicity with both schemes (see the variations of $J$ with $\alpha_S$ plotted in Fig.~\ref{fig:DensiteCourant_parallel_fctAlphaA_plusieurs_qa_MC_ParkingABC_pa_0-10_qb_0-50_beta_1-00}). Accordingly, it is 
possible to study the emergence of such non-monotonicity as a perturbation to the periodic evolution.

More precisely, we will focus here on the non-monotonicity of $J$ as a function of $q_S$ (and $q_F$) rather than as a function of $\alpha_S$ for simplicity (see Annexe \ref{annexe:deterministic_parallel_update} for more details). We start with the parameters $p_S = q_F = \alpha_S = \beta_B = 1$. For $q_S =\infty$ (a $S$ particle  parks immediately), the model becomes deterministic and, listing the possible configurations of the first two sites, we notice that a new particle is, on average, introduced every three time steps due to our update rule sequence, wherein the lane is updated before the parked vehicles. Consequently, the current is $J=1/3$, contrasting with the maximum of $J=1/2$ achieved for $q_S = 0$, i.e., for  \emph{bona fide} parallel-update TASEP. In-between, the current undergoes a dip. This can be explained by the fact that, as $q_S$ departs from $\infty$, sometimes a $S$ particle  will not park but try to move forward instead. This gives rise to a reduced current, elaborated in Appendix \ref{annexe:deterministic_parallel_update}, and establishes the non-monotonic behavior (see Fig.~\ref{fig:Current_parallel_fct_qA_different_qb_MC_ParkingABC_pa_1-00_alphaA_1-00_beta_1-00_a}).   
Similarly, now focusing on the variations with $q_F$ at $q_S \to \infty$, as $q_F$ gets smaller than 1, sometimes a particle parked along the first site will remain parked instead of pulling out as soon as possible. Evaluating all possible configurations where this may happen reveals that it results in the injection of three particles over ten time steps, which is lower than $1/3$. This explains the drop in current when $q_F$ hovers slightly below 1, substantiating the non-monotonic trend. 
The two flow-altering mechanisms specific to SFP explain these results. When $q_F = 1$, as $q_S$ increase from zero, the current decreases because parked cars pull-out into the main road closer to the left boundary and then have higher probability to create a perturbation that will reach it. It also means that the density close to the left boundary increases. Then, a bottleneck is formed and the two mechanisms occur. While $q_S$ further increase, the current start to increase because of a reduction of the bottleneck effect as $S$ particles minimize their time on the road by pulling-in the parking spot as soon as possible and reach $J=1/3$ for $q_S \to \infty$.    

Lowering the hopping probability $p_S$ changes radically the behavior of the current with $q_S$ (Fig.~\ref{fig:Current_parallel_fct_qA_different_qb_MC_ParkingABC_pa_1-00_alphaA_1-00_beta_1-00_b}) but is still well explained with the two flow-altering mechanisms. For low $q_S$, bottleneck effect dominates as the road is mainly occupied by slow particles. As $q_S$ increases, slow particles spend less time on the road, and fast particles can move more freely increasing the current. But increasing $q_S$ also increases the effect on the entrance of parked particles pulling-out into the main road. A compromise might be then found resulting in a peak of the current at an intermediate rate $q_S$ which is observed in Fig.~\ref{fig:Current_parallel_fct_qA_different_qb_MC_ParkingABC_pa_1-00_alphaA_1-00_beta_1-00_b}. 
Exact expression of the current in some specific limiting cases and an analysis of the gap distribution in Annexe \ref{annexe:deterministic_parallel_update} support this explanation.




\subsection{Impact of the boundary conditions}
\label{sub:topology}
\begin{figure}[htbp]
    \centering
    \includegraphics[width=0.8\columnwidth]{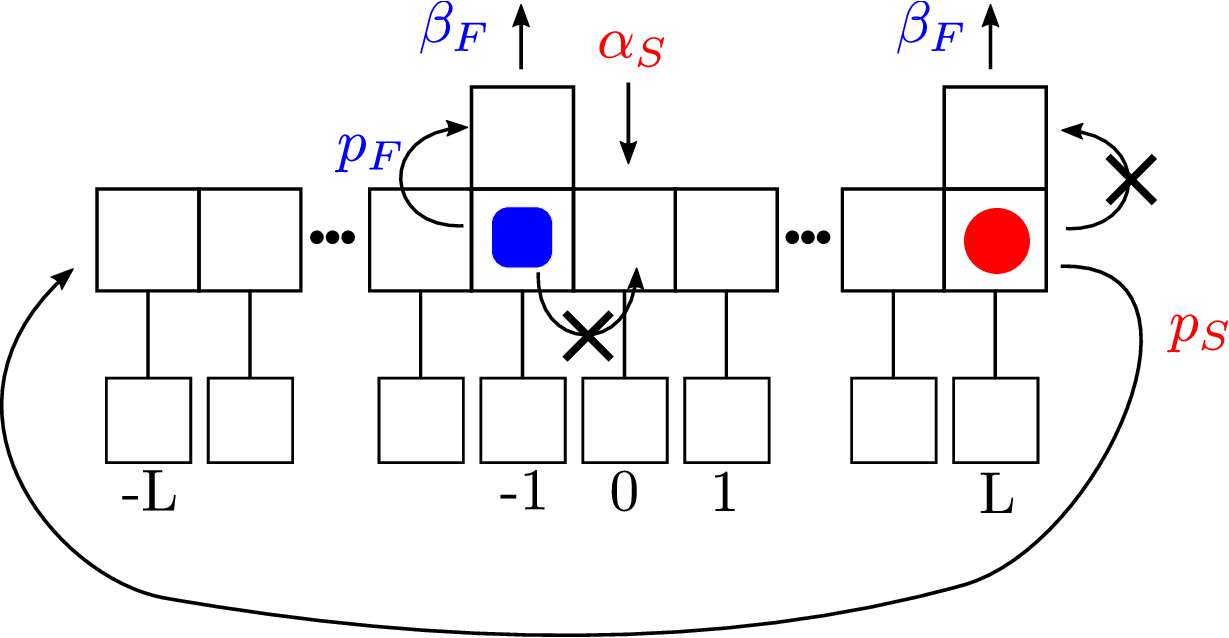}
    \caption{Sketch of the SFP dynamics with periodic boundary conditions, i.e., on a ring. $S$-particles  are injected at site $i=0$ and cannot leave the system.  $F$-particles egress as soon as they encounter an exit, at $i=L$ or $i=-1$.}
    \label{fig:Schema_parking_ABC_ring_topology}
\end{figure}

Here, we show that the non-monotonicity evidenced in SFP with an infinite chain with open boundaries survives, and is actually substantially enhanced, if periodic boundary conditions are imposed instead, hence, a ring topology with a finite number of sites $L$ as displayed on Fig.~\ref{fig:Schema_parking_ABC_ring_topology}. Cruising cars $S$ are injected at the site $0$ with an input rate $\alpha_S$ and will circle around the ring until they can park \cite{shoup_cruising_2006}. There are two exit points. On the other hand, particles $F$ exit the system at the first encountered exit point with a rate of $\beta$. 
For convenience, we consider an update rule in which sites are updated sequentially, but in a random order, every fixed (small) time step $dt$ ($dt=0.1)$; this converges to the random update rule strict speaking only in the mathematical limit $dt\to 0$.

With these boundary conditions, particles $S$ that hop from site $-1$ to site $0$ compete with incoming particles. Intuitively, one might assume that this process would raise the density of $S$ particle  over time, eventually saturating it. However, this is not the case, as the system's dynamics intricately depend non monotonously on the density of particles $S$.

When the propensity to park $q_S$ is large (dark blue dots in Fig.~\ref{fig:Courant_fonctionAlphaA_MC_plusieurs_qa_qb_0-001_pa_0-70_betaB_0-60_RingGeometry_L_200}), the results are identical to the infinite lane as all particles $S$ turned into $F$ particles before their first loop. On the other hand, as $q_S$ decreases, more and more cruising particles circle around the loop leading to larger current through the first site $J_{0\to 1}$ at small input rate and then to a decrease of the injection current $J_{\mathrm{in}}$ at fixed $\alpha_S$ with $q_S$. More importantly, when  $q_S < 0.01$, both the injection current $J_{\mathrm{in}}$ and the current through the first site ($J_{0\to 1}$)  depend non monotonically with $\alpha_S$, for $q_F=0.001$, as shown in Fig.~\ref{fig:Courant_fonctionAlphaA_MC_plusieurs_qa_qb_0-001_pa_0-70_betaB_0-60_RingGeometry_L_200}.

Finally, since only particles $F$ can leave the system, a blocking state can emerge wherein all road sites are occupied, immobilizing the particles. This happens for all $q_S < 0.01$ considered around $\alpha_S \sim 0.5$. Since a blocking state is inescapable and randomness governs the dynamics, calculating averages would need to consider a substantial number of samples which have not been done. 

\section{Discussion and Conclusion}

In conclusion, we have introduced a new variant of TASEP including searching-for-parking (SFP) behavior: two distinct particle species, denoted as $S$ and $F$, drive along the main road; $S$-particles can park on parking spots along the road and parked cars are (irreversibly) converted into $F$ when they leave their parking space and get back on the road. Thus, our SFP model is reminiscent of several TASEP variants, such as TASEP with pockets \cite{humenyuk_separation_2020, bhatia_far_2022}, an on-ramp \cite{xiao_investigation_2013, liu_effect_2016} or a reservoir of particles \cite{ha_macroscopic_2002, adams_far--equilibrium_2008, cook_feedback_2009, greulich_mixed_2012}, but nonetheless it markedly differs from them in that only the total number of particles is conserved, because of the irreversible conversion to $F$ along the road.

This feature sets the model apart: on the basis of Monte-Carlo simulations, supplemented with an analysis resorting to mean-field theory and exact diagonalization,  singular non-monotonic variations of the flow rate with the injection rate, the leaving rate, and the parking rate are observed, for a certain range of parameters. This stands in stark contrast with the monotonic variations obtained with TASEP and it gives rise to re-entrant transitions in the phase diagram, such as LD-MC-LD or LD-HD-LD. Interestingly, these singular features subsist regardless of the update rule used, whether random or fully parallel, and the (open or periodic) boundary conditions, which points to a fairly robust effect.

We have ascribed the non-monotonicity to a `cramming' effect (loosely similar in its cause to a paper jam in a printer) that arises because the large arrival rate of $S$-cars on the first sites blocks the departure of parked cars from their parking spaces, which in turn impedes the parking of newly injected slow ($S$) cars and their conversion to $F$. More quantitatively, the persistence of $S$-cars can be handled as a slow bond at the entrance \cite{kolomeisky_asymmetric_1998, dong_inhomogeneous_2007, greulich_phase_2008, basu_last_2016}. Even when $S$ and $F$ particles move at the same speed, the effect manifests itself in the form of a bump in the density profile of the LD phase and the system's inability to reach the expected maximum current of $J=1/4$ in the MC phase.

Because of the particularly stylized nature of the model, it is worth wondering to what extent the foregoing features may survive in real-world traffic. Their survival should not be taken for granted in any way. First, one should bear in mind that non-monotonicity is only observed for particularly short parking durations ($p_S / q_S \approx 1$), shorter than typical drop-off times 
(These somewhat unrealistic values are however brought much closer to real-world figures for other boundary conditions and topologies, such as the ring that we studied.) Secondly, variations in the model rules that may bring it closer to reality may suppress the singular features. Therefore, it is not granted that one could observe non-monotonic variations of the throughput of a street portion with curbside parking as the arrival rate of cars in search of parking increases.

That being said, we believe that our findings have two significant implications for real-world applications. 
Firstly, they provide compelling evidence that even the very qualitative trends (here, the monotonic flow rate dependence on the arrival rate in TASEP) predicted by idealized traffic models that overlook perturbations  can be upset if disturbances to traffic are duly taken into account (here, related to parking search). The qualitative predictions for idealized traffic should thus be taken with a grain of salt, in some cases. On the brighter side, our paper also shows that, even when some perturbations are included in the model, the general features of its output (possibly distinct from those of the stylized model) are still amenable to theoretical understanding.
Secondly, the `cramming' effect unveiled by the study is not specific to the SFP model. Therefore, this mechanism will still play a role in instances of real traffic, even if its nonmonotonic repercussions in the idealized case are masked by a variety of other factors in the real world. In fact, this effect is probably of much broader scope and may be applicable in other contexts in a generalized form. For instance, in biochemistry, it is related to the alleged possibility of a non-monotonic evolution of the reaction rate with the cell volume (hence, the concentration of reactants), where the growth of the reaction rate at low concentration may be opposed by crowding effects at very high concentrations, which limit the diffusion of reactants \cite{schmit2009lattice}. More tentatively, one may suspect that, in any binary macrochemical reaction with a large transition volume, an oversaturation of reactants will not leave enough room for the reaction to take place, hence producing slower kinetics as the concentration of reactants increases, at very high concentrations.


\section*{Acknowledgement}
We aknowledge financial support from institut de physique du CNRS (Emergence program).

\appendix
\section{Master equations}

In the asymptotic limits considered in the main text, notably for $q_F \rightarrow \infty$ or $q_S \rightarrow \infty$, the master equations of the SFP model should be written with specific expressions, detailed here.

\subsection{Master equation for $q_F \rightarrow \infty$}
\label{annexe:master_equation_qF}
When $q_F \rightarrow \infty$, the transmutation from $S$ particles to $F$ takes place on the road, instantly, without an intermediate $P$ state. 
The master equations then reduce to 
\begin{align*}
    \frac{d \langle n_i^S \rangle}{dt} =& p_S \langle n_{i-1}^S(1-n_i^S-n_i^F) \rangle - p_S \langle n_i^S(1-n_{i+1}^S-n_{i+1}^F) \rangle \\ &- q_S \langle n_i^S \rangle \\
    \frac{d\langle n_i^F \rangle}{dt} =& p_F \langle n_{i-1}^F(1-n_i^S-n_i^F) \rangle - p_F \langle n_i^F(1-n_{i+1}^S-n_{i+1}^F) \rangle \\ &+ q_S \langle n_i^S \rangle \\
\end{align*}
with boundary conditions
\begin{align*}
    \frac{d\langle n_1^S \rangle}{dt} =& \alpha_S(1-\langle n_1^S \rangle-\langle n_1^F \rangle) - p_S\langle n_1^S(1-n_2^S-n_2^F) \rangle \\ &- q_S \langle n_1^S \rangle \\
    \frac{d\langle n_1^F \rangle}{dt} =& \alpha_F(1-\langle n_1^S \rangle-\langle n_1^F \rangle) - p_F\langle n_1^F(1-n_2^S-n_2^F) \rangle \\ &+ q_S \langle n_1^S \rangle\\
    \frac{d\langle n_L^S \rangle}{dt} =& -\beta_S \langle n_L^S \rangle + p_S\langle n_{L-1}^S(1-n_L^S-n_L^F) \rangle- q_S \langle n_L^S\rangle \\
    \frac{d\langle n_L^F \rangle}{dt} =& -\beta_F \langle n_L^F \rangle + p_F\langle n_{L-1}^F(1-n_L^S-n_L^F) \rangle+ q_S \langle n_L^S \rangle
\end{align*}

In the continuous (hydrodynamic) limit $i \to x$, under a mean-field approximation, the equations turn into
\begin{align*}
\frac{d\rho^S}{dt'} =&- p_S \partial_x \rho^S  ( 1-2\rho^S-\rho^F) + p_S \frac{\epsilon}{2}\partial_x^2\rho^S (1-\rho^F ) \\ &+ p_S \rho^S( \partial_x \rho^F + \frac{\epsilon}{2}\partial_x^2\rho^F)- L q_S \rho^S\\
    \frac{d\rho^F}{dt'} =& - p_F\partial_x \rho^F\left[1-2\rho^F-\rho^S \right] +  p_F\frac{\epsilon}{2}\partial_x^2\rho^F(1-\rho^S) \\ &+ p_F \rho^F (\partial_x \rho^S +  \frac{\epsilon}{2}\partial_x^2\rho^S) +  L q_S \rho^S 
\end{align*}

\subsection{Master equation for $q_S \rightarrow \infty$}
\label{annexe:master_equation_qS}

When $q_S \rightarrow \infty$, particles $S$ pull in a parking spot as soon as they encounter a vacant space, which entails the following master equation
\begin{align*}
    \frac{d \langle n_i^S \rangle}{dt} =& p_S \langle n_{i-1}^S(1-n_i^S-n_i^F)n_i^P \rangle \\ &- p_S \langle n_i^S(1-n_{i+1}^S-n_{i+1}^F) \rangle \\
    \frac{d\langle n_i^F \rangle}{dt} =& p_F \langle n_{i-1}^F(1-n_i^S-n_i^F) \rangle \\ &- p_F \langle n_i^F(1-n_{i+1}^S-n_{i+1}^F) \rangle \\ &+ q_F \langle n_i^P(1-n_i^S-n_i^F) \rangle \\
    \frac{d\langle n_i^P \rangle}{dt} =& p_S \langle n_{i-1}^S(1-n_i^S-n_i^F)(1-n_i^P) \rangle  \\ &- q_F \langle n_i^P(1-n_i^S-n_i^F) \rangle
\end{align*}
with $i = 2, ..., L-1$ for  species $S$ and $F$ and $i=2,...,L$ for species $P$. 

The boundary conditions read
\begin{align*}
    \frac{d\langle n_1^S \rangle}{dt} =& \alpha_S\langle(1- n_1^S - n_1^F )n_1^P\rangle  - p_S\langle n_1^S(1-n_2^S-n_2^F) \rangle \\
    \frac{d\langle n_1^F \rangle}{dt} =& \alpha_F(1-\langle n_1^S \rangle-\langle n_1^F \rangle) - p_F\langle n_1^F(1-n_2^S-n_2^F) \rangle \\ &+ q_F \langle n_1^P(1-n_1^S-n_1^F) \rangle\\
    \frac{d\langle n_1^P \rangle}{dt} =& \alpha_S\langle(1- n_1^S - n_1^F )(1-n_1^P)\rangle \\ &- q_F \langle n_1^P(1-n_1^S-n_1^F) \rangle\\
    \frac{d\langle n_L^S \rangle}{dt} =& -\beta_S \langle n_L^S \rangle + p_S\langle n_{L-1}^S(1-n_L^S-n_L^F)n_L^P \rangle \\
    \frac{d\langle n_L^F \rangle}{dt} =& -\beta_F \langle n_L^F \rangle + p_F\langle n_{L-1}^F(1-n_L^S-n_L^F) \rangle \\ &+ q_F \langle n_L^P(1-n_L^S-n_L^F) \rangle
\end{align*}

In the hydrodynamic limit $i \to x$, the mean-field equations can then be written as
\begin{align*}
    \frac{d\rho^S}{dt'} =& Lp_S\rho^S\left[1-\rho^S-\rho^F \right] \\ &+ p_S \rho^S( \partial_x \rho^F + \frac{\epsilon}{2}\partial_x^2\rho^F)(\rho^P-1) \\ &- p_S \partial_x \rho^S  ( \left[1-\rho^S-\rho^F \right] \rho^P - \rho^S)  \\ &+ p_S \frac{\epsilon}{2}\partial_x^2\rho^S ( \left[1-\rho^S-\rho^F \right] \rho^P + \rho^S) \\
    \frac{d\rho^F}{dt'} =& - p_F\left[1-2\rho^F-\rho^S \right]\partial_x \rho^F +  p_F \rho^F (\partial_x \rho^S + \frac{\epsilon}{2}\partial_x^2\rho^S) \\ &+ \frac{\epsilon}{2}p_F(1-\rho^S)\partial_x^2\rho^F  + L q_F \rho^P(1-\rho^S-\rho^F) \\
    \frac{d\rho^P}{dt'} =&  Lp_S(\rho^S-\epsilon \partial_x \rho^S + \frac{\epsilon^2}{2}\partial_x^2\rho^S)\left[1-\rho^S-\rho^F \right](1-\rho^P) \\ &- L q_F \rho^P(1-\rho^S-\rho^F).
\end{align*}


\section{Case of transmutation into an identical particle ($q_F\to \infty$)}
\label{app:Identical_transmutation}
In this Appendix, we consider the limit of vanishing parking duration, $q_F\to \infty$, which is equivalent to having $S$-particles turn into $F$ on the road, while $S$ and $F$ have identical properties ($p_S=p_F$). Obviously, the global dynamics are those of a single-species TASEP; only the distribution of the labels ($S$ or $F$) of the particles along the road deserve further investigation.

Applying the mean-field (MF) approximation of Sec.~\ref{annexe:master_equation_qF}, the occupation of species $S$ in the steady state is related to the ensemble or time-averaged total occupation $\rho_i=\langle n_i \rangle$ by
\begin{equation}
\rho_i^S = \frac{\rho_{i-1}^S  (1- \rho_i)}{1-\rho_{i+1} +q_S}.
\end{equation}
It follows that, in LD, where we expect $\rho_i = \rho_{i+1} =\rho_b$ even at the left boundary, the density of $S$ particles decreases as
\begin{equation} \rho_i^S = C \rho_{i-1}^S
\end{equation}
where $C = \frac{1-\rho_b}{1-\rho_b +q_S}$ is a constant.

\begin{figure}[htbp]
    \centering
    \includegraphics[width=0.8\columnwidth]{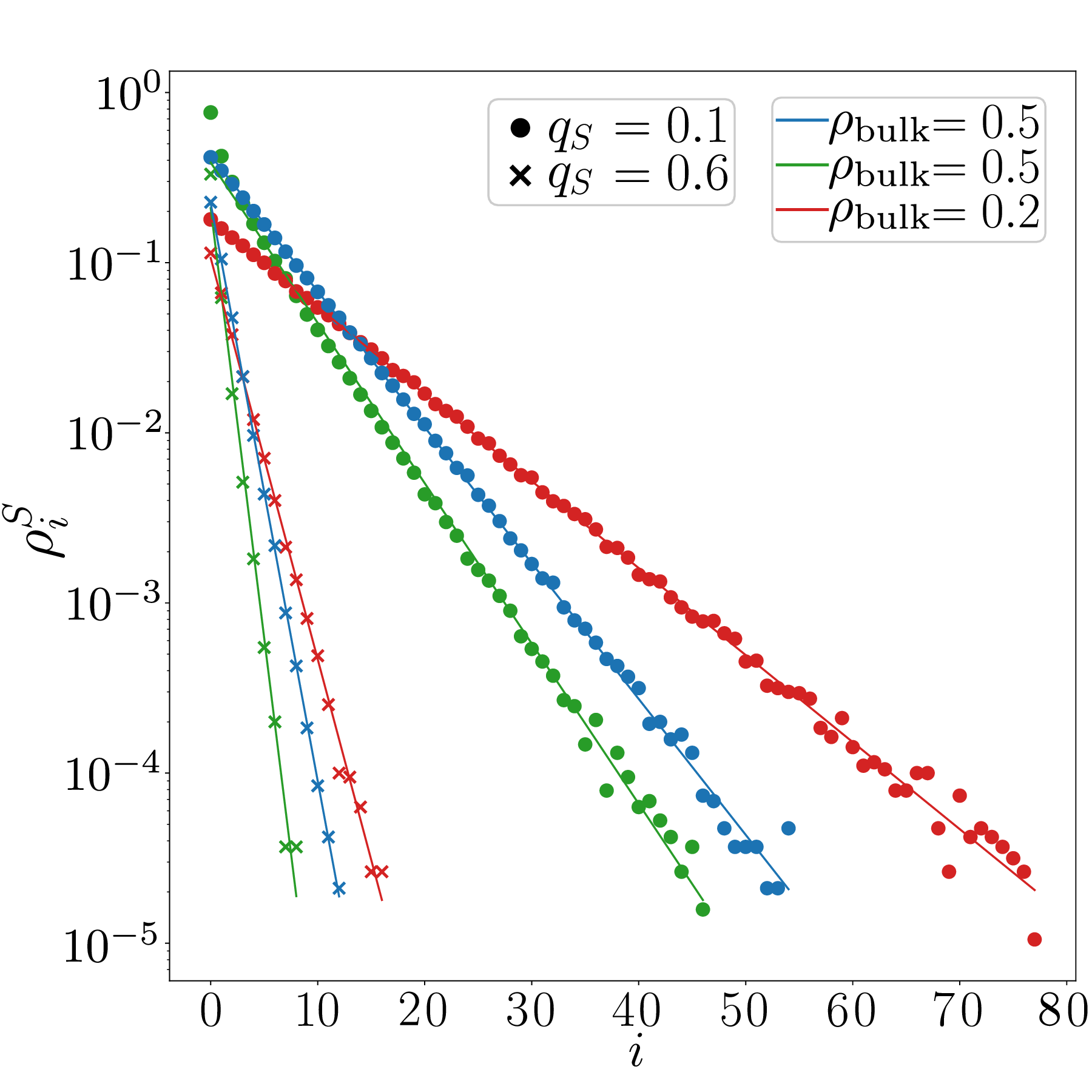}
    \caption{Density profile $\rho_i^S = \langle n_i^S\rangle$ of particles $S$ for a cruising speed $p_S = 1$, in the MC phase (blue: $\alpha_S = \beta_B = 0.5$; green:  $\alpha_S = 100, \beta_B =0.5$) and in the LD phase (red, $\alpha_S = 0.2$). Dots: $q_S = 0.1$; crosses: $q_S = 0.6$.}
    \label{fig:Evolution_DensiteA_MC_LD_pa_1-00_qb_100000-00}
\end{figure}

The semi-logarithmic plot of Fig.~ \ref{fig:Evolution_DensiteA_MC_LD_pa_1-00_qb_100000-00}
shows that these MF expectations are generally supported
by kMC simulations, with a linear decrease of $\ln \rho_i^S$ with coordinate $i$. 
Furthermore, the fitted slopes are in very good agreement with the MF prediction for $C$, in the LD phase (in red) as well as in MC (in blue). It is not surprising as the phase diagram of TASEP with MF is identical to its true phase diagram. Near the first-order transition or in MC with $\alpha > 1/2$, the approximation of uniform density at the left boundary fails and so is the simple relation $\rho_i^S = C \rho_{i-1}^S$. Instead, one should use the density variation close to the boundaries exactly known in TASEP \cite{derrida_exact_1993}.

\section{Deterministic parallel update}
\label{annexe:deterministic_parallel_update}

This Appendix details our findings using a parallel update scheme (presented in Sec.~\ref{sub:parallel_update}) in a deterministic limit for the implementation of SFP.
More precisely, the parallel update makes the algorithm deterministic (strictly predictable) when $\alpha_S = \beta = p_S = 1$, $q_S \in {0,\infty}$, and $q_F \in {0,1}$. These situations result in
cyclic output, whose perturbative study sheds light on the mechanisms that can alter the flow.

\begin{figure}
\centering
\includegraphics[width=0.8\columnwidth]{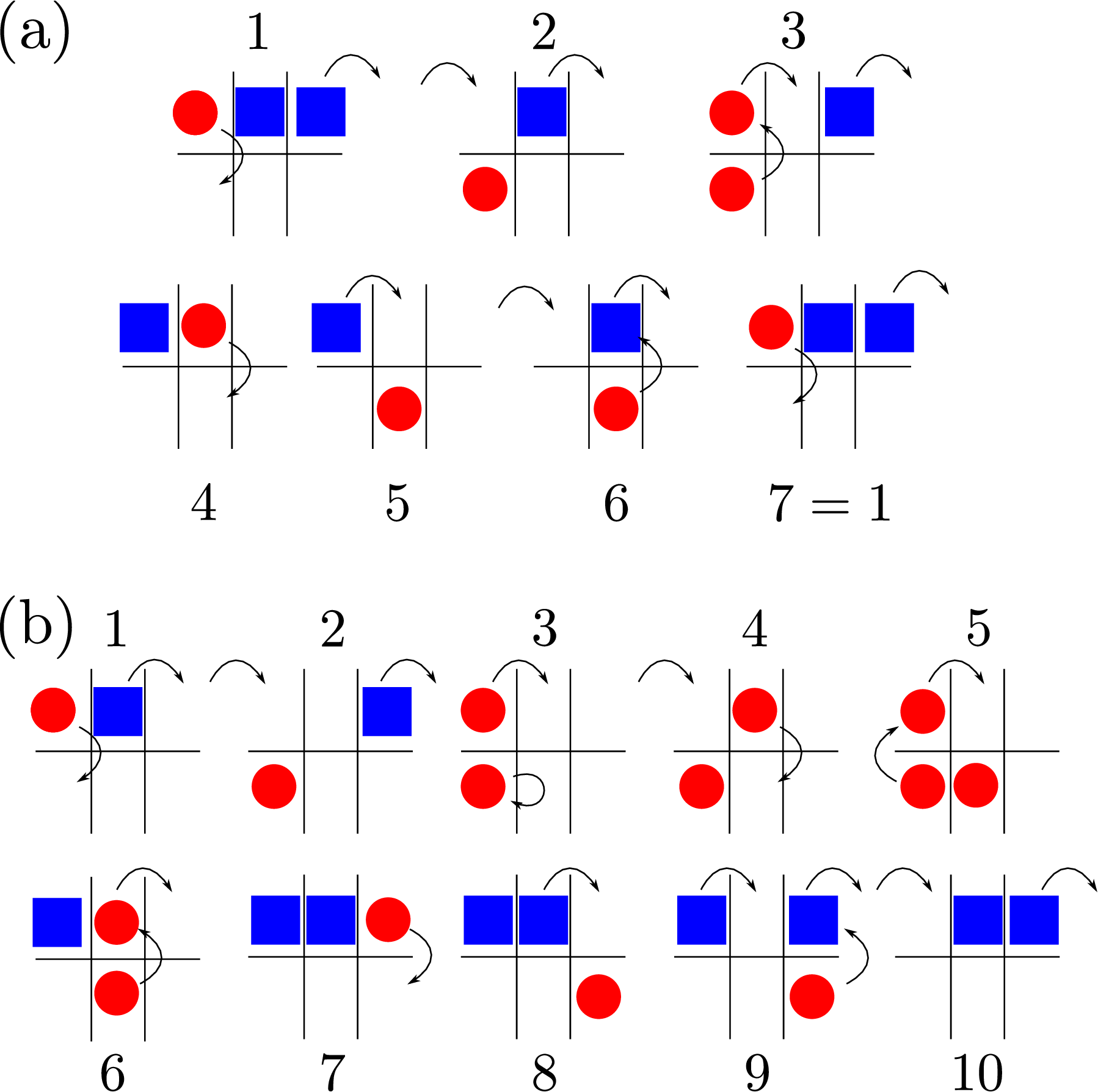}
\caption{Deterministic cycle occurring with a parallel update scheme (a) when $q_F = 1$ (b) when the parked particle fails to pull out at step 3, due to a small perturbation $q_F = 1 - \epsilon$. Red circles (blue square) are particles of species $S$ ($F$, respectively). Fixed parameters : $\alpha_S = \beta = p_S = 1$ and $q_S \rightarrow \infty$. }
    \label{fig:Schema_update_deterministic_qb1}
\end{figure}

The cases $q_S = 0$ and/or $q_F = 0$ are straightforward as they exactly correspond to TASEP, possibly after a transient time when $q_F = 0$. We focus on the more interesting case $q_S \rightarrow \infty$ and $q_F = 1$, in which $S$-cars pull in (and out) as soon as possible, i.e., when the space on the parking lane or on the road is vacant. The current can then be computed exactly by exhaustively listing all possible configuration types. In fact, inspecting the three first sites is sufficient, for there are only $F$ particles on sites $i>3$. The dynamics in this scenario consist of a cycle of six distinct states, shown in Fig.~\ref{fig:Schema_update_deterministic_qb1}a. Two particles are injected per cycle, so that the total current is $J = 1/3$.

Let us now consider the influence of two small perturbations with respect to the deterministic case. Firstly, if $q_F = 1 -\epsilon$ with $\epsilon \rightarrow 0$, a parked particle might choose not to pull out even if the road is free. In the cycle in Fig.~\ref{fig:Schema_update_deterministic_qb1}a, this can occur 
\begin{itemize}
    \item either at step 6: the particle then remains parked on the second site; this perturbation does not alter the current because the $F$ particle on site 2 will be unable to move at the next step anyway;
    \item or at step 3: the particle then remains parked on the first site.
\end{itemize}
 The second possibility leads to the scenario of Fig.~\ref{fig:Schema_update_deterministic_qb1}b. The modified cycle involves 10 steps, including the injection of 3 particles, resulting in a current of $J = 3/10$, which is lower than the baseline $1/3$. Consequently, lowering $q_F$ perturbatively yields a lower current.
 
 Secondly, if $q_S$ is finite but large, a particle $S$ might try to hop instead of parking. Considering the cycle in Fig.~\ref{fig:Schema_update_deterministic_qb1}a,  this event could occur
 \begin{itemize}
     \item at step 1:  the $S$ particle is then on the first site; it tries to hop but actually cannot move because the second site is occupied. This makes the cycle one step longer and leads to a current $J = 2/7$;
     \item at step 4: the $S$ particle is on the second site; the current remains unaffected since the $S$ particle hops to the third site, causing no obstruction.
 \end{itemize}
 Therefore, if $q_S$ is finite but considerable, the current diminishes.
 
Leaving the deterministic realm, let us now inspect the case $p_S < p_F = 1$, in particular $p_S = 0.7$. In the main text, we showed that the current is then non-monotonic with respect to $q_S$ for all inverse parking durations $q_F$ considered with a peak occurring beyond the extreme values. 
For $q_S = 0$, the current is known exactly, $J = \frac{1-\sqrt{1-p_S}}{2}\simeq 0.23$ \cite{schadschneider_cellular_1993} and is independent of $q_F$ (note that the first data point is $q_S = 0.05$). For $q_S \to \infty$ and $q_F=1$, the current can also be computed exactly and is equal to $ J = \frac{2}{5+1/p_S}$, in perfect agreement with the dotted dark blue line in Fig.~\ref{fig:Current_parallel_fct_qA_different_qb_MC_ParkingABC_pa_1-00_alphaA_1-00_beta_1-00_b}.

\begin{figure}
\centering
\begin{subfigure}{.48\textwidth}
  \centering
    \includegraphics[width=0.8\textwidth]{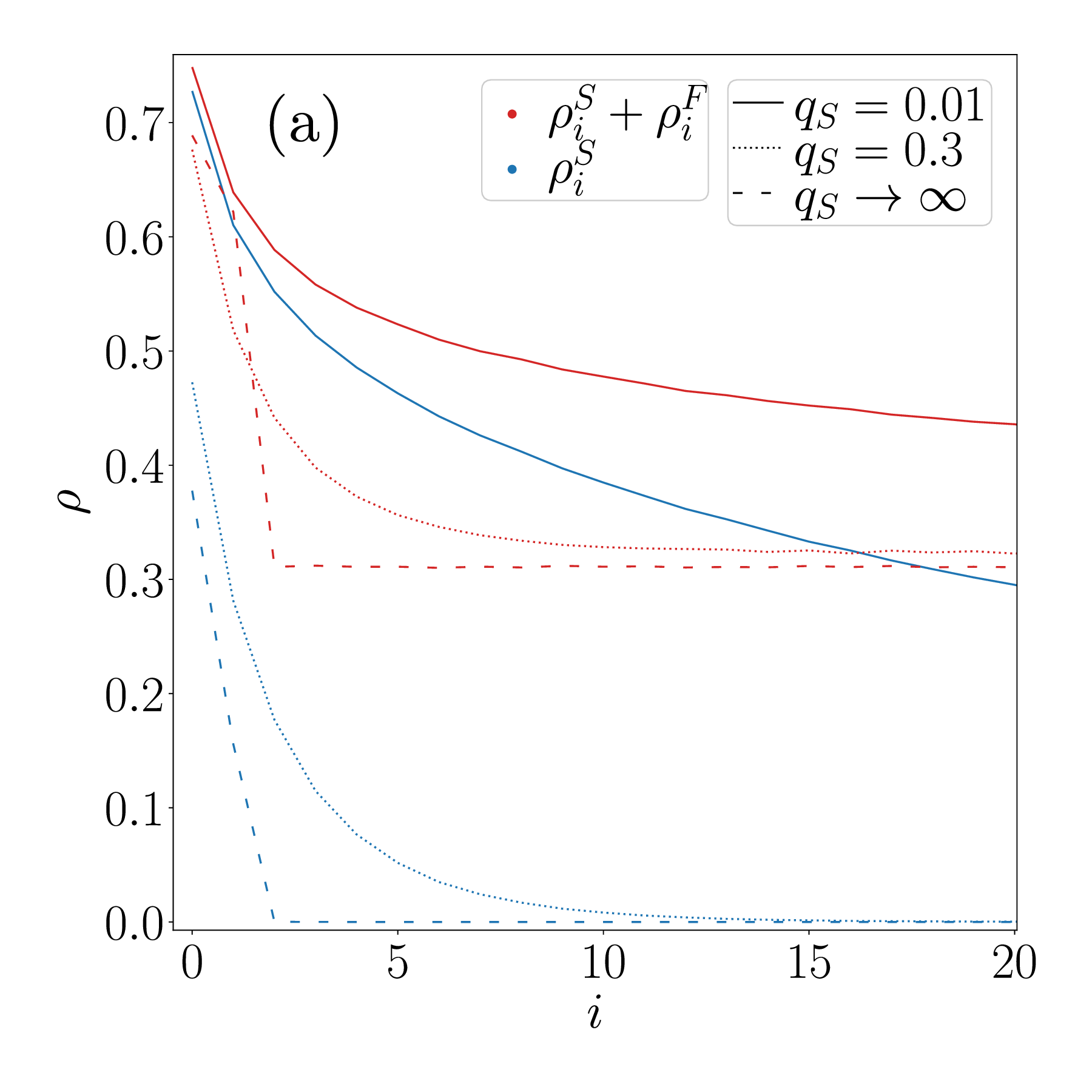}
    \phantomsubcaption
    \label{fig:Densite_sites_MC_ParallelParkingABC_plusieurs_qa_pa_0-70_qb_1-0_alphaA_1-00_beta_1-0_a}
\end{subfigure}
\begin{subfigure}{.48\textwidth}
  \centering
  \includegraphics[width=0.8\textwidth]{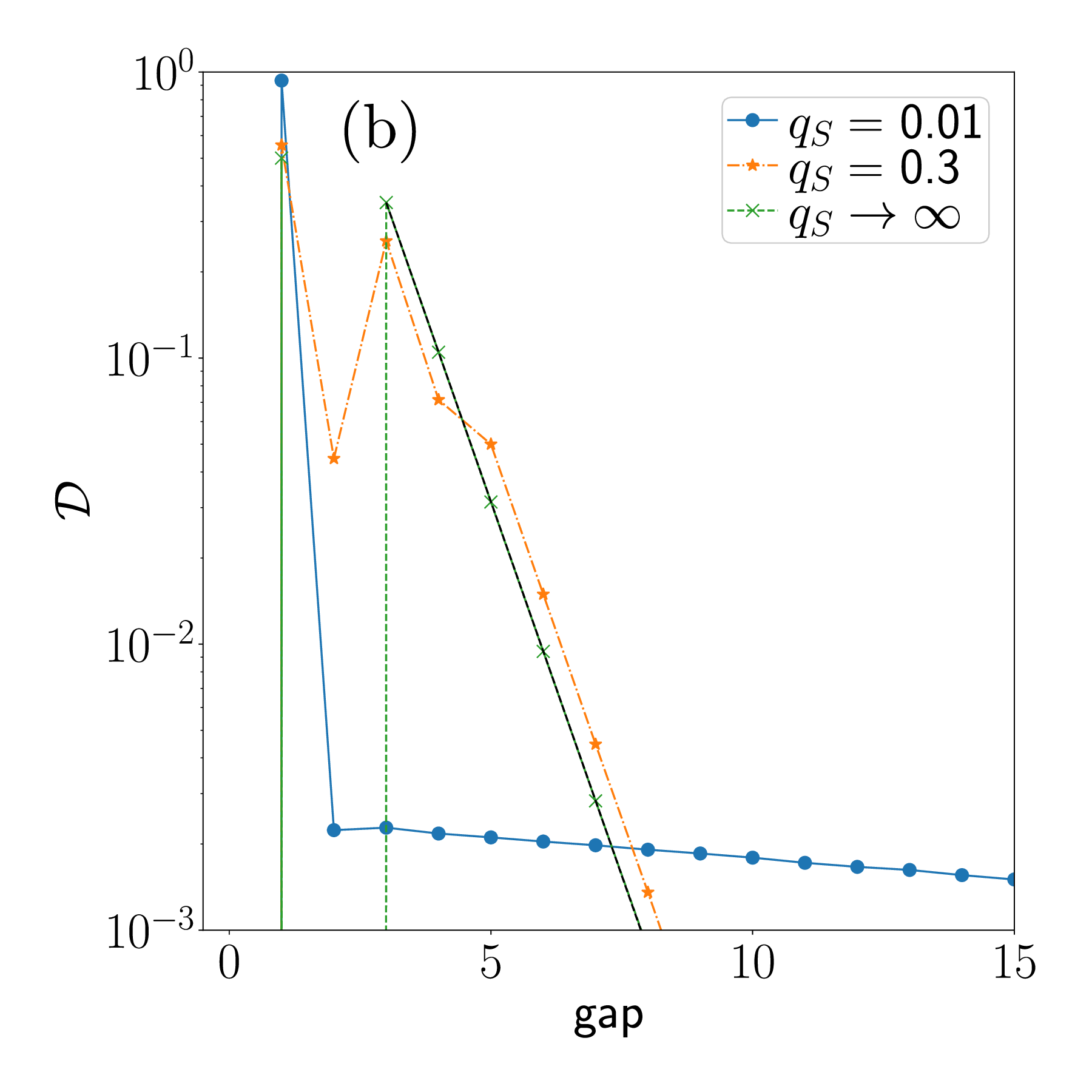}
  \phantomsubcaption
  \label{fig:Densite_sites_MC_ParallelParkingABC_plusieurs_qa_pa_0-70_qb_1-0_alphaA_1-00_beta_1-0_b}
\end{subfigure}
\hfill
\caption{(a) Density profile and (b) steady-state distribution of gaps  $\mathcal{D}(g)$ in the
SFP model with parallel update. The current is maximized at an intermediate parking rate $q_S^{\star} \simeq 0.3$; at this value, the extended spatial decrease of $S$ particles, over $\sim 10$ sites (dotted blue line), minimizes the effect of pull-out perturbations on the left boundary, while limiting the impact of the survival of slow $S$ particles. 
In panel (b), gaps of size $1$ correspond to the separation between cars within a `cluster'. Parameters: $p_S = 0.7, q_F = 1, \alpha_S = 1$ and $ \beta = 1$. }
    \label{fig:Densite_sites_MC_ParallelParkingABC_plusieurs_qa_pa_0-70_qb_1-0_alphaA_1-00_beta_1-0}
\end{figure}

The peak in the current at an intermediate rate $q_S$ can be understood as an optimal trade-off
between the two flow-altering mechanisms exposed in the main text, namely, longer survival of slow cruising particles at small $q_S$ and the disturbance due to the pulling-out maneuvers of parked vehicles which is more likely to block the injection at high $q_S$, where cars park near the entrance.
This is supported  by the density profile for low, intermediate and infinite parking rates $q_S$  plotted in Fig.~\ref{fig:Densite_sites_MC_ParallelParkingABC_plusieurs_qa_pa_0-70_qb_1-0_alphaA_1-00_beta_1-0}(a). For $q_S = 0.01$ (full line), the long life time of slow particles generates a bottleneck (high density) close to the left boundary. At the other extreme, for $q_S \to \infty$ (long dashed line), the density of $S$ particles is minimal, but the parked particles that get back on the road on the first site systematically hamper the injection. For $q_S = 0.3$ (dotted line), a trade-off is found where slow particles can park further than the two first sites without going to far away. It reduces the total density at the first site and allows a higher current.  

A more detailed picture of the dynamics is obtained by inspecting the gaps $g$ between successive particles (downstream of the last $S$ particle), whose 
probability distribution $\mathcal{D}(g)$ is plotted in Fig.~\ref{fig:Densite_sites_MC_ParallelParkingABC_plusieurs_qa_pa_0-70_qb_1-0_alphaA_1-00_beta_1-0_b}. 
The current can readily be derived from the mean (first cumulant) of this distribution, viz., $J = 1/\sum_{i}^\infty \mathcal{D}(i)(i+1)$.
Note that, for $q_S \rightarrow \infty$, the probability distribution is analytically derived and the black curve confirms the predicted values for $g>1$ : $\mathcal{D}(g) = \frac{p_S(1-p_S)^{g-3}}{2}$. At the other extreme, for $q_S = 0.01$,  we observe a peak at $g=1$, followed by a nearly uniform distribution. This means that clusters of $F$-particles form behind a slow particle; once the latter parks, the cluster structure remains unchanged, as $p_F = 1$. Steady-state configurations thus consist of clusters of particles interspersed by a single site ($g=1$), each separated from the next cluster by a gap $g>1$.  The broad distribution of these gaps between clusters reduces the total current below that found for $q_S \rightarrow \infty$. The distribution for $q_S = 0.3$ lies in-between these limit-cases. Its tail scales similarly to $q_S \rightarrow \infty$-one, without any incidence of the possibly stronger bottleneck of cruising cars due to the lower $q_S$ value. Furthermore, the probability of having gaps of $1$ or $2$ is higher compared to $q_S \rightarrow \infty$, resulting in a higher current.

The position of the peak around $q_S^{\star} \simeq 0.3 \simeq  1-p_S $ should be understood in the context of the parallel update rule with a chosen hopping probabilty $p_S = 0.7$. For $q_S < q_S^{\star}$, an $S$ particle may stay idle even if it could hop forward or park, whence the current for $q_F=1$ is not maximal. For $q_S > q_S^{\star}$, since $p_S + q_S > 1$, the renormalization of the probabilities leads to a reduced hopping probability if there is a free parking spot; the probability to park close to the left boundary is thus strongly enhanced when further increasing $q_S$, with the bottleneck that ensues. As a result, an optimum balance between reducing the bottleneck and pulling-in further way should be found around $q_S^{\star} = 1-p_S$.

\begin{figure}
    \centering
    \includegraphics[width= 0.8\columnwidth]{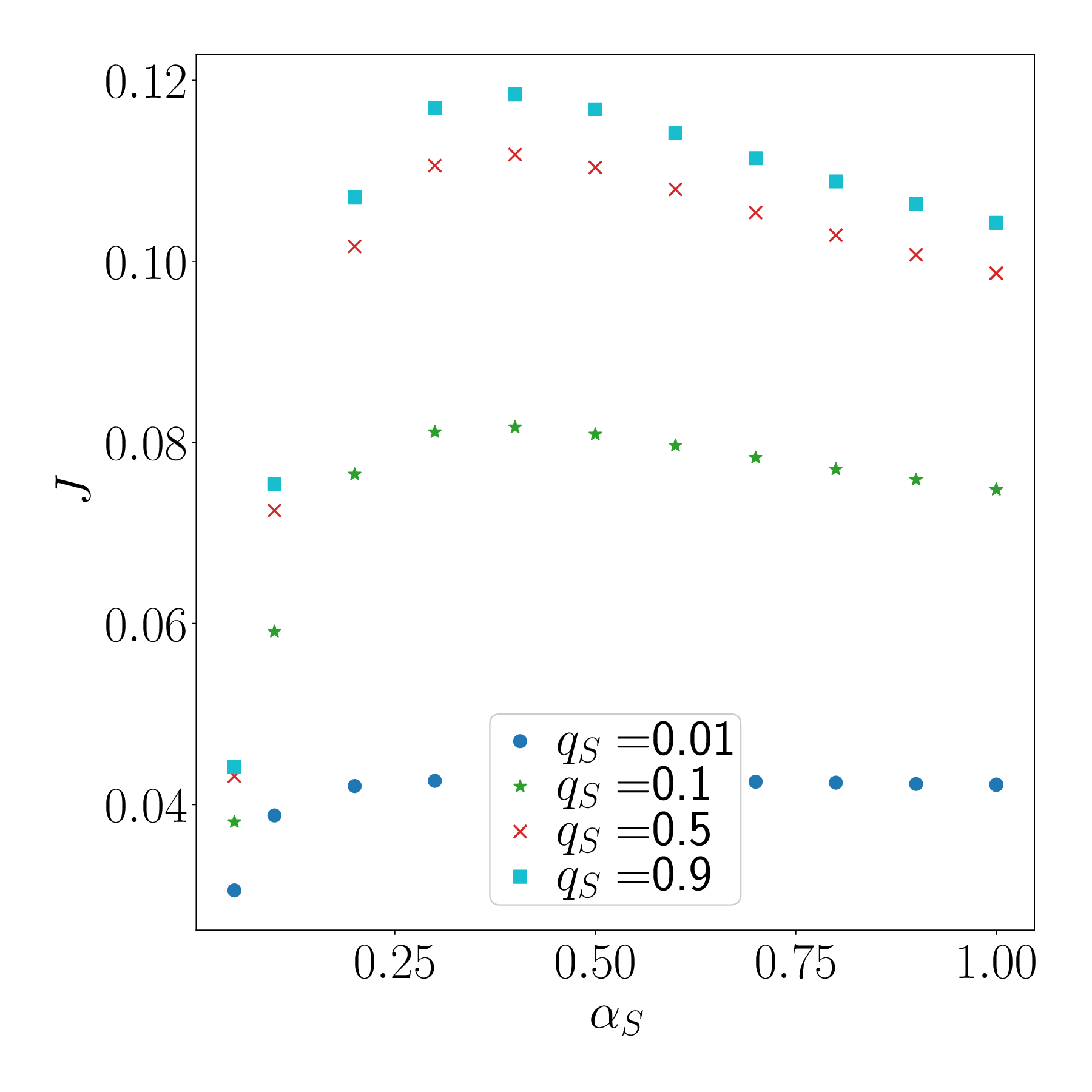}
    \caption{Non-monotonic evolution of the current $J$ with the input probability $\alpha_S$ using the parallel update rule for finite parking propensity and duration. Parameters : $p_S = 0.1$, $\beta = 1$, $q_F = 0.5$.}
    \label{fig:DensiteCourant_parallel_fctAlphaA_plusieurs_qa_MC_ParkingABC_pa_0-10_qb_0-50_beta_1-00}
\end{figure}

We close this section by considering the variation of the current as a function of the input rate $\alpha_S$. We do not make an exhaustive study as previously because the non-monotonic trend with $\alpha_S$ cannot be analytically derived on the sole basis of the deterministic cases. Nevertheless, relying on numerical simulations, we show in Fig.~\ref{fig:DensiteCourant_parallel_fctAlphaA_plusieurs_qa_MC_ParkingABC_pa_0-10_qb_0-50_beta_1-00} that non-monotonicity is also observed with the parallel update rule, pointing to the robustness of this effect.

\section{Propagation of a local (rectangular) jam}
\label{annexe:rectangular_bump}

In this Appendix, we investigate the propagation of a density perturbation along the road. To this end, we artificially generate a `jam' in a TASEP flow, i.e., starting with a flat density profile $\rho^-$, we introduce a rectangular `bump' (or step) in the density profile with a value of $\rho^+$ ($\rho^+=1$, in particular), by filling the gaps between positions $x_0$ and $x_1$ . We will first examine the case when $\rho^- < 1/2$, indicating that the system is in the Low-Density (LD) phase. Our goal here is to determine the minimum position $x_0^{\min}$ reached by regions of density higher than $\rho^-$, for all times, to see if the perturbation affects the injection.

\begin{figure}
    \centering
    \includegraphics[width=0.8\columnwidth]{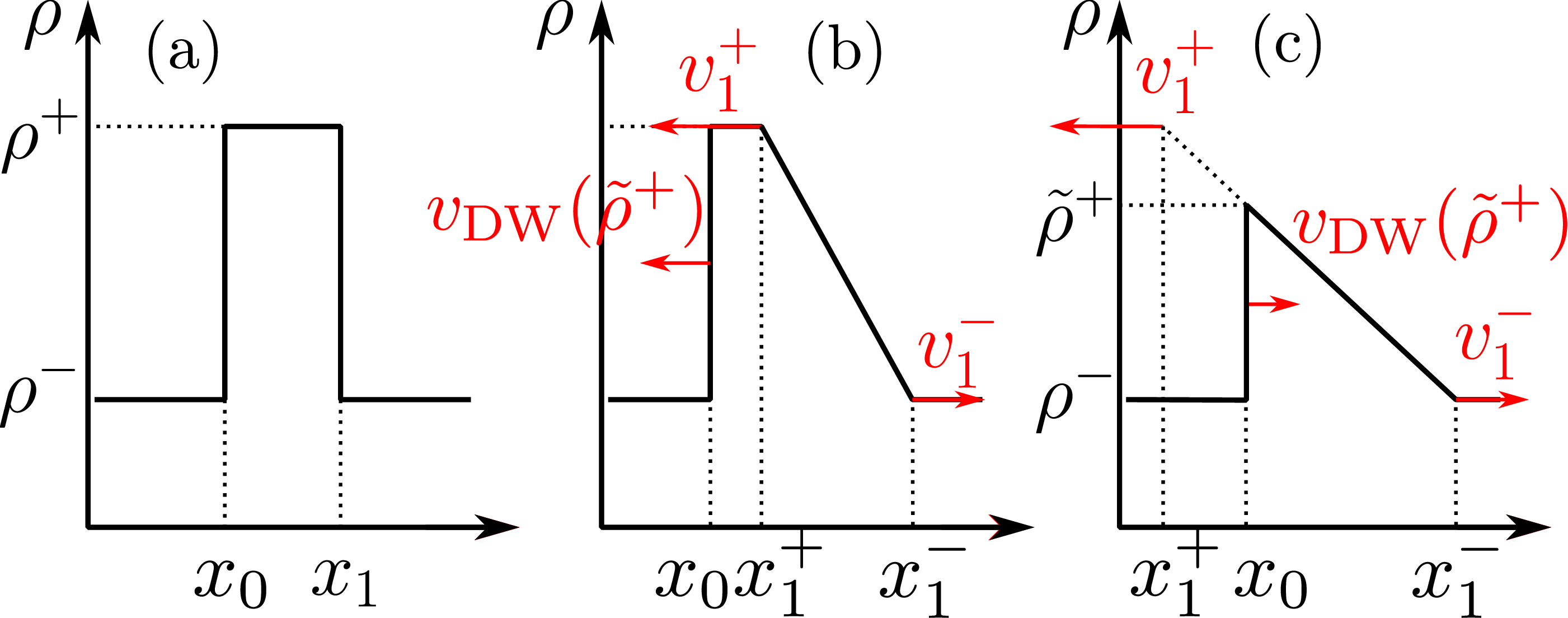}
    \caption{Sketch of the evolution of an artificially generated jam (rectangular bump in the density profile). (a) Initial condition. (b) Independent motion of the two `domain walls'. (c) Ultimate reduction
    of the bump density, when the diffusion of the top of the right step $x_1^-$ crosses the left step $x_0$; the reduction in the density $\Tilde{\rho}^+$  will gradually reduce the speed $v_{\mathrm{DM}}(\Tilde{\rho}^+)$, until it fully reverses the direction of propagation if $\Tilde{\rho}^+ < 1-\rho^-$.  }
    \label{fig:DomainWall_RectangularBump}
\end{figure}

Analytically, this problem can be approached by considering a domain wall initially separating two density regions: $\rho^-$ on the left and $\rho^+$ on the right. There are two possible scenarios:
\begin{itemize}
    \item When $\rho^- < \rho^+$, the domain wall moves at a velocity \cite{kolomeisky_asymmetric_1998}
    $$ v_{\mathrm{DW}} = \frac{J(\rho^+) - J (\rho^-)}{\rho^+ - \rho^-}$$
    where $J(\rho) = \rho(1-\rho)$ for TASEP.
    \item When $\rho^- > \rho^+$ the density decreases linearly between positions $x_1^-(t) = v_{\mathrm{coll}}(\rho^-)t + x_1$ and $x_1^+(t) = v_{\mathrm{coll}}(\rho^+)t + x_1$ with $v_{\mathrm{coll}}(\rho)$ the collective velocity which is equal to $1-2\rho$ for TASEP \cite{corwin_limit_2010}.
\end{itemize}
Starting with a rectangular bump, these two scenarios are present, the first one at the left border of the step, the second one at the right, as depicted in Fig.~\ref{fig:DomainWall_RectangularBump}a. 

Determining the leftmost reach $x_0^{\min}$ of the perturbation becomes non-trivial when the left domain wall moves toward the left, which occurs if $\rho^+ > 1-\rho^-$. When these two densities match, the left domain wall comes to a halt.
The speed of the left domain wall will decrease if $x_1^+(t) < x_0(t)$, viz., if $|v_{\mathrm{coll}}(\rho^+)| > |v_{\mathrm{DW}}| $. This condition is met because $\rho^+ = 1$ and then $|v_{\mathrm{coll}}(\rho^+)| = 1$. 

When $x_1^+(t) < x_0(t)$, the height of the step of the left domain wall wanes and the density at the right side of the step is denoted $\Tilde{\rho}^+ < \rho^+$. This leads to lowering its speed, until it changes sign, see Fig.~\ref{fig:DomainWall_RectangularBump}c. 

We make the assumption that the two speeds $v_1^- = v_{\mathrm{coll}}(\rho^-)$ and $v_1^+ = v_{\mathrm{coll}}(\rho^+)$ are constant. Although not immediately obvious, our numerical simulations support this claim. Subsequently, at time $t$, the density at position $x \in [x_1^+(t), x_1^-(t)]$ is given by:
$$ \rho(x,t) = f(x,t) = \frac{\rho^- - \rho^+}{(v_1^--v_1^+)t}(x-v_1^+t-x_1) + \rho^+.$$

We can implement the following simple algorithm :
\begin{enumerate}
    \item Initialization $x_0 = x_0, \Tilde{\rho}^+ = \rho^+,  x_1^-= x_1^+=x_1, t=0$.
    \item Compute $f(x_0, t)$. If $f(x_0, t) > 1$ then $\Tilde{\rho}^+ = 1$ otherwise $\Tilde{\rho}^+ = f(x,t)$.
    \item Update positions $x_0 = x_0+v_{\mathrm{DW}}(\Tilde{\rho}^+), x_1^- = x_1^-+v_1^-, x_1^+ = x_1^+ + v_1^+$ and the time $t=t+1$.
    \item Return to step 2 until $\Tilde{\rho}^+ \leq 1-\rho^-$.
\end{enumerate}
At the end of the algorithm, $x_0$ corresponds to the desired minimum value, $x_0^{\min}$. 

\begin{figure}
    \centering
    \includegraphics[width=0.8\columnwidth]{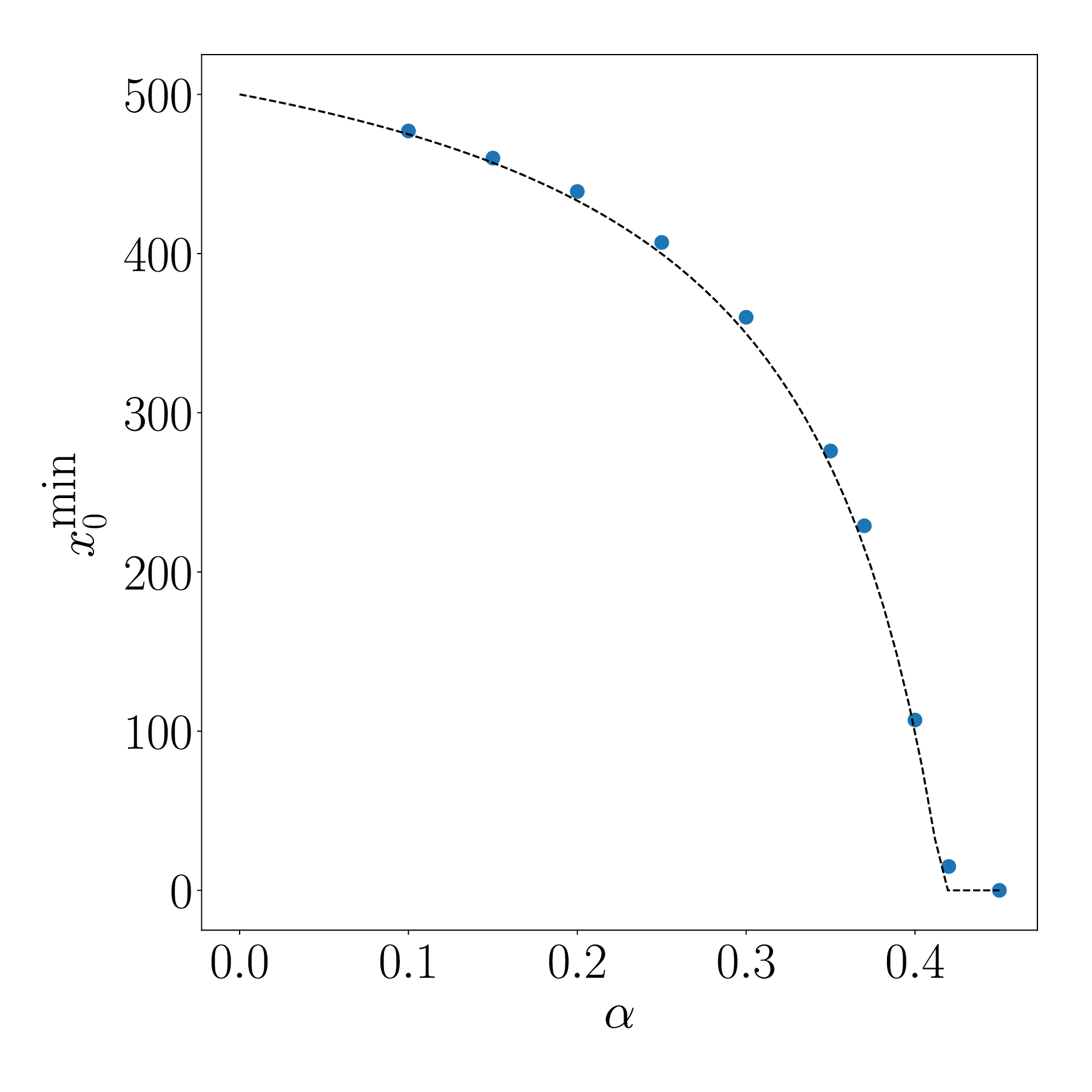}
    \caption{Leftmost position $x_0^{\min}$reached by the density perturbation as a function of the input rate $\alpha$ for an initial jam between $x_0 = 500$ and $x_0 = 699$ and $\beta = 1/2$. Dotted lines represent the theoretical predictions; symbols represent the Monte-Carlo simulations with $L=1000$ using $2000$ different realizations. Here, $x_0^{\min}$ was defined as the minimum position with $\rho \geq 1/2$.   }
    \label{fig:PosMin_DensityBump_pls_alphaA_pa_1_beta_0-50_L_1000_posPertub_500_timeStep_s_10-00000_blocSize_200}
\end{figure}

Figure~\ref{fig:PosMin_DensityBump_pls_alphaA_pa_1_beta_0-50_L_1000_posPertub_500_timeStep_s_10-00000_blocSize_200} proves that these theoretical predictions for $x_0^{\min}$ agree well with the direct numerical simulations, in which $x_0^{\min}$ is defined as the leftmost position with $\rho \geq 1/2$. (Using a value other than $1/2$, but still bigger than $\alpha$, only slightly shifts the results, because the bump density on the left side is not perfectly step-like, as obvious from in the red curve in Fig. \ref{fig:EvolutionTemporelleDensite_pa_1_alphaA_0-5_beta_0-50_perturbation_500_L_1000_posPertub_300_timeStep_s_10-00000_blocSize_50}.)

When $\rho^- = 1/2$, corresponding to the MC phase, $v_{\mathrm{DW}}$ cannot be positive as $J (\rho^-) = 1/4$ is the maximum possible current. Then, theoretically, $x_0^{\min}$ decreases indefinitely but infinitely slowly as the height of the step goes to $1/2$.

\section{Importance of correlations : range of validity of the Mean-Field (MF) approximation and Exact Diagonalization (ED) }
\label{annexe:comparaison_MF_ED}
The phase diagram of TASEP is correctly predicted by MF approximations \cite{krug_boundary-induced_1991, derrida_exact_1992}.  As a result, most of the TASEP variants outlined in the introduction have undergone scrutiny through this approximation or by employing a more sophisticated refined MF theory. In this section, we will compare the Monte-Carlo simulation with the mean-field approximation; deviations between the two will point to the importance of correlations on the flow. Furthermore, we will also compare those results with those of  ED to gauge  finite-size effects.

\begin{figure}
    \centering
    \includegraphics[width=0.8\columnwidth]{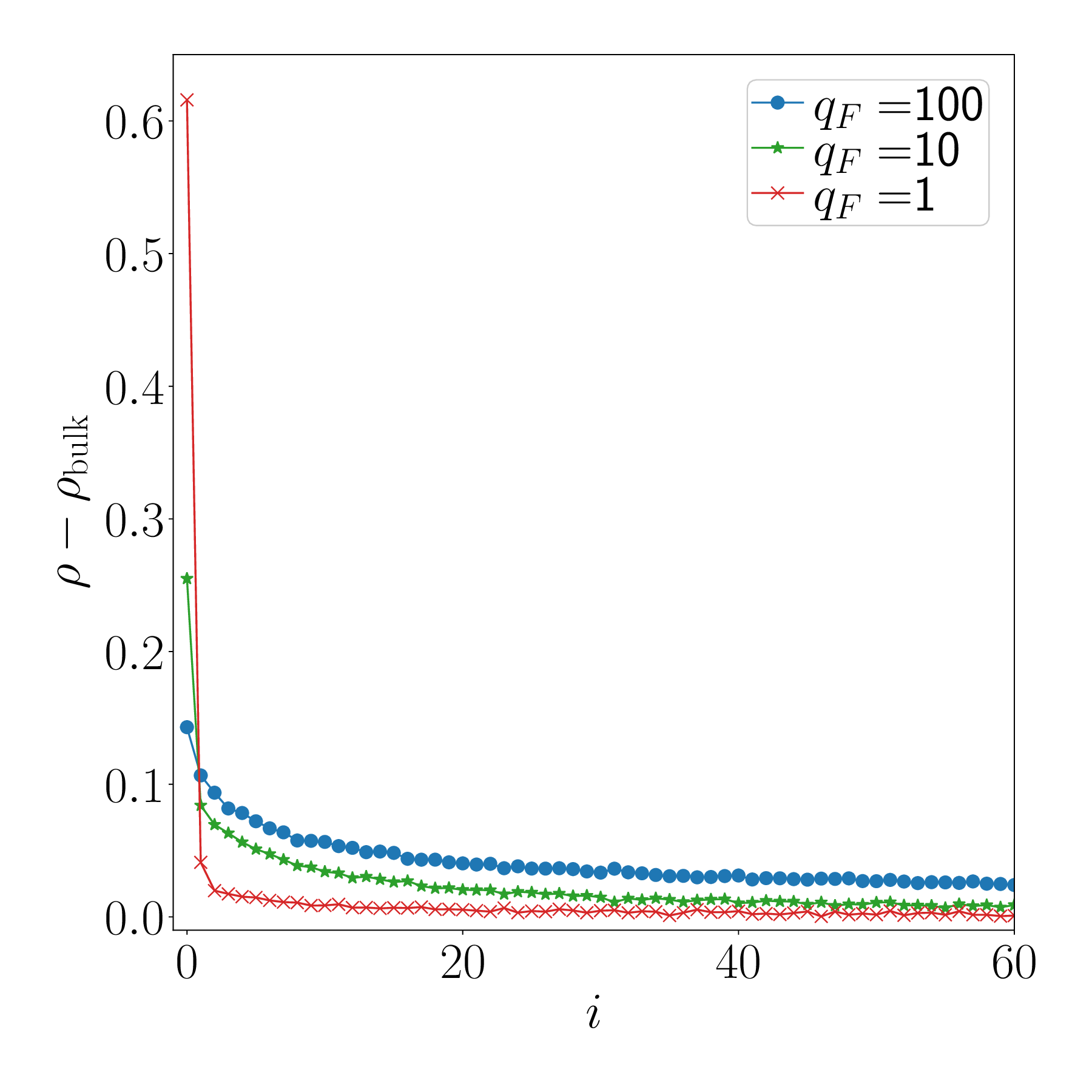}
    \caption{Rescaled total density profiles $\rho - \rho_{bulk}$ as a function of the position $i$ for $q_S \rightarrow \infty, \beta = 0.6, p_S = 0.1$ and $\alpha_S = 0.7$ and different $q_F$. $i=0$ corresponds to the left boundary where particles are injected. $\rho_{bulk}$ is the total density at the middle of the system. Large density tails ($i>10$) are obtained for $q_F = 100, 10$, in which case ED is not expected to be accurate; see Fig. \ref{fig:Courant_fonctionAlphaA_MC_MF_ED_ParkingABC_pa_0-10_qaINF_betaB_0-60}. Symbols are kMC simulations with $L=1000$.   }
    \label{fig:Densite_sites_MC_ParkingABC_pa_0-10_plusieursqb_qa_INF_alphaA_0-70_beta_0-60.eps}
\end{figure}

ED is numerically tractable only for a limited number of sites $L_1$ as the dimension of the matrix grows exponentially as $6^{L_1}$. In the LD phase, the value of the exit rate $\beta$ does not change the current nor the density profile sufficiently far away from the left boundary.
When considering a small system size, the current can take values that can vary by a few percent depending on $\beta$ in LD. To minimize these finite-size effects, we consider an exit rate $\beta_{diag}$ which is not a constant, as exposed in Sec.~\ref{section:methodology_exactDiag}. In LD, this methodology provides a unique current, in very good agreement with kMC, as shown in Figs. \ref{fig:Courant_fonctionAlphaA_MC_MF_ED_ParkingABC_pa_0-10_qaINF_betaB_0-60} and \ref{fig:EvolutionJ_qa_differents_pa_MC_MF_ED_ParkingABC_qbINF_alphaA_1-00_betaB_0-60}. Discrepancies occur when the system size used for ED becomes too small to guarantee a flat or slowly changing density profile at the right boundary. It happens specifically in the MC phase where the density decreases (increases) at the first order as $1/n^{1/2}$ at the left (right) boundary with $n$ the site number \cite{derrida_exact_1993}. Unlike in TASEP, it can happen also in the LD phase where the density profile can have a large tail near the left boundary. For instance, rescaled density profiles close to the left boundary for $q_S \rightarrow \infty, \beta = 0.6, p_S = 0.1$ and $\alpha_S = 0.7$ and different $q_F$ are shown in Fig.~\ref{fig:Densite_sites_MC_ParkingABC_pa_0-10_plusieursqb_qa_INF_alphaA_0-70_beta_0-60.eps}. This choice of parameters and the color mirror the ones used in Fig.~\ref{fig:Courant_fonctionAlphaA_MC_MF_ED_ParkingABC_pa_0-10_qaINF_betaB_0-60}. Discrepancies between ED and kMC in the MC phase (blue) and in LD for $q_F = 10$ (green) stem for strong tail in the density profile. However, with a lower parking rate (red), the density profile flattens past a few sites, and ED becomes accurate.

This methodology no longer works in the HD phase. Nonetheless, using the current $J$ obtained without setting $\beta_{\mathrm{diag}}$, i.e. in the LD phase,  enables us to determine the effective input rate $\alpha_{eff}$ for a TASEP model, viz., 
\begin{equation}
    J = \alpha_{eff}(1-\alpha_{eff}).
\end{equation}
Consequently,  a transition should occur at $\beta_{\mathrm{diag}} = \alpha_{eff}$. Figure \ref{fig:DiagPhase_MC_ED_ParkingABC_qaINF_pa_0-10_qb_100-00} shows the phase diagram resulting from this argument; boundaries are in agreement with kMC simulations, bar those close to the MC phase. 

\begin{figure}
    \centering
    \includegraphics[width=0.8\columnwidth]{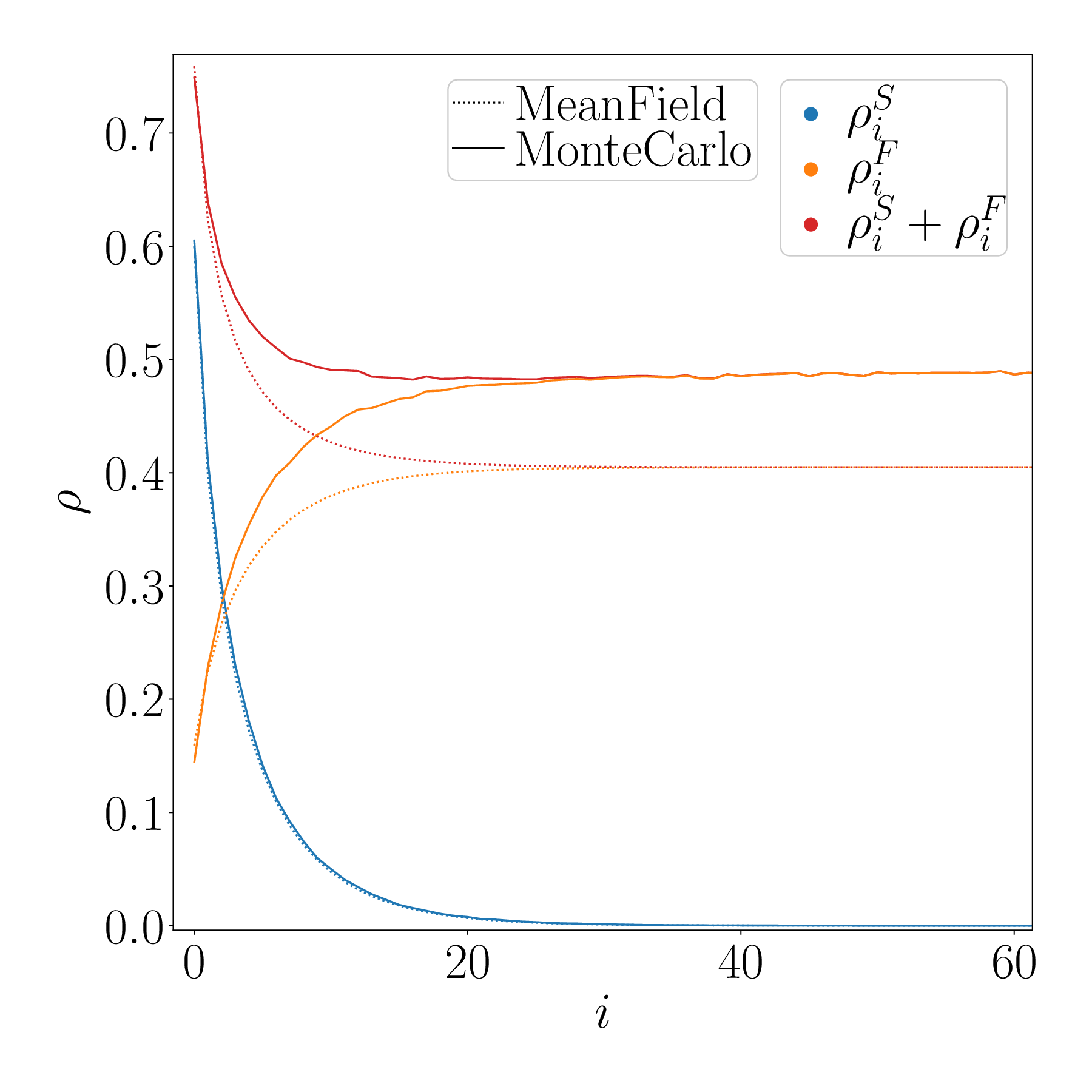}
    \caption{Density profile predicted by MF vs. obtained with kMC simulations. When there are no effective parking spots ($q_F \rightarrow \infty$), correlations still need to be taken into account if particles $S$ and $F$ do not have the same speed. 
    Parameters : $q_F \rightarrow \infty, \alpha_S = 1, p_S = 0.8, q_S = 0.1$ and $\beta=0.6$.}
    \label{fig:Densite_sites_MC_MF_ParkingABC_pa_0-80_qb_INF_qa_0-10_alphaA_1-00_beta_0-60}
\end{figure}

In contrast to ED, MF does not provide accurate results most of the time. In Figs. \ref{fig:Courant_fonctionAlphaA_MC_MF_ED_ParkingABC_pa_0-10_qaINF_betaB_0-60} and \ref{fig:EvolutionJ_qa_differents_pa_MC_MF_ED_ParkingABC_qbINF_alphaA_1-00_betaB_0-60}, MF estimations of the current differ from Monte-Carlo ones. There are two mains reasons. Firstly, the new dynamics are mainly governed by the first site(s). Then potential errors in the MF approximation are enhanced. Secondly, it is known that in TASEP-like models with distinct species' speeds, MF does not provide good results and a refined MF is necessary \cite{bottero_analysis_2017}. Since there is no number conservation per species in our model, a refined MF approximation encapsulating core properties has not been identified. However, a simple MF approach falters when the two particle species interact for extended periods even when no parked particles need to be taken into account, i.e. $q_F \to \infty$. Figure~\ref{fig:Densite_sites_MC_MF_ParkingABC_pa_0-80_qb_INF_qa_0-10_alphaA_1-00_beta_0-60} shows the density profile of particles $S$ and $F$ for $q_F \rightarrow \infty, \alpha_S = 1, p_S = 0.8$ and $\beta = 0.6$. When the inverse parking duration $q_F$ is infinite, SFP operates on a 1D line, with particles $S$ turning into $F$ at rate $q_S$ which is identical to \cite{bottero_analysis_2017} if $q_S = 0$. The decrease in the density of slow particles $\rho^S$ (blue) is nearly identical between MF and kMC, but differs for $\rho^F$ (orange). This disparity could be ascribed to fast particle that have higher probability to be behind a slow particle than expected by MF approximation. This led Bottero \textit{and al.} to consider a refined MF approach accounting for such correlations by handling particles of different sizes like slow particles, potentially accompanied by many fast particles.
MF failure highlights the importance of correlations between slow and fast particles in the main road, as well as between road and parking spot particles. 
Nevertheless, even with  MF, a core feature of SFP - the non-monotonic nature of the current - persists across all $q_F$, as evidenced in Fig.~\ref{fig:Courant_fonctionAlphaA_MC_MF_ED_ParkingABC_pa_0-10_qaINF_betaB_0-60}. 

%


\begin{thebibliography}{0}%
\makeatletter
\providecommand \@ifxundefined [1]{%
 \@ifx{#1\undefined}
}%
\providecommand \@ifnum [1]{%
 \ifnum #1\expandafter \@firstoftwo
 \else \expandafter \@secondoftwo
 \fi
}%
\providecommand \@ifx [1]{%
 \ifx #1\expandafter \@firstoftwo
 \else \expandafter \@secondoftwo
 \fi
}%
\providecommand \natexlab [1]{#1}%
\providecommand \enquote  [1]{``#1''}%
\providecommand \bibnamefont  [1]{#1}%
\providecommand \bibfnamefont [1]{#1}%
\providecommand \citenamefont [1]{#1}%
\providecommand \href@noop [0]{\@secondoftwo}%
\providecommand \href [0]{\begingroup \@sanitize@url \@href}%
\providecommand \@href[1]{\@@startlink{#1}\@@href}%
\providecommand \@@href[1]{\endgroup#1\@@endlink}%
\providecommand \@sanitize@url [0]{\catcode `\\12\catcode `\$12\catcode
  `\&12\catcode `\#12\catcode `\^12\catcode `\_12\catcode `\%12\relax}%
\providecommand \@@startlink[1]{}%
\providecommand \@@endlink[0]{}%
\providecommand \url  [0]{\begingroup\@sanitize@url \@url }%
\providecommand \@url [1]{\endgroup\@href {#1}{\urlprefix }}%
\providecommand \urlprefix  [0]{URL }%
\providecommand \Eprint [0]{\href }%
\providecommand \doibase [0]{https://doi.org/}%
\providecommand \selectlanguage [0]{\@gobble}%
\providecommand \bibinfo  [0]{\@secondoftwo}%
\providecommand \bibfield  [0]{\@secondoftwo}%
\providecommand \translation [1]{[#1]}%
\providecommand \BibitemOpen [0]{}%
\providecommand \bibitemStop [0]{}%
\providecommand \bibitemNoStop [0]{.\EOS\space}%
\providecommand \EOS [0]{\spacefactor3000\relax}%
\providecommand \BibitemShut  [1]{\csname bibitem#1\endcsname}%
\let\auto@bib@innerbib\@empty
\end{thebibliography}%


\begin{thebibliography}{80}%
\makeatletter
\providecommand \@ifxundefined [1]{%
 \@ifx{#1\undefined}
}%
\providecommand \@ifnum [1]{%
 \ifnum #1\expandafter \@firstoftwo
 \else \expandafter \@secondoftwo
 \fi
}%
\providecommand \@ifx [1]{%
 \ifx #1\expandafter \@firstoftwo
 \else \expandafter \@secondoftwo
 \fi
}%
\providecommand \natexlab [1]{#1}%
\providecommand \enquote  [1]{``#1''}%
\providecommand \bibnamefont  [1]{#1}%
\providecommand \bibfnamefont [1]{#1}%
\providecommand \citenamefont [1]{#1}%
\providecommand \href@noop [0]{\@secondoftwo}%
\providecommand \href [0]{\begingroup \@sanitize@url \@href}%
\providecommand \@href[1]{\@@startlink{#1}\@@href}%
\providecommand \@@href[1]{\endgroup#1\@@endlink}%
\providecommand \@sanitize@url [0]{\catcode `\\12\catcode `\$12\catcode
  `\&12\catcode `\#12\catcode `\^12\catcode `\_12\catcode `\%12\relax}%
\providecommand \@@startlink[1]{}%
\providecommand \@@endlink[0]{}%
\providecommand \url  [0]{\begingroup\@sanitize@url \@url }%
\providecommand \@url [1]{\endgroup\@href {#1}{\urlprefix }}%
\providecommand \urlprefix  [0]{URL }%
\providecommand \Eprint [0]{\href }%
\providecommand \doibase [0]{https://doi.org/}%
\providecommand \selectlanguage [0]{\@gobble}%
\providecommand \bibinfo  [0]{\@secondoftwo}%
\providecommand \bibfield  [0]{\@secondoftwo}%
\providecommand \translation [1]{[#1]}%
\providecommand \BibitemOpen [0]{}%
\providecommand \bibitemStop [0]{}%
\providecommand \bibitemNoStop [0]{.\EOS\space}%
\providecommand \EOS [0]{\spacefactor3000\relax}%
\providecommand \BibitemShut  [1]{\csname bibitem#1\endcsname}%
\let\auto@bib@innerbib\@empty
\bibitem [{\citenamefont {Kerner}(1999)}]{kerner1999physics}%
  \BibitemOpen
  \bibfield  {author} {\bibinfo {author} {\bibfnamefont {B.~S.}\ \bibnamefont
  {Kerner}},\ }\bibfield  {title} {\bibinfo {title} {The physics of traffic},\
  }\href@noop {} {\bibfield  {journal} {\bibinfo  {journal} {Physics World}\
  }\textbf {\bibinfo {volume} {12}},\ \bibinfo {pages} {25} (\bibinfo {year}
  {1999})}\BibitemShut {NoStop}%
\bibitem [{\citenamefont {Schadschneider}\ \emph {et~al.}()\citenamefont
  {Schadschneider}, \citenamefont {Chowdhury},\ and\ \citenamefont
  {Nishinari}}]{schadschneider_stochastic_2010}%
  \BibitemOpen
  \bibfield  {author} {\bibinfo {author} {\bibfnamefont {A.}~\bibnamefont
  {Schadschneider}}, \bibinfo {author} {\bibfnamefont {D.}~\bibnamefont
  {Chowdhury}},\ and\ \bibinfo {author} {\bibfnamefont {K.}~\bibnamefont
  {Nishinari}},\ }\href@noop {} {\emph {\bibinfo {title} {Stochastic transport
  in complex systems: from molecules to vehicles}}}\ (\bibinfo  {publisher}
  {Elsevier})\BibitemShut {NoStop}%
\bibitem [{\citenamefont {Schutz}()}]{schutz_exactly_2000}%
  \BibitemOpen
  \bibfield  {author} {\bibinfo {author} {\bibfnamefont {G.}~\bibnamefont
  {Schutz}},\ }\bibfield  {title} {\bibinfo {title} {Exactly solvable models
  for many-body systems far from equilibrium},\ }\href@noop {} {\ }\BibitemShut
  {NoStop}%
\bibitem [{\citenamefont {Wolff}()}]{wolff_collective_1989}%
  \BibitemOpen
  \bibfield  {author} {\bibinfo {author} {\bibfnamefont {U.}~\bibnamefont
  {Wolff}},\ }\bibfield  {title} {\bibinfo {title} {Collective monte carlo
  updating for spin systems},\ }\href
  {https://doi.org/10.1103/PhysRevLett.62.361} {\ \textbf {\bibinfo {volume}
  {62}},\ \bibinfo {pages} {361}},\ \bibinfo {note} {publisher: American
  Physical Society}\BibitemShut {NoStop}%
\bibitem [{\citenamefont {Chowdhury}\ \emph {et~al.}()\citenamefont
  {Chowdhury}, \citenamefont {Santen},\ and\ \citenamefont
  {Schadschneider}}]{chowdhury_statistical_2000}%
  \BibitemOpen
  \bibfield  {author} {\bibinfo {author} {\bibfnamefont {D.}~\bibnamefont
  {Chowdhury}}, \bibinfo {author} {\bibfnamefont {L.}~\bibnamefont {Santen}},\
  and\ \bibinfo {author} {\bibfnamefont {A.}~\bibnamefont {Schadschneider}},\
  }\bibfield  {title} {\bibinfo {title} {Statistical physics of vehicular
  traffic and some related systems},\ }\href
  {https://doi.org/10.1016/S0370-1573(99)00117-9} {\ \textbf {\bibinfo {volume}
  {329}},\ \bibinfo {pages} {199}}\BibitemShut {NoStop}%
\bibitem [{\citenamefont {Schadschneider}()}]{schadschneider_traffic_2002}%
  \BibitemOpen
  \bibfield  {author} {\bibinfo {author} {\bibfnamefont {A.}~\bibnamefont
  {Schadschneider}},\ }\bibfield  {title} {\bibinfo {title} {Traffic flow: a
  statistical physics point of view},\ }\href
  {https://doi.org/10.1016/S0378-4371(02)01036-1} {\ \bibinfo {series}
  {Fundamental Problems in Statistical Physics},\ \textbf {\bibinfo {volume}
  {313}},\ \bibinfo {pages} {153}}\BibitemShut {NoStop}%
\bibitem [{\citenamefont {Seo}\ \emph {et~al.}()\citenamefont {Seo},
  \citenamefont {Bayen}, \citenamefont {Kusakabe},\ and\ \citenamefont
  {Asakura}}]{seo_traffic_2017}%
  \BibitemOpen
  \bibfield  {author} {\bibinfo {author} {\bibfnamefont {T.}~\bibnamefont
  {Seo}}, \bibinfo {author} {\bibfnamefont {A.~M.}\ \bibnamefont {Bayen}},
  \bibinfo {author} {\bibfnamefont {T.}~\bibnamefont {Kusakabe}},\ and\
  \bibinfo {author} {\bibfnamefont {Y.}~\bibnamefont {Asakura}},\ }\bibfield
  {title} {\bibinfo {title} {Traffic state estimation on highway: A
  comprehensive survey},\ }\href
  {https://doi.org/10.1016/j.arcontrol.2017.03.005} {\ \textbf {\bibinfo
  {volume} {43}},\ \bibinfo {pages} {128}}\BibitemShut {NoStop}%
\bibitem [{\citenamefont {Lighthill}\ and\ \citenamefont
  {Whitham}()}]{lighthill_kinematic_1955}%
  \BibitemOpen
  \bibfield  {author} {\bibinfo {author} {\bibfnamefont {M.~J.}\ \bibnamefont
  {Lighthill}}\ and\ \bibinfo {author} {\bibfnamefont {G.~B.}\ \bibnamefont
  {Whitham}},\ }\bibfield  {title} {\bibinfo {title} {On kinematic waves {II}.
  a theory of traffic flow on long crowded roads},\ }\href@noop {} {\ \textbf
  {\bibinfo {volume} {229}},\ \bibinfo {pages} {317}},\ \bibinfo {note}
  {publisher: The Royal Society London}\BibitemShut {NoStop}%
\bibitem [{\citenamefont {Kerner}\ and\ \citenamefont
  {Konhäuser}()}]{kerner_cluster_1993}%
  \BibitemOpen
  \bibfield  {author} {\bibinfo {author} {\bibfnamefont {B.~S.}\ \bibnamefont
  {Kerner}}\ and\ \bibinfo {author} {\bibfnamefont {P.}~\bibnamefont
  {Konhäuser}},\ }\bibfield  {title} {\bibinfo {title} {Cluster effect in
  initially homogeneous traffic flow},\ }\href
  {https://doi.org/10.1103/PhysRevE.48.R2335} {\ \textbf {\bibinfo {volume}
  {48}},\ \bibinfo {pages} {R2335}},\ \bibinfo {note} {publisher: American
  Physical Society}\BibitemShut {NoStop}%
\bibitem [{\citenamefont {Helbing}()}]{helbing_improved_1995}%
  \BibitemOpen
  \bibfield  {author} {\bibinfo {author} {\bibfnamefont {D.}~\bibnamefont
  {Helbing}},\ }\bibfield  {title} {\bibinfo {title} {Improved fluid-dynamic
  model for vehicular traffic},\ }\href
  {https://doi.org/10.1103/PhysRevE.51.3164} {\ \textbf {\bibinfo {volume}
  {51}},\ \bibinfo {pages} {3164}},\ \bibinfo {note} {publisher: American
  Physical Society}\BibitemShut {NoStop}%
\bibitem [{\citenamefont {Bando}\ \emph {et~al.}()\citenamefont {Bando},
  \citenamefont {Hasebe}, \citenamefont {Nakayama}, \citenamefont {Shibata},\
  and\ \citenamefont {Sugiyama}}]{bando_dynamical_1995}%
  \BibitemOpen
  \bibfield  {author} {\bibinfo {author} {\bibfnamefont {M.}~\bibnamefont
  {Bando}}, \bibinfo {author} {\bibfnamefont {K.}~\bibnamefont {Hasebe}},
  \bibinfo {author} {\bibfnamefont {A.}~\bibnamefont {Nakayama}}, \bibinfo
  {author} {\bibfnamefont {A.}~\bibnamefont {Shibata}},\ and\ \bibinfo {author}
  {\bibfnamefont {Y.}~\bibnamefont {Sugiyama}},\ }\bibfield  {title} {\bibinfo
  {title} {Dynamical model of traffic congestion and numerical simulation},\
  }\href {https://doi.org/10.1103/PhysRevE.51.1035} {\ \textbf {\bibinfo
  {volume} {51}},\ \bibinfo {pages} {1035}},\ \bibinfo {note} {publisher:
  American Physical Society}\BibitemShut {NoStop}%
\bibitem [{\citenamefont {Nagel}\ and\ \citenamefont
  {Schreckenberg}()}]{nagel_cellular_1992}%
  \BibitemOpen
  \bibfield  {author} {\bibinfo {author} {\bibfnamefont {K.}~\bibnamefont
  {Nagel}}\ and\ \bibinfo {author} {\bibfnamefont {M.}~\bibnamefont
  {Schreckenberg}},\ }\bibfield  {title} {\bibinfo {title} {A cellular
  automaton model for freeway traffic},\ }\href@noop {} {\ \textbf {\bibinfo
  {volume} {2}},\ \bibinfo {pages} {2221}},\ \bibinfo {note} {publisher: {EDP}
  Sciences}\BibitemShut {NoStop}%
\bibitem [{\citenamefont {Schreckenberg}\ \emph {et~al.}()\citenamefont
  {Schreckenberg}, \citenamefont {Schadschneider}, \citenamefont {Nagel},\ and\
  \citenamefont {Ito}}]{schreckenberg_discrete_1995}%
  \BibitemOpen
  \bibfield  {author} {\bibinfo {author} {\bibfnamefont {M.}~\bibnamefont
  {Schreckenberg}}, \bibinfo {author} {\bibfnamefont {A.}~\bibnamefont
  {Schadschneider}}, \bibinfo {author} {\bibfnamefont {K.}~\bibnamefont
  {Nagel}},\ and\ \bibinfo {author} {\bibfnamefont {N.}~\bibnamefont {Ito}},\
  }\bibfield  {title} {\bibinfo {title} {Discrete stochastic models for traffic
  flow},\ }\href {https://doi.org/10.1103/PhysRevE.51.2939} {\ \textbf
  {\bibinfo {volume} {51}},\ \bibinfo {pages} {2939}},\ \bibinfo {note}
  {publisher: American Physical Society}\BibitemShut {NoStop}%
\bibitem [{\citenamefont {Nagel}()}]{nagel_particle_1996}%
  \BibitemOpen
  \bibfield  {author} {\bibinfo {author} {\bibfnamefont {K.}~\bibnamefont
  {Nagel}},\ }\bibfield  {title} {\bibinfo {title} {Particle hopping models and
  traffic flow theory},\ }\href {https://doi.org/10.1103/PhysRevE.53.4655} {\
  \textbf {\bibinfo {volume} {53}},\ \bibinfo {pages} {4655}},\ \bibinfo {note}
  {publisher: American Physical Society}\BibitemShut {NoStop}%
\bibitem [{\citenamefont {Rickert}\ \emph {et~al.}()\citenamefont {Rickert},
  \citenamefont {Nagel}, \citenamefont {Schreckenberg},\ and\ \citenamefont
  {Latour}}]{rickert_two_1996}%
  \BibitemOpen
  \bibfield  {author} {\bibinfo {author} {\bibfnamefont {M.}~\bibnamefont
  {Rickert}}, \bibinfo {author} {\bibfnamefont {K.}~\bibnamefont {Nagel}},
  \bibinfo {author} {\bibfnamefont {M.}~\bibnamefont {Schreckenberg}},\ and\
  \bibinfo {author} {\bibfnamefont {A.}~\bibnamefont {Latour}},\ }\bibfield
  {title} {\bibinfo {title} {Two lane traffic simulations using cellular
  automata},\ }\href {https://doi.org/10.1016/0378-4371(95)00442-4} {\ \textbf
  {\bibinfo {volume} {231}},\ \bibinfo {pages} {534}}\BibitemShut {NoStop}%
\bibitem [{\citenamefont {Derrida}\ \emph {et~al.}({\natexlab{a}})\citenamefont
  {Derrida}, \citenamefont {Domany},\ and\ \citenamefont
  {Mukamel}}]{derrida_exact_1992}%
  \BibitemOpen
  \bibfield  {author} {\bibinfo {author} {\bibfnamefont {B.}~\bibnamefont
  {Derrida}}, \bibinfo {author} {\bibfnamefont {E.}~\bibnamefont {Domany}},\
  and\ \bibinfo {author} {\bibfnamefont {D.}~\bibnamefont {Mukamel}},\
  }\bibfield  {title} {\bibinfo {title} {An exact solution of a one-dimensional
  asymmetric exclusion model with open boundaries},\ }\href
  {https://doi.org/10.1007/BF01050430} {\ \textbf {\bibinfo {volume} {69}},\
  \bibinfo {pages} {667} ({\natexlab{a}})}\BibitemShut {NoStop}%
\bibitem [{\citenamefont {Derrida}\ \emph {et~al.}({\natexlab{b}})\citenamefont
  {Derrida}, \citenamefont {Evans}, \citenamefont {Hakim},\ and\ \citenamefont
  {Pasquier}}]{derrida_exact_1993}%
  \BibitemOpen
  \bibfield  {author} {\bibinfo {author} {\bibfnamefont {B.}~\bibnamefont
  {Derrida}}, \bibinfo {author} {\bibfnamefont {M.~R.}\ \bibnamefont {Evans}},
  \bibinfo {author} {\bibfnamefont {V.}~\bibnamefont {Hakim}},\ and\ \bibinfo
  {author} {\bibfnamefont {V.}~\bibnamefont {Pasquier}},\ }\bibfield  {title}
  {\bibinfo {title} {Exact solution of a 1d asymmetric exclusion model using a
  matrix formulation},\ }\href {https://doi.org/10.1088/0305-4470/26/7/011} {\
  \textbf {\bibinfo {volume} {26}},\ \bibinfo {pages} {1493}
  ({\natexlab{b}})}\BibitemShut {NoStop}%
\bibitem [{\citenamefont {Schütz}\ and\ \citenamefont
  {Domany}()}]{schutz_phase_1993}%
  \BibitemOpen
  \bibfield  {author} {\bibinfo {author} {\bibfnamefont {G.}~\bibnamefont
  {Schütz}}\ and\ \bibinfo {author} {\bibfnamefont {E.}~\bibnamefont
  {Domany}},\ }\bibfield  {title} {\bibinfo {title} {Phase transitions in an
  exactly soluble one-dimensional exclusion process},\ }\href
  {https://doi.org/10.1007/BF01048050} {\ \textbf {\bibinfo {volume} {72}},\
  \bibinfo {pages} {277}}\BibitemShut {NoStop}%
\bibitem [{\citenamefont {Rajewsky}\ \emph {et~al.}()\citenamefont {Rajewsky},
  \citenamefont {Santen}, \citenamefont {Schadschneider},\ and\ \citenamefont
  {Schreckenberg}}]{rajewsky_asymmetric_1997}%
  \BibitemOpen
  \bibfield  {author} {\bibinfo {author} {\bibfnamefont {N.}~\bibnamefont
  {Rajewsky}}, \bibinfo {author} {\bibfnamefont {L.}~\bibnamefont {Santen}},
  \bibinfo {author} {\bibfnamefont {A.}~\bibnamefont {Schadschneider}},\ and\
  \bibinfo {author} {\bibfnamefont {M.}~\bibnamefont {Schreckenberg}},\ }\href
  {http://arxiv.org/abs/cond-mat/9710316} {\bibinfo {title} {The asymmetric
  exclusion process: Comparison of update procedures}},\ \Eprint
  {https://arxiv.org/abs/cond-mat/9710316} {cond-mat/9710316} \BibitemShut
  {NoStop}%
\bibitem [{\citenamefont {de~Gier}\ and\ \citenamefont
  {Essler}()}]{de_gier_bethe_2005}%
  \BibitemOpen
  \bibfield  {author} {\bibinfo {author} {\bibfnamefont {J.}~\bibnamefont
  {de~Gier}}\ and\ \bibinfo {author} {\bibfnamefont {F.~H.~L.}\ \bibnamefont
  {Essler}},\ }\bibfield  {title} {\bibinfo {title} {Bethe ansatz solution of
  the asymmetric exclusion process with open boundaries},\ }\href
  {https://doi.org/10.1103/PhysRevLett.95.240601} {\ \textbf {\bibinfo {volume}
  {95}},\ \bibinfo {pages} {240601}}\BibitemShut {NoStop}%
\bibitem [{\citenamefont {Brankov}\ \emph {et~al.}()\citenamefont {Brankov},
  \citenamefont {Pesheva},\ and\ \citenamefont
  {Bunzarova}}]{brankov_totally_2004}%
  \BibitemOpen
  \bibfield  {author} {\bibinfo {author} {\bibfnamefont {J.}~\bibnamefont
  {Brankov}}, \bibinfo {author} {\bibfnamefont {N.}~\bibnamefont {Pesheva}},\
  and\ \bibinfo {author} {\bibfnamefont {N.}~\bibnamefont {Bunzarova}},\
  }\bibfield  {title} {\bibinfo {title} {Totally asymmetric exclusion process
  on chains with a double-chain section in the middle: Computer simulations and
  a simple theory},\ }\href {https://doi.org/10.1103/PhysRevE.69.066128} {\
  \textbf {\bibinfo {volume} {69}},\ \bibinfo {pages} {066128}},\ \bibinfo
  {note} {publisher: American Physical Society}\BibitemShut {NoStop}%
\bibitem [{\citenamefont {Pronina}\ and\ \citenamefont
  {Kolomeisky}({\natexlab{a}})}]{pronina_two-channel_2004}%
  \BibitemOpen
  \bibfield  {author} {\bibinfo {author} {\bibfnamefont {E.}~\bibnamefont
  {Pronina}}\ and\ \bibinfo {author} {\bibfnamefont {A.~B.}\ \bibnamefont
  {Kolomeisky}},\ }\bibfield  {title} {\bibinfo {title} {Two-channel totally
  asymmetric simple exclusion processes},\ }\href
  {https://doi.org/10.1088/0305-4470/37/42/005} {\ \textbf {\bibinfo {volume}
  {37}},\ \bibinfo {pages} {9907} ({\natexlab{a}})}\BibitemShut {NoStop}%
\bibitem [{\citenamefont {Mitsudo}\ and\ \citenamefont
  {Hayakawa}()}]{mitsudo_synchronization_2005}%
  \BibitemOpen
  \bibfield  {author} {\bibinfo {author} {\bibfnamefont {T.}~\bibnamefont
  {Mitsudo}}\ and\ \bibinfo {author} {\bibfnamefont {H.}~\bibnamefont
  {Hayakawa}},\ }\bibfield  {title} {\bibinfo {title} {Synchronization of kinks
  in the two-lane totally asymmetric simple exclusion process with open
  boundary conditions},\ }\href {https://doi.org/10.1088/0305-4470/38/14/002}
  {\ \textbf {\bibinfo {volume} {38}},\ \bibinfo {pages} {3087}}\BibitemShut
  {NoStop}%
\bibitem [{\citenamefont {Pronina}\ and\ \citenamefont
  {Kolomeisky}({\natexlab{b}})}]{pronina_spontaneous_2007}%
  \BibitemOpen
  \bibfield  {author} {\bibinfo {author} {\bibfnamefont {E.}~\bibnamefont
  {Pronina}}\ and\ \bibinfo {author} {\bibfnamefont {A.~B.}\ \bibnamefont
  {Kolomeisky}},\ }\bibfield  {title} {\bibinfo {title} {Spontaneous symmetry
  breaking in two-channel asymmetric exclusion processes with narrow
  entrances},\ }\href {https://doi.org/10.1088/1751-8113/40/10/004} {\ \textbf
  {\bibinfo {volume} {40}},\ \bibinfo {pages} {2275}
  ({\natexlab{b}})}\BibitemShut {NoStop}%
\bibitem [{\citenamefont {Tsekouras}\ and\ \citenamefont
  {Kolomeisky}()}]{tsekouras_inhomogeneous_2008}%
  \BibitemOpen
  \bibfield  {author} {\bibinfo {author} {\bibfnamefont {K.}~\bibnamefont
  {Tsekouras}}\ and\ \bibinfo {author} {\bibfnamefont {A.~B.}\ \bibnamefont
  {Kolomeisky}},\ }\bibfield  {title} {\bibinfo {title} {Inhomogeneous coupling
  in two-channel asymmetric simple exclusion processes},\ }\href
  {https://doi.org/10.1088/1751-8113/41/9/095002} {\ \textbf {\bibinfo {volume}
  {41}},\ \bibinfo {pages} {095002}}\BibitemShut {NoStop}%
\bibitem [{\citenamefont {Yuan}\ \emph {et~al.}()\citenamefont {Yuan},
  \citenamefont {Jiang}, \citenamefont {Wang}, \citenamefont {Hu},\ and\
  \citenamefont {Wu}}]{yuan_totally_2007}%
  \BibitemOpen
  \bibfield  {author} {\bibinfo {author} {\bibfnamefont {Y.-M.}\ \bibnamefont
  {Yuan}}, \bibinfo {author} {\bibfnamefont {R.}~\bibnamefont {Jiang}},
  \bibinfo {author} {\bibfnamefont {R.}~\bibnamefont {Wang}}, \bibinfo {author}
  {\bibfnamefont {M.-B.}\ \bibnamefont {Hu}},\ and\ \bibinfo {author}
  {\bibfnamefont {Q.-S.}\ \bibnamefont {Wu}},\ }\bibfield  {title} {\bibinfo
  {title} {Totally asymmetric simple exclusion process with a shortcut},\
  }\href {https://doi.org/10.1088/1751-8113/40/41/006} {\ \textbf {\bibinfo
  {volume} {40}},\ \bibinfo {pages} {12351}}\BibitemShut {NoStop}%
\bibitem [{\citenamefont {Bunzarova}\ \emph {et~al.}()\citenamefont
  {Bunzarova}, \citenamefont {Pesheva},\ and\ \citenamefont
  {Brankov}}]{bunzarova_asymmetric_2014}%
  \BibitemOpen
  \bibfield  {author} {\bibinfo {author} {\bibfnamefont {N.}~\bibnamefont
  {Bunzarova}}, \bibinfo {author} {\bibfnamefont {N.}~\bibnamefont {Pesheva}},\
  and\ \bibinfo {author} {\bibfnamefont {J.}~\bibnamefont {Brankov}},\
  }\bibfield  {title} {\bibinfo {title} {Asymmetric simple exclusion process on
  chains with a shortcut},\ }\href {https://doi.org/10.1103/PhysRevE.89.032125}
  {\ \textbf {\bibinfo {volume} {89}},\ \bibinfo {pages} {032125}},\ \bibinfo
  {note} {publisher: American Physical Society}\BibitemShut {NoStop}%
\bibitem [{\citenamefont {Xiao}\ \emph {et~al.}({\natexlab{a}})\citenamefont
  {Xiao}, \citenamefont {Chen}, \citenamefont {Liang},\ and\ \citenamefont
  {Liu}}]{xiao_shortcut_2017}%
  \BibitemOpen
  \bibfield  {author} {\bibinfo {author} {\bibfnamefont {S.}~\bibnamefont
  {Xiao}}, \bibinfo {author} {\bibfnamefont {X.}~\bibnamefont {Chen}}, \bibinfo
  {author} {\bibfnamefont {R.}~\bibnamefont {Liang}},\ and\ \bibinfo {author}
  {\bibfnamefont {Y.}~\bibnamefont {Liu}},\ }\bibfield  {title} {\bibinfo
  {title} {Shortcut with pool under parallel update rule on one-dimensional
  lattice totally asymmetric simple exclusion process},\ }\href
  {https://doi.org/10.1016/j.physleta.2017.10.016} {\ \textbf {\bibinfo
  {volume} {381}},\ \bibinfo {pages} {3940} ({\natexlab{a}})}\BibitemShut
  {NoStop}%
\bibitem [{\citenamefont {Evans}\ \emph {et~al.}({\natexlab{a}})\citenamefont
  {Evans}, \citenamefont {Juhász},\ and\ \citenamefont
  {Santen}}]{evans_shock_2003}%
  \BibitemOpen
  \bibfield  {author} {\bibinfo {author} {\bibfnamefont {M.~R.}\ \bibnamefont
  {Evans}}, \bibinfo {author} {\bibfnamefont {R.}~\bibnamefont {Juhász}},\
  and\ \bibinfo {author} {\bibfnamefont {L.}~\bibnamefont {Santen}},\
  }\bibfield  {title} {\bibinfo {title} {Shock formation in an exclusion
  process with creation and annihilation},\ }\href
  {https://doi.org/10.1103/PhysRevE.68.026117} {\ \textbf {\bibinfo {volume}
  {68}},\ \bibinfo {pages} {026117} ({\natexlab{a}})},\ \bibinfo {note}
  {publisher: American Physical Society}\BibitemShut {NoStop}%
\bibitem [{\citenamefont {Jiang}\ \emph {et~al.}()\citenamefont {Jiang},
  \citenamefont {Wang},\ and\ \citenamefont {Wu}}]{jiang_two-lane_2007}%
  \BibitemOpen
  \bibfield  {author} {\bibinfo {author} {\bibfnamefont {R.}~\bibnamefont
  {Jiang}}, \bibinfo {author} {\bibfnamefont {R.}~\bibnamefont {Wang}},\ and\
  \bibinfo {author} {\bibfnamefont {Q.-S.}\ \bibnamefont {Wu}},\ }\bibfield
  {title} {\bibinfo {title} {Two-lane totally asymmetric exclusion processes
  with particle creation and annihilation},\ }\href
  {https://doi.org/10.1016/j.physa.2006.08.025} {\ \textbf {\bibinfo {volume}
  {375}},\ \bibinfo {pages} {247}}\BibitemShut {NoStop}%
\bibitem [{\citenamefont {Wang}\ \emph {et~al.}()\citenamefont {Wang},
  \citenamefont {Jiang}, \citenamefont {Liu}, \citenamefont {Liu},\ and\
  \citenamefont {Wu}}]{wang_effects_2007}%
  \BibitemOpen
  \bibfield  {author} {\bibinfo {author} {\bibfnamefont {R.}~\bibnamefont
  {Wang}}, \bibinfo {author} {\bibfnamefont {R.}~\bibnamefont {Jiang}},
  \bibinfo {author} {\bibfnamefont {M.}~\bibnamefont {Liu}}, \bibinfo {author}
  {\bibfnamefont {J.}~\bibnamefont {Liu}},\ and\ \bibinfo {author}
  {\bibfnamefont {Q.-S.}\ \bibnamefont {Wu}},\ }\bibfield  {title} {\bibinfo
  {title} {Effects of langmuir kinetics on two-lane totally asymmetric
  exclusion processes of molecular motor traffic},\ }\href
  {https://doi.org/10.1142/S0129183107011479} {\ \textbf {\bibinfo {volume}
  {18}},\ \bibinfo {pages} {1483}},\ \bibinfo {note} {publisher: World
  Scientific Publishing Co.}\BibitemShut {Stop}%
\bibitem [{\citenamefont {Gupta}\ and\ \citenamefont
  {Dhiman}()}]{gupta_asymmetric_2014}%
  \BibitemOpen
  \bibfield  {author} {\bibinfo {author} {\bibfnamefont {A.~K.}\ \bibnamefont
  {Gupta}}\ and\ \bibinfo {author} {\bibfnamefont {I.}~\bibnamefont {Dhiman}},\
  }\bibfield  {title} {\bibinfo {title} {Asymmetric coupling in two-lane simple
  exclusion processes with langmuir kinetics: Phase diagrams and boundary
  layers},\ }\href {https://doi.org/10.1103/PhysRevE.89.022131} {\ \textbf
  {\bibinfo {volume} {89}},\ \bibinfo {pages} {022131}},\ \bibinfo {note}
  {publisher: American Physical Society}\BibitemShut {NoStop}%
\bibitem [{\citenamefont {Botto}\ \emph {et~al.}()\citenamefont {Botto},
  \citenamefont {Pelizzola}, \citenamefont {Pretti},\ and\ \citenamefont
  {Zamparo}}]{botto_dynamical_2019}%
  \BibitemOpen
  \bibfield  {author} {\bibinfo {author} {\bibfnamefont {D.}~\bibnamefont
  {Botto}}, \bibinfo {author} {\bibfnamefont {A.}~\bibnamefont {Pelizzola}},
  \bibinfo {author} {\bibfnamefont {M.}~\bibnamefont {Pretti}},\ and\ \bibinfo
  {author} {\bibfnamefont {M.}~\bibnamefont {Zamparo}},\ }\bibfield  {title}
  {\bibinfo {title} {Dynamical transition in the {TASEP} with langmuir
  kinetics: mean-field theory},\ }\href
  {https://doi.org/10.1088/1751-8121/aaf1f8} {\ \textbf {\bibinfo {volume}
  {52}},\ \bibinfo {pages} {045001}},\ \Eprint
  {https://arxiv.org/abs/1809.03231 [cond-mat, physics:math-ph, physics:nlin]}
  {1809.03231 [cond-mat, physics:math-ph, physics:nlin]} \BibitemShut {NoStop}%
\bibitem [{\citenamefont {Evans}\ \emph {et~al.}({\natexlab{b}})\citenamefont
  {Evans}, \citenamefont {Foster}, \citenamefont {Godrèche},\ and\
  \citenamefont {Mukamel}}]{evans_asymmetric_1995}%
  \BibitemOpen
  \bibfield  {author} {\bibinfo {author} {\bibfnamefont {M.~R.}\ \bibnamefont
  {Evans}}, \bibinfo {author} {\bibfnamefont {D.~P.}\ \bibnamefont {Foster}},
  \bibinfo {author} {\bibfnamefont {C.}~\bibnamefont {Godrèche}},\ and\
  \bibinfo {author} {\bibfnamefont {D.}~\bibnamefont {Mukamel}},\ }\bibfield
  {title} {\bibinfo {title} {Asymmetric exclusion model with two species:
  Spontaneous symmetry breaking},\ }\href {https://doi.org/10.1007/BF02178354}
  {\ \textbf {\bibinfo {volume} {80}},\ \bibinfo {pages} {69}
  ({\natexlab{b}})}\BibitemShut {NoStop}%
\bibitem [{\citenamefont {Schütz}()}]{schutz_critical_2003}%
  \BibitemOpen
  \bibfield  {author} {\bibinfo {author} {\bibfnamefont {G.~M.}\ \bibnamefont
  {Schütz}},\ }\bibfield  {title} {\bibinfo {title} {Critical phenomena and
  universal dynamics in one-dimensional driven diffusive systems with two
  species of particles},\ }\href {https://doi.org/10.1088/0305-4470/36/36/201}
  {\ \textbf {\bibinfo {volume} {36}},\ \bibinfo {pages} {R339}}\BibitemShut
  {NoStop}%
\bibitem [{\citenamefont {Muhuri}\ and\ \citenamefont
  {Pagonabarraga}()}]{muhuri_collective_2008}%
  \BibitemOpen
  \bibfield  {author} {\bibinfo {author} {\bibfnamefont {S.}~\bibnamefont
  {Muhuri}}\ and\ \bibinfo {author} {\bibfnamefont {I.}~\bibnamefont
  {Pagonabarraga}},\ }\bibfield  {title} {\bibinfo {title} {Collective vesicle
  transport on biofilaments carried by competing molecular motors},\ }\href
  {https://doi.org/10.1209/0295-5075/84/58009} {\ \textbf {\bibinfo {volume}
  {84}},\ \bibinfo {pages} {58009}}\BibitemShut {NoStop}%
\bibitem [{\citenamefont {Ayyer}\ \emph {et~al.}()\citenamefont {Ayyer},
  \citenamefont {Lebowitz},\ and\ \citenamefont {Speer}}]{ayyer_classes_2010}%
  \BibitemOpen
  \bibfield  {author} {\bibinfo {author} {\bibfnamefont {A.}~\bibnamefont
  {Ayyer}}, \bibinfo {author} {\bibfnamefont {J.~L.}\ \bibnamefont
  {Lebowitz}},\ and\ \bibinfo {author} {\bibfnamefont {E.~R.}\ \bibnamefont
  {Speer}},\ }\href {https://doi.org/10.48550/arXiv.1008.4721} {\bibinfo
  {title} {On some classes of open two-species exclusion processes}},\ \Eprint
  {https://arxiv.org/abs/1008.4721 [cond-mat]} {1008.4721 [cond-mat]}
  \BibitemShut {NoStop}%
\bibitem [{\citenamefont {Crampe}\ \emph {et~al.}()\citenamefont {Crampe},
  \citenamefont {Evans}, \citenamefont {Mallick}, \citenamefont {Ragoucy},\
  and\ \citenamefont {Vanicat}}]{crampe_matrix_2016}%
  \BibitemOpen
  \bibfield  {author} {\bibinfo {author} {\bibfnamefont {N.}~\bibnamefont
  {Crampe}}, \bibinfo {author} {\bibfnamefont {M.~R.}\ \bibnamefont {Evans}},
  \bibinfo {author} {\bibfnamefont {K.}~\bibnamefont {Mallick}}, \bibinfo
  {author} {\bibfnamefont {E.}~\bibnamefont {Ragoucy}},\ and\ \bibinfo {author}
  {\bibfnamefont {M.}~\bibnamefont {Vanicat}},\ }\bibfield  {title} {\bibinfo
  {title} {Matrix product solution to a 2-species {TASEP} with open integrable
  boundaries},\ }\href {https://doi.org/10.1088/1751-8113/49/47/475001} {\
  \textbf {\bibinfo {volume} {49}},\ \bibinfo {pages} {475001}},\ \bibinfo
  {note} {publisher: {IOP} Publishing}\BibitemShut {NoStop}%
\bibitem [{\citenamefont {Bottero}\ and\ \citenamefont
  {Frey}(2017)}]{bottero_analysis_2017}%
  \BibitemOpen
  \bibfield  {author} {\bibinfo {author} {\bibfnamefont {A.}~\bibnamefont
  {Bottero}}\ and\ \bibinfo {author} {\bibfnamefont {E.}~\bibnamefont {Frey}},\
  }\bibfield  {title} {\bibinfo {title} {Analysis of a two species tasep as a
  model for heterogeneous transport on microtubules},\ }\href@noop {}
  {\bibfield  {journal} {\bibinfo  {journal} {Theoretical and Mathematical
  Physics}\ } (\bibinfo {year} {2017})}\BibitemShut {NoStop}%
\bibitem [{\citenamefont {Bonnin}\ \emph {et~al.}()\citenamefont {Bonnin},
  \citenamefont {Stansfield}, \citenamefont {Romano},\ and\ \citenamefont
  {Kern}}]{bonnin_two-species_2022}%
  \BibitemOpen
  \bibfield  {author} {\bibinfo {author} {\bibfnamefont {P.}~\bibnamefont
  {Bonnin}}, \bibinfo {author} {\bibfnamefont {I.}~\bibnamefont {Stansfield}},
  \bibinfo {author} {\bibfnamefont {M.~C.}\ \bibnamefont {Romano}},\ and\
  \bibinfo {author} {\bibfnamefont {N.}~\bibnamefont {Kern}},\ }\bibfield
  {title} {\bibinfo {title} {Two-species totally asymmetric simple exclusion
  process model: From a simple description to intermittency and traveling
  traffic jams},\ }\href {https://doi.org/10.1103/PhysRevE.105.034117} {\
  \textbf {\bibinfo {volume} {105}},\ \bibinfo {pages} {034117}},\ \bibinfo
  {note} {publisher: American Physical Society}\BibitemShut {NoStop}%
\bibitem [{\citenamefont {Kolomeisky}()}]{kolomeisky_asymmetric_1998}%
  \BibitemOpen
  \bibfield  {author} {\bibinfo {author} {\bibfnamefont {A.~B.}\ \bibnamefont
  {Kolomeisky}},\ }\bibfield  {title} {\bibinfo {title} {Asymmetric simple
  exclusion model with local inhomogeneity},\ }\href
  {https://doi.org/10.1088/0305-4470/31/4/006} {\ \textbf {\bibinfo {volume}
  {31}},\ \bibinfo {pages} {1153}}\BibitemShut {NoStop}%
\bibitem [{\citenamefont {Ha}\ \emph {et~al.}()\citenamefont {Ha},
  \citenamefont {Timonen},\ and\ \citenamefont {Nijs}}]{ha_queuing_2003}%
  \BibitemOpen
  \bibfield  {author} {\bibinfo {author} {\bibfnamefont {M.}~\bibnamefont
  {Ha}}, \bibinfo {author} {\bibfnamefont {J.}~\bibnamefont {Timonen}},\ and\
  \bibinfo {author} {\bibfnamefont {M.~d.}\ \bibnamefont {Nijs}},\ }\bibfield
  {title} {\bibinfo {title} {Queuing transitions in the asymmetric simple
  exclusion process},\ }\href {https://doi.org/10.1103/PhysRevE.68.056122} {\
  \textbf {\bibinfo {volume} {68}},\ \bibinfo {pages} {056122}},\ \Eprint
  {https://arxiv.org/abs/cond-mat/0307403} {cond-mat/0307403} \BibitemShut
  {NoStop}%
\bibitem [{\citenamefont {Chou}\ and\ \citenamefont
  {Lakatos}()}]{chou_clustered_2004}%
  \BibitemOpen
  \bibfield  {author} {\bibinfo {author} {\bibfnamefont {T.}~\bibnamefont
  {Chou}}\ and\ \bibinfo {author} {\bibfnamefont {G.}~\bibnamefont {Lakatos}},\
  }\bibfield  {title} {\bibinfo {title} {Clustered bottlenecks in {mRNA}
  translation and protein synthesis},\ }\href
  {https://doi.org/10.1103/PhysRevLett.93.198101} {\ \textbf {\bibinfo {volume}
  {93}},\ \bibinfo {pages} {198101}},\ \bibinfo {note} {publisher: American
  Physical Society}\BibitemShut {NoStop}%
\bibitem [{\citenamefont {Pierobon}\ \emph {et~al.}()\citenamefont {Pierobon},
  \citenamefont {Mobilia}, \citenamefont {Kouyos},\ and\ \citenamefont
  {Frey}}]{pierobon_bottleneck-induced_2006}%
  \BibitemOpen
  \bibfield  {author} {\bibinfo {author} {\bibfnamefont {P.}~\bibnamefont
  {Pierobon}}, \bibinfo {author} {\bibfnamefont {M.}~\bibnamefont {Mobilia}},
  \bibinfo {author} {\bibfnamefont {R.}~\bibnamefont {Kouyos}},\ and\ \bibinfo
  {author} {\bibfnamefont {E.}~\bibnamefont {Frey}},\ }\bibfield  {title}
  {\bibinfo {title} {Bottleneck-induced transitions in a minimal model for
  intracellular transport},\ }\href
  {https://doi.org/10.1103/PhysRevE.74.031906} {\ \textbf {\bibinfo {volume}
  {74}},\ \bibinfo {pages} {031906}},\ \bibinfo {note} {publisher: American
  Physical Society}\BibitemShut {NoStop}%
\bibitem [{\citenamefont {Dong}\ \emph {et~al.}({\natexlab{a}})\citenamefont
  {Dong}, \citenamefont {Schmittmann},\ and\ \citenamefont
  {Zia}}]{dong_towards_2007}%
  \BibitemOpen
  \bibfield  {author} {\bibinfo {author} {\bibfnamefont {J.~J.}\ \bibnamefont
  {Dong}}, \bibinfo {author} {\bibfnamefont {B.}~\bibnamefont {Schmittmann}},\
  and\ \bibinfo {author} {\bibfnamefont {R.~K.~P.}\ \bibnamefont {Zia}},\
  }\bibfield  {title} {\bibinfo {title} {Towards a model for protein production
  rates},\ }\href {https://doi.org/10.1007/s10955-006-9134-7} {\ \textbf
  {\bibinfo {volume} {128}},\ \bibinfo {pages} {21}
  ({\natexlab{a}})}\BibitemShut {NoStop}%
\bibitem [{\citenamefont {Greulich}\ and\ \citenamefont
  {Schadschneider}()}]{greulich_phase_2008}%
  \BibitemOpen
  \bibfield  {author} {\bibinfo {author} {\bibfnamefont {P.}~\bibnamefont
  {Greulich}}\ and\ \bibinfo {author} {\bibfnamefont {A.}~\bibnamefont
  {Schadschneider}},\ }\bibfield  {title} {\bibinfo {title} {Phase diagram and
  edge effects in the {ASEP} with bottlenecks},\ }\href
  {https://doi.org/10.1016/j.physa.2007.11.037} {\ \textbf {\bibinfo {volume}
  {387}},\ \bibinfo {pages} {1972}}\BibitemShut {NoStop}%
\bibitem [{\citenamefont {Schmidt}\ \emph {et~al.}()\citenamefont {Schmidt},
  \citenamefont {Popkov},\ and\ \citenamefont
  {Schadschneider}}]{schmidt_defect-induced_2015}%
  \BibitemOpen
  \bibfield  {author} {\bibinfo {author} {\bibfnamefont {J.}~\bibnamefont
  {Schmidt}}, \bibinfo {author} {\bibfnamefont {V.}~\bibnamefont {Popkov}},\
  and\ \bibinfo {author} {\bibfnamefont {A.}~\bibnamefont {Schadschneider}},\
  }\bibfield  {title} {\bibinfo {title} {Defect-induced phase transition in the
  asymmetric simple exclusion process},\ }\href
  {https://doi.org/10.1209/0295-5075/110/20008} {\ \textbf {\bibinfo {volume}
  {110}},\ \bibinfo {pages} {20008}},\ \bibinfo {note} {publisher: {EDP}
  Sciences, {IOP} Publishing and Società Italiana di Fisica}\BibitemShut
  {NoStop}%
\bibitem [{\citenamefont {Shoup}()}]{shoup_cruising_2006}%
  \BibitemOpen
  \bibfield  {author} {\bibinfo {author} {\bibfnamefont {D.~C.}\ \bibnamefont
  {Shoup}},\ }\bibfield  {title} {\bibinfo {title} {Cruising for parking},\
  }\href {https://doi.org/10.1016/j.tranpol.2006.05.005} {\ \bibinfo {series}
  {Parking},\ \textbf {\bibinfo {volume} {13}},\ \bibinfo {pages}
  {479}}\BibitemShut {NoStop}%
\bibitem [{\citenamefont {Dowling}\ \emph {et~al.}(2019)\citenamefont
  {Dowling}, \citenamefont {Ratliff},\ and\ \citenamefont
  {Zhang}}]{dowling2019modeling}%
  \BibitemOpen
  \bibfield  {author} {\bibinfo {author} {\bibfnamefont {C.~P.}\ \bibnamefont
  {Dowling}}, \bibinfo {author} {\bibfnamefont {L.~J.}\ \bibnamefont
  {Ratliff}},\ and\ \bibinfo {author} {\bibfnamefont {B.}~\bibnamefont
  {Zhang}},\ }\bibfield  {title} {\bibinfo {title} {Modeling curbside parking
  as a network of finite capacity queues},\ }\href@noop {} {\bibfield
  {journal} {\bibinfo  {journal} {IEEE Transactions on Intelligent
  Transportation Systems}\ }\textbf {\bibinfo {volume} {21}},\ \bibinfo {pages}
  {1011} (\bibinfo {year} {2019})}\BibitemShut {NoStop}%
\bibitem [{\citenamefont {Ha}\ and\ \citenamefont {den
  Nijs}()}]{ha_macroscopic_2002}%
  \BibitemOpen
  \bibfield  {author} {\bibinfo {author} {\bibfnamefont {M.}~\bibnamefont
  {Ha}}\ and\ \bibinfo {author} {\bibfnamefont {M.}~\bibnamefont {den Nijs}},\
  }\bibfield  {title} {\bibinfo {title} {Macroscopic car condensation in a
  parking garage},\ }\href {https://doi.org/10.1103/PhysRevE.66.036118} {\
  \textbf {\bibinfo {volume} {66}},\ \bibinfo {pages} {036118}},\ \bibinfo
  {note} {publisher: American Physical Society}\BibitemShut {NoStop}%
\bibitem [{\citenamefont {Adams}\ \emph {et~al.}()\citenamefont {Adams},
  \citenamefont {Schmittmann},\ and\ \citenamefont
  {Zia}}]{adams_far--equilibrium_2008}%
  \BibitemOpen
  \bibfield  {author} {\bibinfo {author} {\bibfnamefont {D.~A.}\ \bibnamefont
  {Adams}}, \bibinfo {author} {\bibfnamefont {B.}~\bibnamefont {Schmittmann}},\
  and\ \bibinfo {author} {\bibfnamefont {R.~K.~P.}\ \bibnamefont {Zia}},\
  }\bibfield  {title} {\bibinfo {title} {Far-from-equilibrium transport with
  constrained resources},\ }\href
  {https://doi.org/10.1088/1742-5468/2008/06/P06009} {\ \textbf {\bibinfo
  {volume} {2008}},\ \bibinfo {pages} {P06009}},\ \Eprint
  {https://arxiv.org/abs/0804.2717 [cond-mat]} {0804.2717 [cond-mat]}
  \BibitemShut {NoStop}%
\bibitem [{\citenamefont {Cook}\ and\ \citenamefont
  {Zia}()}]{cook_feedback_2009}%
  \BibitemOpen
  \bibfield  {author} {\bibinfo {author} {\bibfnamefont {L.~J.}\ \bibnamefont
  {Cook}}\ and\ \bibinfo {author} {\bibfnamefont {R.~K.~P.}\ \bibnamefont
  {Zia}},\ }\bibfield  {title} {\bibinfo {title} {Feedback and fluctuations in
  a totally asymmetric simple exclusion process with finite resources},\ }\href
  {https://doi.org/10.1088/1742-5468/2009/02/P02012} {\ \textbf {\bibinfo
  {volume} {2009}},\ \bibinfo {pages} {P02012}}\BibitemShut {NoStop}%
\bibitem [{\citenamefont {Cook}\ \emph {et~al.}()\citenamefont {Cook},
  \citenamefont {Zia},\ and\ \citenamefont
  {Schmittmann}}]{cook_competition_2009}%
  \BibitemOpen
  \bibfield  {author} {\bibinfo {author} {\bibfnamefont {L.~J.}\ \bibnamefont
  {Cook}}, \bibinfo {author} {\bibfnamefont {R.~K.~P.}\ \bibnamefont {Zia}},\
  and\ \bibinfo {author} {\bibfnamefont {B.}~\bibnamefont {Schmittmann}},\
  }\bibfield  {title} {\bibinfo {title} {Competition between multiple totally
  asymmetric simple exclusion processes for a finite pool of resources},\
  }\href {https://doi.org/10.1103/PhysRevE.80.031142} {\ \textbf {\bibinfo
  {volume} {80}},\ \bibinfo {pages} {031142}},\ \bibinfo {note} {publisher:
  American Physical Society}\BibitemShut {NoStop}%
\bibitem [{\citenamefont {Greulich}\ \emph {et~al.}()\citenamefont {Greulich},
  \citenamefont {Ciandrini}, \citenamefont {Allen},\ and\ \citenamefont
  {Romano}}]{greulich_mixed_2012}%
  \BibitemOpen
  \bibfield  {author} {\bibinfo {author} {\bibfnamefont {P.}~\bibnamefont
  {Greulich}}, \bibinfo {author} {\bibfnamefont {L.}~\bibnamefont {Ciandrini}},
  \bibinfo {author} {\bibfnamefont {R.~J.}\ \bibnamefont {Allen}},\ and\
  \bibinfo {author} {\bibfnamefont {M.~C.}\ \bibnamefont {Romano}},\ }\bibfield
   {title} {\bibinfo {title} {Mixed population of competing totally asymmetric
  simple exclusion processes with a shared reservoir of particles},\ }\href
  {https://doi.org/10.1103/PhysRevE.85.011142} {\ \textbf {\bibinfo {volume}
  {85}},\ \bibinfo {pages} {011142}},\ \bibinfo {note} {publisher: American
  Physical Society}\BibitemShut {NoStop}%
\bibitem [{\citenamefont {Levy}\ \emph {et~al.}()\citenamefont {Levy},
  \citenamefont {Martens},\ and\ \citenamefont
  {Benenson}}]{levy_exploring_2013}%
  \BibitemOpen
  \bibfield  {author} {\bibinfo {author} {\bibfnamefont {N.}~\bibnamefont
  {Levy}}, \bibinfo {author} {\bibfnamefont {K.}~\bibnamefont {Martens}},\ and\
  \bibinfo {author} {\bibfnamefont {I.}~\bibnamefont {Benenson}},\ }\bibfield
  {title} {\bibinfo {title} {Exploring cruising using agent-based and
  analytical models of parking},\ }\href
  {https://doi.org/10.1080/18128602.2012.664575} {\ \textbf {\bibinfo {volume}
  {9}},\ \bibinfo {pages} {773}},\ \bibinfo {note} {publisher: Taylor \&
  Francis \_eprint: https://doi.org/10.1080/18128602.2012.664575}\BibitemShut
  {NoStop}%
\bibitem [{\citenamefont {Xiao}\ \emph {et~al.}({\natexlab{b}})\citenamefont
  {Xiao}, \citenamefont {Lou},\ and\ \citenamefont {Frisby}}]{xiao_how_2018}%
  \BibitemOpen
  \bibfield  {author} {\bibinfo {author} {\bibfnamefont {J.}~\bibnamefont
  {Xiao}}, \bibinfo {author} {\bibfnamefont {Y.}~\bibnamefont {Lou}},\ and\
  \bibinfo {author} {\bibfnamefont {J.}~\bibnamefont {Frisby}},\ }\bibfield
  {title} {\bibinfo {title} {How likely am i to find parking? – a practical
  model-based framework for predicting parking availability},\ }\href
  {https://doi.org/10.1016/j.trb.2018.04.001} {\ \textbf {\bibinfo {volume}
  {112}},\ \bibinfo {pages} {19} ({\natexlab{b}})}\BibitemShut {NoStop}%
\bibitem [{\citenamefont {Fulman}\ \emph {et~al.}()\citenamefont {Fulman},
  \citenamefont {Benenson},\ and\ \citenamefont
  {Ben-Elia}}]{fulman_modeling_2020}%
  \BibitemOpen
  \bibfield  {author} {\bibinfo {author} {\bibfnamefont {N.}~\bibnamefont
  {Fulman}}, \bibinfo {author} {\bibfnamefont {I.}~\bibnamefont {Benenson}},\
  and\ \bibinfo {author} {\bibfnamefont {E.}~\bibnamefont {Ben-Elia}},\
  }\bibfield  {title} {\bibinfo {title} {Modeling parking search behavior in
  the city center: A game-based approach},\ }\href
  {https://doi.org/10.1016/j.trc.2020.102800} {\ \textbf {\bibinfo {volume}
  {120}},\ \bibinfo {pages} {102800}}\BibitemShut {NoStop}%
\bibitem [{\citenamefont {Lubrich}()}]{lubrich_analysis_2023}%
  \BibitemOpen
  \bibfield  {author} {\bibinfo {author} {\bibfnamefont {P.}~\bibnamefont
  {Lubrich}},\ }\bibfield  {title} {\bibinfo {title} {Analysis of parking
  traffic in cologne, germany, based on an extended macroscopic transport model
  and parking {API} data},\ }\href {https://doi.org/10.1016/j.cstp.2022.100940}
  {\ \textbf {\bibinfo {volume} {11}},\ \bibinfo {pages} {100940}}\BibitemShut
  {NoStop}%
\bibitem [{\citenamefont {Dutta}\ \emph {et~al.}()\citenamefont {Dutta},
  \citenamefont {Charlottin},\ and\ \citenamefont
  {Nicolas}}]{dutta_parking_2023}%
  \BibitemOpen
  \bibfield  {author} {\bibinfo {author} {\bibfnamefont {N.}~\bibnamefont
  {Dutta}}, \bibinfo {author} {\bibfnamefont {T.}~\bibnamefont {Charlottin}},\
  and\ \bibinfo {author} {\bibfnamefont {A.}~\bibnamefont {Nicolas}},\ }\href
  {https://doi.org/10.48550/arXiv.2202.00258} {\bibinfo {title} {Parking search
  in the physical world: Calculating the search time by leveraging physical and
  graph theoretical methods}},\ \Eprint {https://arxiv.org/abs/2202.00258
  [cond-mat, physics:physics]} {2202.00258 [cond-mat, physics:physics]}
  \BibitemShut {NoStop}%
\bibitem [{\citenamefont {Humenyuk}\ \emph {et~al.}()\citenamefont {Humenyuk},
  \citenamefont {Kotrla}, \citenamefont {Netočný},\ and\ \citenamefont
  {Slanina}}]{humenyuk_separation_2020}%
  \BibitemOpen
  \bibfield  {author} {\bibinfo {author} {\bibfnamefont {Y.~A.}\ \bibnamefont
  {Humenyuk}}, \bibinfo {author} {\bibfnamefont {M.}~\bibnamefont {Kotrla}},
  \bibinfo {author} {\bibfnamefont {K.}~\bibnamefont {Netočný}},\ and\
  \bibinfo {author} {\bibfnamefont {F.}~\bibnamefont {Slanina}},\ }\bibfield
  {title} {\bibinfo {title} {Separation of dense colloidal suspensions in
  narrow channels: A stochastic model},\ }\href
  {https://doi.org/10.1103/PhysRevE.101.032608} {\ \textbf {\bibinfo {volume}
  {101}},\ \bibinfo {pages} {032608}},\ \bibinfo {note} {publisher: American
  Physical Society}\BibitemShut {NoStop}%
\bibitem [{\citenamefont {Bhatia}\ and\ \citenamefont
  {Gupta}()}]{bhatia_far_2022}%
  \BibitemOpen
  \bibfield  {author} {\bibinfo {author} {\bibfnamefont {N.}~\bibnamefont
  {Bhatia}}\ and\ \bibinfo {author} {\bibfnamefont {A.~K.}\ \bibnamefont
  {Gupta}},\ }\bibfield  {title} {\bibinfo {title} {Far from equilibrium
  transport on {TASEP} with pockets},\ }\href
  {https://doi.org/10.1140/epjp/s13360-022-03119-2} {\ \textbf {\bibinfo
  {volume} {137}},\ \bibinfo {pages} {892}}\BibitemShut {NoStop}%
\bibitem [{\citenamefont {Brackley}\ \emph {et~al.}()\citenamefont {Brackley},
  \citenamefont {Ciandrini},\ and\ \citenamefont
  {Romano}}]{brackley_multiple_2012}%
  \BibitemOpen
  \bibfield  {author} {\bibinfo {author} {\bibfnamefont {C.~A.}\ \bibnamefont
  {Brackley}}, \bibinfo {author} {\bibfnamefont {L.}~\bibnamefont
  {Ciandrini}},\ and\ \bibinfo {author} {\bibfnamefont {M.~C.}\ \bibnamefont
  {Romano}},\ }\bibfield  {title} {\bibinfo {title} {Multiple phase transitions
  in a system of exclusion processes with limited reservoirs of particles and
  fuel carriers},\ }\href {https://doi.org/10.1088/1742-5468/2012/03/P03002} {\
  \textbf {\bibinfo {volume} {2012}},\ \bibinfo {pages} {P03002}}\BibitemShut
  {NoStop}%
\bibitem [{\citenamefont {Verma}\ and\ \citenamefont
  {Gupta}()}]{verma_stochastic_2019}%
  \BibitemOpen
  \bibfield  {author} {\bibinfo {author} {\bibfnamefont {A.~K.}\ \bibnamefont
  {Verma}}\ and\ \bibinfo {author} {\bibfnamefont {A.~K.}\ \bibnamefont
  {Gupta}},\ }\bibfield  {title} {\bibinfo {title} {Stochastic transport on
  flexible lattice under limited resources},\ }\href
  {https://doi.org/10.1088/1742-5468/ab417c} {\ \textbf {\bibinfo {volume}
  {2019}},\ \bibinfo {pages} {103210}},\ \bibinfo {note} {publisher: {IOP}
  Publishing and {SISSA}}\BibitemShut {NoStop}%
\bibitem [{\citenamefont {Banerjee}\ and\ \citenamefont
  {Basu}()}]{banerjee_smooth_2020}%
  \BibitemOpen
  \bibfield  {author} {\bibinfo {author} {\bibfnamefont {T.}~\bibnamefont
  {Banerjee}}\ and\ \bibinfo {author} {\bibfnamefont {A.}~\bibnamefont
  {Basu}},\ }\bibfield  {title} {\bibinfo {title} {Smooth or shock:
  Universality in closed inhomogeneous driven single file motions},\ }\href
  {https://doi.org/10.1103/PhysRevResearch.2.013025} {\ \textbf {\bibinfo
  {volume} {2}},\ \bibinfo {pages} {013025}},\ \bibinfo {note} {publisher:
  American Physical Society}\BibitemShut {NoStop}%
\bibitem [{\citenamefont {S}\ and\ \citenamefont
  {Verma}({\natexlab{a}})}]{s_role_2022}%
  \BibitemOpen
  \bibfield  {author} {\bibinfo {author} {\bibfnamefont {T.}~\bibnamefont {S}}\
  and\ \bibinfo {author} {\bibfnamefont {A.~K.}\ \bibnamefont {Verma}},\
  }\bibfield  {title} {\bibinfo {title} {Role of extended coupling in
  bidirectional transport system},\ }\href
  {https://doi.org/10.1103/PhysRevE.106.014120} {\ \textbf {\bibinfo {volume}
  {106}},\ \bibinfo {pages} {014120} ({\natexlab{a}})},\ \bibinfo {note}
  {publisher: American Physical Society}\BibitemShut {NoStop}%
\bibitem [{\citenamefont {S}\ and\ \citenamefont
  {Verma}({\natexlab{b}})}]{s_multiple_2023}%
  \BibitemOpen
  \bibfield  {author} {\bibinfo {author} {\bibfnamefont {T.}~\bibnamefont {S}}\
  and\ \bibinfo {author} {\bibfnamefont {A.~K.}\ \bibnamefont {Verma}},\
  }\bibfield  {title} {\bibinfo {title} {Multiple reentrance transitions in
  exclusion process with finite reservoir},\ }\href
  {https://doi.org/10.1103/PhysRevE.107.044133} {\ \textbf {\bibinfo {volume}
  {107}},\ \bibinfo {pages} {044133} ({\natexlab{b}})},\ \bibinfo {note}
  {publisher: American Physical Society}\BibitemShut {NoStop}%
\bibitem [{\citenamefont {Bortz}\ \emph {et~al.}()\citenamefont {Bortz},
  \citenamefont {Kalos},\ and\ \citenamefont {Lebowitz}}]{bortz_new_1975}%
  \BibitemOpen
  \bibfield  {author} {\bibinfo {author} {\bibfnamefont {A.~B.}\ \bibnamefont
  {Bortz}}, \bibinfo {author} {\bibfnamefont {M.~H.}\ \bibnamefont {Kalos}},\
  and\ \bibinfo {author} {\bibfnamefont {J.~L.}\ \bibnamefont {Lebowitz}},\
  }\bibfield  {title} {\bibinfo {title} {A new algorithm for monte carlo
  simulation of ising spin systems},\ }\href
  {https://doi.org/10.1016/0021-9991(75)90060-1} {\ \textbf {\bibinfo {volume}
  {17}},\ \bibinfo {pages} {10}}\BibitemShut {NoStop}%
\bibitem [{\citenamefont {Sokal}()}]{sokal_monte_1997}%
  \BibitemOpen
  \bibfield  {author} {\bibinfo {author} {\bibfnamefont {A.}~\bibnamefont
  {Sokal}},\ }\bibfield  {title} {\bibinfo {title} {Monte carlo methods in
  statistical mechanics: Foundations and new algorithms},\ }in\ \href
  {https://doi.org/10.1007/978-1-4899-0319-8_6} {\emph {\bibinfo {booktitle}
  {Functional Integration: Basics and Applications}}},\ \bibinfo {editor}
  {edited by\ \bibinfo {editor} {\bibfnamefont {C.}~\bibnamefont
  {{DeWitt}-Morette}}, \bibinfo {editor} {\bibfnamefont {P.}~\bibnamefont
  {Cartier}},\ and\ \bibinfo {editor} {\bibfnamefont {A.}~\bibnamefont
  {Folacci}}}\ (\bibinfo  {publisher} {Springer {US}})\ pp.\ \bibinfo {pages}
  {131--192}\BibitemShut {NoStop}%
\bibitem [{\citenamefont {Sch{\"u}tz}(2001)}]{schutz2001exactly}%
  \BibitemOpen
  \bibfield  {author} {\bibinfo {author} {\bibfnamefont {G.~M.}\ \bibnamefont
  {Sch{\"u}tz}},\ }\bibfield  {title} {\bibinfo {title} {Exactly solvable
  models for many-body systems far from equilibrium},\ }in\ \href@noop {}
  {\emph {\bibinfo {booktitle} {Phase transitions and critical phenomena}}},\
  Vol.~\bibinfo {volume} {19}\ (\bibinfo  {publisher} {Elsevier},\ \bibinfo
  {year} {2001})\ pp.\ \bibinfo {pages} {1--251}\BibitemShut {NoStop}%
\bibitem [{\citenamefont {Biham}\ \emph {et~al.}()\citenamefont {Biham},
  \citenamefont {Middleton},\ and\ \citenamefont
  {Levine}}]{biham_self-organization_1992}%
  \BibitemOpen
  \bibfield  {author} {\bibinfo {author} {\bibfnamefont {O.}~\bibnamefont
  {Biham}}, \bibinfo {author} {\bibfnamefont {A.~A.}\ \bibnamefont
  {Middleton}},\ and\ \bibinfo {author} {\bibfnamefont {D.}~\bibnamefont
  {Levine}},\ }\bibfield  {title} {\bibinfo {title} {Self-organization and a
  dynamical transition in traffic-flow models},\ }\href
  {https://doi.org/10.1103/PhysRevA.46.R6124} {\ \textbf {\bibinfo {volume}
  {46}},\ \bibinfo {pages} {R6124}},\ \bibinfo {note} {publisher: American
  Physical Society}\BibitemShut {NoStop}%
\bibitem [{\citenamefont {de~Gier}\ and\ \citenamefont
  {Nienhuis}(1999)}]{de_gier_exact_1999}%
  \BibitemOpen
  \bibfield  {author} {\bibinfo {author} {\bibfnamefont {J.}~\bibnamefont
  {de~Gier}}\ and\ \bibinfo {author} {\bibfnamefont {B.}~\bibnamefont
  {Nienhuis}},\ }\bibfield  {title} {\bibinfo {title} {Exact stationary state
  for an asymmetric exclusion process with fully parallel dynamics},\ }\href
  {https://doi.org/10.1103/PhysRevE.59.4899} {\bibfield  {journal} {\bibinfo
  {journal} {Phys. Rev. E}\ }\textbf {\bibinfo {volume} {59}},\ \bibinfo
  {pages} {4899} (\bibinfo {year} {1999})}\BibitemShut {NoStop}%
\bibitem [{\citenamefont {Basu}\ \emph {et~al.}()\citenamefont {Basu},
  \citenamefont {Sidoravicius},\ and\ \citenamefont {Sly}}]{basu_last_2016}%
  \BibitemOpen
  \bibfield  {author} {\bibinfo {author} {\bibfnamefont {R.}~\bibnamefont
  {Basu}}, \bibinfo {author} {\bibfnamefont {V.}~\bibnamefont {Sidoravicius}},\
  and\ \bibinfo {author} {\bibfnamefont {A.}~\bibnamefont {Sly}},\ }\href
  {https://doi.org/10.48550/arXiv.1408.3464} {\bibinfo {title} {Last passage
  percolation with a defect line and the solution of the slow bond problem}},\
  \Eprint {https://arxiv.org/abs/1408.3464 [math-ph]} {1408.3464 [math-ph]}
  \BibitemShut {NoStop}%
\bibitem [{\citenamefont {Xiao}\ and\ \citenamefont
  {Bai}()}]{xiao_investigation_2013}%
  \BibitemOpen
  \bibfield  {author} {\bibinfo {author} {\bibfnamefont {S.}~\bibnamefont
  {Xiao}}\ and\ \bibinfo {author} {\bibfnamefont {J.-Y.}\ \bibnamefont {Bai}},\
  }\bibfield  {title} {\bibinfo {title} {Investigation of asymmetric simple
  exclusion processes with zoned inhomogeneity and on-ramp},\ }\href
  {https://doi.org/10.1142/S0217984913500620} {\ \textbf {\bibinfo {volume}
  {27}},\ \bibinfo {pages} {1350062}},\ \bibinfo {note} {publisher: World
  Scientific Publishing Co.}\BibitemShut {Stop}%
\bibitem [{\citenamefont {Daganzo}()}]{daganzo_cell_1994}%
  \BibitemOpen
  \bibfield  {author} {\bibinfo {author} {\bibfnamefont {C.~F.}\ \bibnamefont
  {Daganzo}},\ }\bibfield  {title} {\bibinfo {title} {The cell transmission
  model: A dynamic representation of highway traffic consistent with the
  hydrodynamic theory},\ }\href {https://doi.org/10.1016/0191-2615(94)90002-7}
  {\ \textbf {\bibinfo {volume} {28}},\ \bibinfo {pages} {269}}\BibitemShut
  {NoStop}%
\bibitem [{\citenamefont {Corwin}\ \emph {et~al.}()\citenamefont {Corwin},
  \citenamefont {Ferrari},\ and\ \citenamefont {Péché}}]{corwin_limit_2010}%
  \BibitemOpen
  \bibfield  {author} {\bibinfo {author} {\bibfnamefont {I.}~\bibnamefont
  {Corwin}}, \bibinfo {author} {\bibfnamefont {P.~L.}\ \bibnamefont
  {Ferrari}},\ and\ \bibinfo {author} {\bibfnamefont {S.}~\bibnamefont
  {Péché}},\ }\bibfield  {title} {\bibinfo {title} {Limit processes for
  {TASEP} with shocks and rarefaction fans},\ }\href
  {https://doi.org/10.1007/s10955-010-9995-7} {\ \textbf {\bibinfo {volume}
  {140}},\ \bibinfo {pages} {232}}\BibitemShut {NoStop}%
\bibitem [{\citenamefont {Liu}\ \emph {et~al.}()\citenamefont {Liu},
  \citenamefont {Xiao}, \citenamefont {Dong}, \citenamefont {Zhang},\ and\
  \citenamefont {Xiao}}]{liu_effect_2016}%
  \BibitemOpen
  \bibfield  {author} {\bibinfo {author} {\bibfnamefont {Y.}~\bibnamefont
  {Liu}}, \bibinfo {author} {\bibfnamefont {W.}~\bibnamefont {Xiao}}, \bibinfo
  {author} {\bibfnamefont {P.}~\bibnamefont {Dong}}, \bibinfo {author}
  {\bibfnamefont {Y.}~\bibnamefont {Zhang}},\ and\ \bibinfo {author}
  {\bibfnamefont {S.}~\bibnamefont {Xiao}},\ }\bibfield  {title} {\bibinfo
  {title} {The effect of single on-ramp with constrained resources on the
  density of asymmetric simple exclusion processes},\ }\href
  {https://doi.org/10.1016/j.rser.2016.05.038} {\ \textbf {\bibinfo {volume}
  {62}},\ \bibinfo {pages} {815}}\BibitemShut {NoStop}%
\bibitem [{\citenamefont {Dong}\ \emph {et~al.}({\natexlab{b}})\citenamefont
  {Dong}, \citenamefont {Schmittmann},\ and\ \citenamefont
  {Zia}}]{dong_inhomogeneous_2007}%
  \BibitemOpen
  \bibfield  {author} {\bibinfo {author} {\bibfnamefont {J.~J.}\ \bibnamefont
  {Dong}}, \bibinfo {author} {\bibfnamefont {B.}~\bibnamefont {Schmittmann}},\
  and\ \bibinfo {author} {\bibfnamefont {R.~K.~P.}\ \bibnamefont {Zia}},\
  }\bibfield  {title} {\bibinfo {title} {Inhomogeneous exclusion processes with
  extended objects: The effect of defect locations},\ }\href
  {https://doi.org/10.1103/PhysRevE.76.051113} {\ \textbf {\bibinfo {volume}
  {76}},\ \bibinfo {pages} {051113} ({\natexlab{b}})},\ \bibinfo {note}
  {publisher: American Physical Society}\BibitemShut {NoStop}%
\bibitem [{\citenamefont {Schmit}\ \emph {et~al.}(2009)\citenamefont {Schmit},
  \citenamefont {Kamber},\ and\ \citenamefont {Kondev}}]{schmit2009lattice}%
  \BibitemOpen
  \bibfield  {author} {\bibinfo {author} {\bibfnamefont {J.~D.}\ \bibnamefont
  {Schmit}}, \bibinfo {author} {\bibfnamefont {E.}~\bibnamefont {Kamber}},\
  and\ \bibinfo {author} {\bibfnamefont {J.}~\bibnamefont {Kondev}},\
  }\bibfield  {title} {\bibinfo {title} {Lattice model of diffusion-limited
  bimolecular chemical reactions in confined environments},\ }\href@noop {}
  {\bibfield  {journal} {\bibinfo  {journal} {Physical review letters}\
  }\textbf {\bibinfo {volume} {102}},\ \bibinfo {pages} {218302} (\bibinfo
  {year} {2009})}\BibitemShut {NoStop}%
\bibitem [{\citenamefont {Schadschneider}\ and\ \citenamefont
  {Schreckenberg}()}]{schadschneider_cellular_1993}%
  \BibitemOpen
  \bibfield  {author} {\bibinfo {author} {\bibfnamefont {A.}~\bibnamefont
  {Schadschneider}}\ and\ \bibinfo {author} {\bibfnamefont {M.}~\bibnamefont
  {Schreckenberg}},\ }\bibfield  {title} {\bibinfo {title} {Cellular automation
  models and traffic flow},\ }\href
  {https://doi.org/10.1088/0305-4470/26/15/011} {\ \textbf {\bibinfo {volume}
  {26}},\ \bibinfo {pages} {L679}}\BibitemShut {NoStop}%
\bibitem [{\citenamefont {Krug}()}]{krug_boundary-induced_1991}%
  \BibitemOpen
  \bibfield  {author} {\bibinfo {author} {\bibfnamefont {J.}~\bibnamefont
  {Krug}},\ }\bibfield  {title} {\bibinfo {title} {Boundary-induced phase
  transitions in driven diffusive systems},\ }\href
  {https://doi.org/10.1103/PhysRevLett.67.1882} {\ \textbf {\bibinfo {volume}
  {67}},\ \bibinfo {pages} {1882}},\ \bibinfo {note} {publisher: American
  Physical Society}\BibitemShut {NoStop}%
\end{thebibliography}
\end{document}